\documentclass[a4paper,11pt]{article}

\usepackage[margin=2.5cm]{geometry}
\usepackage{amsmath}
\usepackage{graphicx}
\usepackage{placeins}
\usepackage[numbers]{natbib}
\usepackage{amssymb}
\usepackage{mathtools}
\usepackage{bm}
\usepackage{amsthm}
\usepackage{subcaption}
\usepackage{multirow}
\usepackage{booktabs}
\usepackage{siunitx}
\usepackage{hyperref}
\hypersetup{
  colorlinks=true,
  linkcolor=red,
  citecolor=blue,
  urlcolor=black
}
\usepackage[capitalise]{cleveref}

\theoremstyle{plain}
\newtheorem{theorem}{Theorem}[section]

\newtheorem{proposition}[theorem]{Proposition}

\theoremstyle{definition}

\theoremstyle{remark}
\newtheorem{remark}[theorem]{Remark}

\newcommand{\avg}[1]{\{\!\{#1\}\!\}}
\newcommand{\jump}[1]{[\![#1]\!]}
\newcommand{\avgbd}[1]{\{\!\{#1\}\!\}_{\Gamma}}
\newcommand{\jumpbd}[1]{[\![#1]\!]_{\Gamma}}
\newcommand{\rmd}{\mathrm{d}}
\newcommand{\rmR}{\mathrm{R}}
\newcommand{\rmG}{\mathrm{G}}
\newcommand{\rmRe}{\mathrm{Re}}
\newcommand{\rmWs}{\mathrm{Ws}}
\newcommand{\rmLs}{\mathrm{Ls}}
\newcommand{\rmLk}{\mathrm{Lk}}
\newcommand{\rmLd}{\mathrm{Ld}}
\newcommand{\rmLb}{\mathrm{Lb}}
\newcommand{\la}{\langle}
\newcommand{\ra}{\rangle}
\newcommand{\Th}{\mathcal{T}_{h}}
\newcommand{\ED}{\mathcal{E}_{h}^{\Gamma}}
\newcommand{\EI}{\mathcal{E}_{h}^{I}}
\newcommand{\EV}{\mathcal{V}_{h}^{\Gamma}}

\allowdisplaybreaks[1]

\title{A thermodynamically consistent variational formulation and structure-preserving finite element methods for non-isothermal liquid--vapor flows with moving contact lines}

\author{\normalsize
Junkai Wang
\thanks{School of Mathematical Sciences, Chengdu University of Technology, Chengdu, Sichuan 610059, China.
(\href{mailto:jkwang@cdut.edu.cn}{\nolinkurl{jkwang@cdut.edu.cn}}).
ORCID: \url{https://orcid.org/0000-0002-1208-7348}.
}
\quad and\quad
Ting Xiao
\thanks{School of Mathematics, Sichuan University, Chengdu, Sichuan 610065, China.
(\href{mailto:xiaoting0122@163.com}{\nolinkurl{xiaoting0122@163.com}}).
}
\quad and\quad
Qiaolin He
\thanks{Corresponding author. School of Mathematics, Sichuan University, Chengdu, Sichuan 610065, China.
(\href{mailto:qlhejenny@scu.edu.cn}{\nolinkurl{qlhejenny@scu.edu.cn}}).
ORCID: \url{https://orcid.org/0000-0002-4182-0473}.
}
}

\date{}
\begin{document}
\begingroup
\makeatletter
\let\author@fnsymbol\@fnsymbol
\def\@fnsymbol#1{\ifnum#1=1 \author@fnsymbol{2}\else
  \ifnum#1=2 \author@fnsymbol{1}\else\author@fnsymbol{#1}\fi\fi}
\makeatother
\maketitle
\endgroup
\begin{abstract}
\noindent
In this paper, we present a variational formulation and structure-preserving finite element methods for non-isothermal liquid--vapor flows with moving contact lines.
Based on the dynamic van der Waals theory with gradient contributions and surface thermodynamics, the governing equations and hydrodynamic boundary conditions are derived from a generalized Hamilton's principle for nonequilibrium thermodynamics.
Two equivalent variational formulations based on internal energy and Helmholtz free energy are developed, using entropy density and temperature, respectively, as the primary thermal variable.
Using the resulting weak forms, compatible spatial operators and discrete thermodynamic chain rules, we construct two thermodynamically consistent fully discrete finite element schemes.
We rigorously prove the following properties for both fully discrete schemes: mass conservation, bound preservation for the number density, positivity of the temperature, an exact total energy balance incorporating external work and heat exchange, and a temperature-weighted weak entropy inequality with nonnegative bulk and boundary entropy production.
Numerical experiments in two and three dimensions assess accuracy and structure preservation, with applications to non-isothermal droplet wetting, pool boiling and droplet ejection.


\end{abstract}

\section{Introduction}
\label{sec:introduction}

Non-isothermal liquid--vapor flows with moving contact lines occur in natural and engineering processes, such as droplet evaporation, spray cooling, pool boiling, coating and inkjet printing \cite{zang2019evaporation,xu2014single}.
The evolution of the liquid--vapor interface couples fluid motion, heat transfer and phase change, while surface wettability and temperature influence the motion of the contact line \cite{anderson1995spreading,bonn2009wetting}.
These processes involve energy exchange and dissipation both in the fluid and at the solid boundary.
A thermodynamically consistent description must account for these processes in accordance with the first and second laws of thermodynamics, ensuring total energy balance and nonnegative entropy production.
An important challenge is to incorporate these requirements into the mathematical model and preserve them after numerical discretization.

Diffuse-interface models describe the liquid--vapor interface as a thin transition region in which the fluid state varies continuously \cite{anderson1998diffuse}.
The thermodynamic basis of this description goes back to the van der Waals theory of capillarity \cite{van1979thermodynamic} and was further developed by Cahn and Hilliard \cite{cahn1958free}.
For a one-component fluid, the number density distinguishes the liquid and vapor phases, without introducing an additional phase variable.
The dynamic van der Waals theory (DVDWT), developed by Onuki \cite{onuki2002phase,onuki2005dynamic,onuki2007dynamic}, incorporates density gradients into the internal energy and entropy and couples these thermodynamic relations to compressible hydrodynamics.
The resulting model describes phase separation and phase change with nonuniform temperatures in liquid--vapor flows.
The gradient contributions enter both the reversible stress and entropy transport and must be coupled consistently with energy balance and nonnegative entropy production.

At the solid surface, the bulk description must be supplemented by boundary conditions for wetting, slip and heat transfer.
For contact line motion, the classical sharp-interface model of immiscible flow with no slip leads to a stress singularity and nonintegrable viscous dissipation \cite{huh1971hydrodynamic}.
Using a phase-field model, Qian, Wang, and Sheng \cite{qian2006variational} derived a relaxation boundary condition and the generalized Navier boundary condition (GNBC) through a variational treatment of surface free energy and dissipation.
For one-component non-isothermal liquid--vapor flows described by DVDWT, Xu and Qian \cite{xu2010contact} derived boundary conditions to describe number density relaxation and velocity slip at solid surfaces.
Xu, Liu, and Qian \cite{liu2012hydrodynamic} subsequently developed hydrodynamic boundary conditions for non-isothermal and heterogeneous solid surfaces.
These conditions account for surface wettability, surface heat conduction and heat exchange across a finite thermal resistance, allowing the fluid temperature at the boundary to differ from the solid temperature \cite{barrat2003kapitza}.
More recently, Qiao et al.\ \cite{qiao2026unified} extended the Onsager variational approach to flows with phase transitions and wetting, including inertial and thermal effects, and developed an associated lattice Boltzmann method.

Preserving thermodynamic consistency after both spatial and temporal discretization presents a further challenge.
Liu et al.\ \cite{liu2015liquid} developed an entropy-stable numerical formulation for liquid--vapor phase transitions and applied it to boiling.
Kou and Sun \cite{kou2018thermodynamically2,kou2018entropy} developed non-isothermal diffuse-interface models for single- and multi-component fluids using the Peng--Robinson equation of state, with gradient coefficients that may depend on temperature.
Using convex--concave splitting and an auxiliary velocity, they constructed time discretizations satisfying an energy balance and, under suitable concavity assumptions with respect to temperature, an entropy inequality.
Guo, Qiao, and Xu \cite{guo2026linear} proposed a linear, decoupled finite difference scheme for DVDWT based on an equivalent reformulation and component-wise convex splitting.
For the gravity-free problem with homogeneous boundary conditions, a fully discrete entropy-production law is established under positivity assumptions.
For non-isothermal moving contact lines, Wang and He \cite{wang2026jcp} developed a thermodynamically consistent temperature formulation of DVDWT and a linear, decoupled, entropy-stable scheme based on the multiple scalar auxiliary variables approach.
They established a modified entropy law for the time discretization and proved mass conservation and number density bounds for the fully discrete scheme.
However, the proposed time discretization does not preserve the total energy balance, and a fully discrete entropy inequality is not established for this scheme.
In this context, one of the objectives of this paper is to construct thermodynamically consistent fully discrete schemes that incorporate the gradient contributions and surface thermodynamics and preserve physical bounds.

Variational principles provide a framework for deriving thermodynamically consistent models and numerical methods.
In classical mechanics, Hamilton's principle yields the equations of motion for conservative systems from the stationarity of an action integral \cite{landau1969mechanics}.
To account for irreversible processes in nonequilibrium thermodynamics, Gay-Balmaz and Yoshimura \cite{gay2017discrete,gay2017lagrangian} extended Hamilton's principle by introducing thermodynamic displacements and paired phenomenological and variational constraints.
The variational framework incorporates viscosity and heat conduction, and the free energy formulation \cite{gay2017freeenergy} uses temperature instead of entropy as the thermal state variable.
Building on this framework, Gawlik and Gay-Balmaz \cite{gawlik2024variational} constructed thermodynamically consistent finite element methods for heat-conducting viscous fluids under insulated, prescribed heat flux or prescribed temperature boundary conditions.
The methods preserve mass and the energy balance, and satisfy a temperature-weighted weak entropy inequality at the fully discrete level.
Further developments include visco-resistive magnetohydrodynamics with thermoelectric coupling \cite{gawlik2025structure}.
Extending this framework to DVDWT with moving contact lines requires the boundary terms arising from density gradients to be coupled consistently with number density relaxation, velocity slip and heat transfer.
Moreover, the continuous equivalence of the entropy and temperature formulations relies on thermodynamic chain rules that are not automatically preserved by numerical discretization.

Based on this variational framework, we develop two constrained variational formulations for DVDWT with moving contact lines, using internal energy and Helmholtz free energy.
The two formulations use entropy density and temperature, respectively, as the thermal state variable.
We extend the generalized Hamilton's principle to include density gradients and surface thermodynamics, and introduce a reversible entropy flux and an associated stress to couple the momentum and thermal equations consistently with the energy and entropy balances.
We specify dissipative constitutive relations to ensure nonnegative entropy production in the bulk and on the boundary.
We note that both formulations yield the same governing equations and hydrodynamic boundary conditions.
The entropy and temperature representations define the $S$- and $T$-models, respectively.

From the dimensionless weak forms, we construct two nonlinear fully discrete finite element schemes, denoted by the $S$- and $T$-schemes.
Both schemes use discontinuous Galerkin (DG) spaces for the scalar variables and $\bm H(\operatorname{div},\Omega)$-conforming finite elements for the velocity, with a DG-type interior penalty treatment of viscosity \cite{cockburn2007note}.
Compatible spatial operators and exact discrete chain rules \cite{mclachlan1999geometric} are designed to preserve the cancellation of energy exchange terms between the mechanical and thermal equations, including boundary contributions.
Both schemes conserve mass locally and globally, satisfy an exact total energy balance including external work and wall heat exchange, and obey a temperature-weighted weak entropy inequality with nonnegative bulk and boundary production.
A logit representation preserves the number density bounds, while entropy-based temperature recovery in the $S$-scheme and a logarithmic temperature variable in the $T$-scheme both ensure positive temperatures \cite{huang2022bound}.

Numerical experiments in two and three dimensions assess the accuracy, number density bounds and temperature positivity of both schemes.
We then simulate non-isothermal droplet wetting, pool boiling and pulse-driven droplet ejection, examining the effects of surface wettability and thermal conditions.
The results also demonstrate mass conservation, total energy balance and nonnegative entropy production.

The remainder of the paper is organized as follows.
In \cref{sec:section}, we develop the two variational formulations, derive the governing equations and hydrodynamic boundary conditions, and establish the first and second laws of thermodynamics.
\cref{sec:The mathematical model} presents the dimensionless models and weak forms, and establishes mass conservation, number density bounds, temperature positivity and thermodynamic consistency at the continuous level.
In \cref{sec:Numerical methods}, we construct the two fully discrete schemes and prove the corresponding discrete properties.
\cref{sec:num_ex} presents numerical tests of accuracy and physical properties, followed by simulations of representative non-isothermal flows.
Finally, \cref{sec:conclusion} summarizes the main results and outlines future work.
Detailed proofs of the two variational formulations are provided in Appendix~\ref{app:variational-proofs}.

\section{A thermodynamically consistent variational formulation}
\label{sec:section}

In this section, we develop two equivalent variational formulations for the DVDWT model, using entropy density and temperature, respectively, as the thermal state variable.
Building on the extension of Hamilton's principle to nonequilibrium thermodynamics \cite{gay2017lagrangian,gay2017freeenergy,gawlik2024variational,gawlik2025structure}, we incorporate gradient contributions and boundary conditions into the variational framework.
We first review the bulk and surface thermodynamic relations.
Then we present the entropy-based principle and its counterpart based on free energy, incorporating the reversible coupling associated with density gradients to derive the bulk governing equations and hydrodynamic boundary conditions.
Finally, we specify the dissipative constitutive relations and establish the energy balance law and nonnegativity of entropy production both in the bulk and on the boundary.

\subsection{Thermodynamic preliminaries}
\label{subsec:Thermodynamic variables}

We summarize the thermodynamic relations underlying the DVDWT \cite{onuki2005dynamic,onuki2007dynamic}.
Let $\Omega\subset\mathbb R^d$, $d=2,3$, be a fixed, bounded, connected domain with smooth boundary $\Gamma\coloneqq\partial\Omega$ and outward unit normal $\bm\nu$.

\paragraph{Bulk thermodynamics.}
We consider a homogeneous one-component van der Waals fluid and let $n$, $e$, $s$, $T$, $p$ and $\mu$ denote the number density, internal energy density, entropy per particle, temperature, pressure and chemical potential per particle, respectively.
The Euler, Gibbs and Gibbs--Duhem relations read \cite{firoozabadi1999thermodynamics,onuki2002phase,onuki2005dynamic}
\begin{equation}\label{eq:threemaineq}
e-Tns+p-n\mu=0,\qquad
\rmd(ns)=\frac{1}{T}\rmd e-\frac{\mu}{T}\rmd n,\qquad
-ns\rmd T+\rmd p-n\rmd\mu=0.
\end{equation}
As in phase-field models, the number density $n$ distinguishes the liquid and vapor phases.
Choosing $n$ and $T$ as the independent state variables, the Helmholtz free energy density $f(n,T)\coloneqq e-Tns$ satisfies $\rmd f=-ns\rmd T+\mu\rmd n$.
With $0<v_0n<1$ and $T>0$, we have \cite{onuki2007dynamic,xu2012droplet,xu2012thermal}
\begin{equation*}
f=\frac{dnk_BT}{2}
-k_BTn\ln\Bigl(
\Bigl(\frac{k_BT}{\epsilon}\Bigr)^{d/2}
\Bigl(\frac{1}{v_0n}-1\Bigr)\Bigr)
-\epsilon v_0n^2,
\end{equation*}
where $k_B$ is the Boltzmann constant, $v_0$ is the molecular volume, $\epsilon>0$ characterizes the attractive interaction and $d$ is the spatial dimension.
The corresponding internal energy density, entropy per particle, and pressure are
\begin{equation*}
e=\frac{dnk_BT}{2}-\epsilon v_0n^2,\qquad
s=k_B\ln\Bigl(
\Bigl(\frac{k_BT}{\epsilon}\Bigr)^{d/2}
\Bigl(\frac{1}{v_0n}-1\Bigr)\Bigr),\qquad
p=\frac{nk_BT}{1-v_0n}-\epsilon v_0n^2.
\end{equation*}
The chemical potential per particle is therefore
\begin{equation*}
\begin{aligned}
\mu(n,T)
&=\partial_n f
=\partial_n e-T\partial_n(ns)\\
&=k_BT\Bigl(
\frac{d}{2}
-\ln\Bigl(
\Bigl(\frac{k_BT}{\epsilon}\Bigr)^{d/2}
\Bigl(\frac{1}{v_0n}-1\Bigr)\Bigr)
+\frac{1}{1-v_0n}
\Bigr)
-2\epsilon v_0n,
\end{aligned}
\end{equation*}
where $\partial_\psi\phi\coloneqq\partial\phi/\partial\psi$ for a function $\phi$ of several variables, with the other independent variables held fixed.
The DVDWT introduces gradient contributions for inhomogeneous fluids as \cite{onuki2007dynamic}
\begin{equation}\label{eq:feSdefine}
    \hat e\coloneqq e+\frac{K(n)}{2}|\nabla n|^2,\qquad
    \hat S\coloneqq ns-\frac{C(n)}{2}|\nabla n|^2,\qquad
    \hat f\coloneqq\hat e-T\hat S=f+\frac{M(n,T)}{2}|\nabla n|^2,
\end{equation}
in which $M(n,T)\coloneqq K(n)+C(n)T$.
When $K$ and $C$ are nonnegative, the gradient terms increase the internal energy and decrease the entropy due to the inhomogeneity of number density.

Let $\mathcal S_b\coloneqq\int_\Omega\hat S\rmd\bm x$ be the bulk entropy.
We regard $\mathcal S_b$ as a functional of $(n,\hat e)$, whose bulk variational derivatives determine the temperature and generalized chemical potential:
\begin{equation}\label{eq:muTdefine}
  \frac{1}{T}
  =\Bigl(\frac{\delta\mathcal S_b}{\delta\hat e}\Bigr)_n,\qquad
  \hat\mu
  \coloneqq-T\Bigl(\frac{\delta\mathcal S_b}{\delta n}\Bigr)_{\hat e}
    =\mu+\frac{\partial_n M}{2}|\nabla n|^2
        -T\operatorname{div}\Bigl(\frac{M}{T}\nabla n\Bigr).
\end{equation}
The subscripts on the variational derivatives indicate the fields held fixed.
Indeed, \eqref{eq:threemaineq} and \eqref{eq:feSdefine} yield
\begin{equation*}
    \delta\mathcal S_b
    =\int_\Omega\Bigl(\frac{\delta\hat e}{T}-\frac{\hat\mu}{T}\delta n\Bigr)\rmd \bm x
    -\int_\Gamma\frac{M}{T}(\nabla n\cdot\bm{\nu})\delta n\rmd A.
\end{equation*}
In the absence of external forces, stationarity of $\mathcal S_b$ under interior variations at fixed particle number $N\coloneqq\int_\Omega n\rmd\bm x$ and internal energy $\mathcal E_b\coloneqq\int_\Omega\hat e\rmd\bm x$ yields the bulk equilibrium conditions: the spatial homogeneity of $T$ and $\hat\mu$ in $\Omega$ \cite{xu2010contact,wang2026jcp}.
At a fixed, spatially homogeneous temperature $T$, minimizing $\mathcal F_b\coloneqq\mathcal E_b-T\mathcal S_b =\int_\Omega\hat f\rmd\bm x$ at fixed $N$ yields the same bulk equilibrium condition on $\hat\mu$.
Furthermore, the generalized pressure is defined by the generalized Euler relation
\begin{equation}\label{eq:generalizedEuler}
  \hat e-T\hat S+\hat p-n\hat\mu=0,
\end{equation}
from which we have $\hat p=p+(n\partial_n M-M)|\nabla n|^2/2 -nT\operatorname{div}(M\nabla n/T)$ \cite{onuki2002phase}.

\paragraph{Bulk balance laws.}
The dynamics of non-isothermal liquid--vapor flows based on the DVDWT are described by three basic physical quantities: number density $n(\bm x,t)$, fluid velocity $\bm u(\bm x,t)$, and temperature $T(\bm x,t)$, where $\bm x\in\overline{\Omega}\coloneqq\Omega\cup\Gamma$ and $t\geq 0$ is time.
We define the mass density by $\rho\coloneqq mn$, where $m$ is the particle mass assumed to be constant, and the total energy density by $e_{\text{tot}}\coloneqq\hat e+\rho|\bm u|^2/2$.
The particle number, momentum, and total energy balances read \cite{onuki2007dynamic}
\begin{subequations}\label{eq:balance}
  \begin{align}
    \partial_t n+\operatorname{div}(n\bm u)
      &=0,\label{eq:massbalance}\\
    \partial_t(\rho\bm u)+\operatorname{div}(\rho\bm u\otimes\bm u)
      &=\operatorname{div}\hat{\bm\sigma}-\rho g\bm e_d,\label{eq:NS}\\
    \partial_t e_{\text{tot}}+\operatorname{div}(e_{\text{tot}}\bm u)
      &=\operatorname{div}(\hat{\bm\sigma}\bm u)
        -\operatorname{div}\bm q-\rho g u_d.\label{eq:energy}
  \end{align}
\end{subequations}
Here $g$ is the gravitational acceleration, $\bm e_d=(0,\ldots,0,1)^{\mathsf T}\in\mathbb R^d$ is the upward
unit vector, $u_d\coloneqq\bm u\cdot\bm e_d$, and $\bm q$ is the heat flux.
The total stress is $\hat{\bm\sigma}\coloneqq-\bm\Pi+\bm\sigma$, where $\bm\Pi$ and $\bm\sigma$ are reversible and viscous stress tensors, respectively.
The total stress is symmetric under the constitutive relations specified below.
Substituting the definition of $e_{\text{tot}}$ into \eqref{eq:energy} and subtracting the kinetic energy balance, we obtain the internal energy equation \cite{onuki2007dynamic,kou2018thermodynamically2,xiao2026thermodynamically}:
\begin{equation}\label{eq:internal}
  \partial_t\hat e+\operatorname{div}(\hat e\bm u)
    =\hat{\bm\sigma}:\nabla\bm u-\operatorname{div}\bm q.
\end{equation}

\paragraph{Surface thermodynamics.}
We assume that $\Gamma$ is a rigid, impermeable and insoluble solid boundary, with prescribed tangential velocity $\bm u_w$ and temperature $T_w>0$.
For a scalar field $\phi$ and a tangential vector $\bm w$ on $\Gamma$, the normal derivative, tangential projector, surface gradient, and surface divergence are defined by
\begin{equation*}
  \nabla_\nu\phi\coloneqq\bm\nu\cdot\nabla\phi,\qquad
  \bm P_\tau\coloneqq\bm I-\bm\nu\otimes\bm\nu,\qquad
  \nabla_\tau\phi\coloneqq\bm P_\tau\nabla\phi,\qquad
  \operatorname{div}_\Gamma\bm w\coloneqq\nabla_\tau\cdot\bm w.
\end{equation*}
The impermeability condition gives $\bm u\cdot\bm\nu=0$ on $\Gamma$.
The tangential fluid velocity on $\Gamma$ is $\bm u_\tau\coloneqq\bm P_\tau\bm u$, and the slip velocity is defined as $\bm u_\tau^{\mathrm{slip}}\coloneqq\bm u_\tau-\bm u_w$ \cite{qian2006molecular}.

We assume that the surface internal energy and entropy densities, $e_s(n)$ and $\sigma_s(n)$, depend only on the number density $n|_\Gamma$ \cite{xu2012droplet}.
We then define the surface internal energy, entropy and Helmholtz free energy by $\mathcal E_s\coloneqq\int_\Gamma e_s\rmd A$, $\mathcal S_s\coloneqq\int_\Gamma\sigma_s\rmd A$ and $\mathcal F_s\coloneqq\int_\Gamma f_s\rmd A$, where $f_s(n,T)\coloneqq e_s-T\sigma_s$ is the surface Helmholtz free energy density.
Notice that the fluid temperature $T|_\Gamma$ need not equal the wall temperature $T_w$.
When the fluid temperature on $\Gamma$ is homogeneous, we also have $\mathcal F_s=\mathcal E_s-T\mathcal S_s$.
Since $\sigma_s=-\partial_T f_s$, the surface Gibbs relations read
\begin{equation}\label{eq:Gibbs-typeequation}
  \rmd f_s=-\sigma_s\rmd T
            +\partial_n f_s\rmd n,\qquad
  \rmd\sigma_s=\frac{1}{T}\rmd e_s
            -\frac{1}{T}\partial_n f_s\rmd n.
\end{equation}
Furthermore, we define a quantity $L\coloneqq\partial_n f_s+M\nabla_\nu n$ on $\Gamma$.
Minimizing the total Helmholtz free energy at fixed $N$ and a fixed, spatially homogeneous $T$ yields the equilibrium boundary condition: $L=0$ on $\Gamma$.
A nonzero $L$ yields the relaxation of $n$ at the solid boundary \cite{qian2006molecular,xu2010contact}.
The hydrodynamic boundary conditions will be derived in \cref{subsec:closure}.

\paragraph{Thermal variables.}
We use two thermal coordinates for the variational formulations of the same constitutive model: (\romannumeral1) $\hat e=\hat e(n,\nabla n,\hat S)$ in the entropy formulation and (\romannumeral2) $\hat f=\hat f(n,\nabla n,T)$ in the free energy formulation.
Using the DVDWT equation of state, we have $\partial_T\hat S(n,\nabla n,T)=-\partial_T^2\hat f=dnk_B/(2T)>0$.
Thus, for fixed $(n,\nabla n)$ with $0<v_0n<1$ and $T>0$, the change of thermal variable between $\hat S$ and $T$ is invertible.
With $\hat S=\hat S(n,\nabla n,T)$, the partial Legendre relation is
\begin{equation*}
  \hat f(n,\nabla n,T)
  =
  \hat e(n,\nabla n,\hat S)-T\hat S,
  \qquad
  \hat S=-\partial_T\hat f.
\end{equation*}
From these thermodynamic relations, we obtain $T=\partial_{\hat S}\hat e$ and $M\nabla n=\partial_{\nabla n}\hat e$.
To prepare for the variational derivation below, we define the generalized chemical potential associated with the internal energy as
\begin{equation*}
  \tilde\mu\coloneqq\partial_n\hat e-\operatorname{div}(M\nabla n),\qquad
  \partial_n\hat e=\mu+\frac{\partial_n M}{2}|\nabla n|^2,
\end{equation*}
which is also the bulk variational derivative of $\mathcal F_b$ with respect to $n$ at a fixed temperature field: $\tilde\mu=\partial_n\hat f-\operatorname{div}(M\nabla n)$.
Notice that $\tilde\mu$ and the generalized chemical potential $\hat\mu$ defined through entropy in \eqref{eq:muTdefine} generally differ when temperature is nonuniform: $\hat\mu=\tilde\mu+M\nabla n\cdot\nabla T/T$.
Furthermore, the reversible stress tensor in \eqref{eq:balance} is defined as
\begin{equation*}
  \bm\Pi\coloneqq\widetilde{\bm\Pi}+\frac{nM\nabla n\cdot\nabla T}{T}\bm I,\qquad
  \widetilde{\bm\Pi}
    \coloneqq(n\tilde\mu+\hat S T-\hat e)\bm I+M\nabla n\otimes\nabla n.
\end{equation*}
Using \eqref{eq:generalizedEuler}, we also have $\bm\Pi=\hat p\bm I+M\nabla n\otimes\nabla n$.
The chain rule gives \cite{wang2026jcp}
\begin{equation}\label{eq:rev stress identity}
  \operatorname{div}\bm\Pi
  =n\nabla\hat\mu+\hat S\nabla T
   +\frac{M\nabla n\cdot\nabla T}{T}\nabla n,
  \qquad
  \operatorname{div}\widetilde{\bm\Pi}
  =n\nabla\tilde\mu+\hat S\nabla T.
\end{equation}
The identity $\operatorname{div}(\bm\Pi/T) =n\nabla(\hat\mu/T)-\hat e\nabla(1/T)$ also follows \cite{onuki2007dynamic}.
 
\paragraph{Thermodynamic transport identity.}
We define the material derivative and conservative transport operator by
\begin{equation*}
  \mathrm D_t\phi\coloneqq\partial_t\phi+\bm u\cdot\nabla\phi,\qquad
  \overline{\mathrm D}_t\phi
    \coloneqq\partial_t\phi+\operatorname{div}(\phi\bm u).
\end{equation*}
The mass balance equation \eqref{eq:massbalance} can be written as $\overline{\mathrm D}_tn=0$ or $\mathrm D_tn=-n\operatorname{div}\bm u$.
Writing the internal energy density as $\hat e=\hat e(n,\nabla n,\hat S)$ and using the identity $\mathrm D_t(\nabla n)=\nabla(\mathrm D_tn) -(\nabla\bm u)^{\mathsf T}\nabla n$, we have
\begin{align*}
  \mathrm D_t\hat e
  =&T\mathrm D_t\hat S
+\partial_n\hat e\mathrm D_tn
+M\nabla n\cdot\mathrm D_t(\nabla n)\\
  =&T\mathrm D_t\hat S
      +\tilde\mu\mathrm D_tn
      -(M\nabla n\otimes\nabla n):\nabla\bm u
      +\operatorname{div}(M\nabla n\mathrm D_tn),
\end{align*}
where we have used $M\nabla n\cdot(\nabla\bm u)^{\mathsf T}\nabla n=(M\nabla n\otimes\nabla n):\nabla\bm u$.
Using $\mathrm D_tn=-n\operatorname{div}\bm u$ and the definitions of $\overline{\mathrm D}_t \phi$ and $\widetilde{\bm\Pi}$, we rewrite the above relation as
\begin{equation}\label{eq:energy-chain}
\begin{aligned}
  \overline{\mathrm D}_t\hat e
    =&T\overline{\mathrm D}_t\hat S+
(\hat e-T\hat S-n\tilde\mu)\operatorname{div}\bm u
-(M\nabla n\otimes\nabla n):\nabla \bm u
+\operatorname{div}(M\nabla n\mathrm D_tn)\\
    =&T\overline{\mathrm D}_t\hat S
       -\widetilde{\bm\Pi}:\nabla\bm u
       +\operatorname{div}(M\nabla n\mathrm D_tn).
\end{aligned}
\end{equation}
This identity motivates the reversible pair of entropy flux and stress introduced below.
In \cref{subsec:free-energy-principle}, we will verify that the derived entropy equation, together with \eqref{eq:energy-chain} and the reversible pair, recovers the internal energy balance \eqref{eq:internal} without introducing an additional flux of gradient energy.

\subsection{Entropy-based variational principle}
\label{subsec:principle}

We extend the Eulerian variational framework for nonequilibrium thermodynamics \cite{gay2017lagrangian,gawlik2024variational} to construct an entropy-based principle incorporating gradient contributions and boundary conditions.
This framework generalizes the Lagrange--d'Alembert principle through paired phenomenological and variational constraints \cite{gay2019perspective}.

Consider smooth fields on $\overline{\Omega}\times[t_0,t_1]$, with $0<v_0n<1$ and $T>0$.
Let $\operatorname{Diff}(\Omega)$ denote the group of smooth diffeomorphisms of $\overline{\Omega}$ onto itself that preserve the boundary $\Gamma$ as a set.
The fluid motion is described by $\bm x=\mathcal{X}(\bm X,t)$, where $\bm X\in\overline{\Omega}$ is the reference position of a particle and $\bm x$ is the position at time $t$.
For each fixed $t\in[t_0,t_1]$, we denote $\mathcal X_t\coloneqq\mathcal X(\cdot,t)$.
Thus $t\mapsto\mathcal{X}_t$ is a smooth curve in $\operatorname{Diff}(\Omega)$, with $\mathcal{X}_{t_0}=\operatorname{id}$.
Notice that the boundary is preserved as a set rather than pointwise fixed, allowing tangential motion along it.

Mass conservation implies that the Eulerian number density is the push-forward of the fixed reference number density:
\begin{equation*}
  n(\bm x,t)\rmd\bm x=(\mathcal{X}_t)_\ast(n_{\mathrm{ref}}\rmd\bm X),\qquad
  n(\mathcal{X}_t(\bm X),t)J_{\mathcal{X}}(\bm X,t)=n_{\mathrm{ref}}(\bm X).
\end{equation*}
Here $n_{\mathrm{ref}}$ is the reference number density, and $(\mathcal{X}_t)_\ast$ denotes the push-forward by the flow map $\mathcal{X}_t$.
The second relation is the coordinate expression, with $J_{\mathcal{X}}\coloneqq\det\nabla_{\bm X}\mathcal{X}_t>0$.
In contrast, the entropy density $\hat S$ is varied independently, and its evolution follows from the constrained variational principle below.

We consider a smooth family of perturbed motions $\mathcal{X}_{t}^\varepsilon\in\operatorname{Diff}(\Omega)$ with $\mathcal{X}_t^0=\mathcal{X}_t$, the same $n_{\mathrm{ref}}$, and fixed temporal endpoints $\mathcal{X}_{t_i}^\varepsilon=\mathcal{X}_{t_i}$, $i=0,1$.
Define $\delta\mathcal{X}_t\coloneqq\left.\partial_\varepsilon \mathcal{X}_{t}^\varepsilon\right|_{\varepsilon=0}$.
The Eulerian velocity and virtual displacement are then given by \cite{gay2019perspective}
\begin{equation*}
 \bm u=\partial_t\mathcal{X}_t\circ\mathcal{X}_t^{-1},\qquad
 \bm v\coloneqq\delta\mathcal{X}_t\circ\mathcal{X}_t^{-1}.
\end{equation*}
The boundary preservation and the endpoint conditions imply $\bm u\cdot\bm\nu=\bm v\cdot\bm\nu=0$ on $\Gamma$ and $\bm v(t_0)=\bm v(t_1)=0$, respectively.
Here $\delta$ denotes the variation of Eulerian fields at fixed $\bm x$ and $t$.
Differentiating the density push-forward relation in time yields $\overline{\mathrm D}_tn=0$, while variation of the flow map gives the Euler--Poincar\'e constraints \cite{gay2017lagrangian}
\begin{equation*}
 \delta n=-\operatorname{div}(n\bm v),\qquad 
 \delta\bm u=\partial_t\bm v+[\bm u,\bm v],
\end{equation*}
where $[\bm u,\bm v]\coloneqq (\bm u\cdot\nabla)\bm v-(\bm v\cdot\nabla)\bm u$ is the Lie bracket.
Consistently with $\mathrm D_t$ and $\overline{\mathrm D}_t$ introduced in \cref{subsec:Thermodynamic variables}, we furthermore define the material variations as \cite{gay2019perspective}
\begin{equation*}
 \mathrm D_\delta\phi\coloneqq\delta\phi+\bm v\cdot\nabla\phi,\qquad
 \overline{\mathrm D}_\delta\phi\coloneqq\delta\phi+\operatorname{div}(\phi\bm v).
\end{equation*}
The first compares scalar fields along perturbed particle positions, whereas the second additionally accounts for the volume change associated with a density.
In particular, we have $\overline{\mathrm D}_\delta n=0$.

Since $\Gamma$ is fixed as a geometric surface, we also define
\begin{alignat*}{2}
 \mathrm D_t^\Gamma\phi
 &\coloneqq\partial_t\phi+\bm u_\tau\cdot\nabla_\tau\phi,\qquad
 &\overline{\mathrm D}_t^\Gamma\phi
 &\coloneqq\partial_t\phi+\operatorname{div}_\Gamma(\phi\bm u_\tau),\\
 \mathrm D_\delta^\Gamma\phi
 &\coloneqq\delta\phi+\bm v_\tau\cdot\nabla_\tau\phi,\qquad
 &\overline{\mathrm D}_\delta^\Gamma\phi
 &\coloneqq\delta\phi+\operatorname{div}_\Gamma(\phi\bm v_\tau),
\end{alignat*}
where $\bm v_\tau\coloneqq\bm P_\tau\bm v$.
Notice that a barred surface operator is not, in general, the trace of its bulk counterpart.
Using mass conservation and its kinematic variation, we obtain the following relations on $\Gamma$:
\begin{equation}\label{eq:wall-rates}
 \mathrm D_t^\Gamma n
  =\partial_t n+\bm u_\tau\cdot\nabla_\tau n
   =-(n\operatorname{div}\bm u)|_\Gamma,\qquad
 \mathrm D_\delta^\Gamma n
  =\delta n+\bm v_\tau\cdot\nabla_\tau n
   =-(n\operatorname{div}\bm v)|_\Gamma.
\end{equation}
Here, $\operatorname{div}$ is the bulk divergence evaluated on $\Gamma$.

\paragraph{Lagrangian and variational condition.}
We introduce the thermal displacement $\Upsilon$ as an Eulerian scalar field, whose material rate $\mathrm D_t\Upsilon$ will be identified with $T$ by stationarity \cite{gay2019perspective,gawlik2024variational}.
Since $\bm u\cdot\bm\nu=\bm v\cdot\bm\nu=0$ on $\Gamma$, we have $\mathrm D_t^\Gamma\Upsilon=(\mathrm D_t\Upsilon)|_\Gamma$ and $\mathrm D_\delta^\Gamma\Upsilon=(\mathrm D_\delta\Upsilon)|_\Gamma$.
In this formulation, $\hat S$ is the independent thermal state variable.
Let $\Sigma$ and $\Sigma_\Gamma$ be auxiliary entropy densities in $\Omega$ and on $\Gamma$, respectively, distinct from the physical entropy densities $\hat S$ and $\sigma_s$.
Unlike $n|_\Gamma$, $T|_\Gamma$ and $\Upsilon|_\Gamma$, which are traces of bulk fields on $\Gamma$, $\Sigma_\Gamma$ is an independent surface field.

We define the Lagrangian by
\begin{equation*}
 \mathcal L_{S}(\bm u,n,\hat S)
 \coloneqq\int_\Omega
 \Bigl(\frac{mn}{2}|\bm u|^2-\hat e(n,\nabla n,\hat S)\Bigr)\rmd\bm x
 -\int_\Gamma e_s(n)\rmd A.
\end{equation*}
With an external force density $\bm f_{\mathrm{ex}}$, the variational condition is
\begin{equation}\label{eq:action}
 \delta\int_{t_0}^{t_1}\biggl\{
 \mathcal L_{S}(\bm u,n,\hat S)
 +\int_\Omega(\hat S-\Sigma)\mathrm D_t\Upsilon\rmd\bm x
 +\int_\Gamma(\sigma_s-\Sigma_\Gamma)
          \mathrm D_t^\Gamma\Upsilon\rmd A\biggr\}\rmd t+\int_{t_0}^{t_1}\int_\Omega\bm f_{\mathrm{ex}}\cdot\bm v\rmd\bm x\rmd t=0,
\end{equation}
subject to the phenomenological and variational constraints \eqref{eq:bulk-constraints} in $\Omega$ and \eqref{eq:wall-constraints} on $\Gamma$.
The variations satisfy $\delta\Upsilon(t_0)=\delta\Upsilon(t_1)=0$, while $\delta\hat S$ is arbitrary.
The auxiliary entropy variations $\delta\Sigma$ and $\delta\Sigma_\Gamma$ are determined by the variational constraints.

\paragraph{Bulk thermodynamic constraints.}
Motivated by \eqref{eq:energy-chain}, we introduce the following reversible pair of entropy flux and stress for the DVDWT model \cite{onuki2007dynamic}:
\begin{equation}\label{eq:rev-pair}
  \bm j_{\mathrm{rev}}
    \coloneqq\frac{M\nabla n}{T}\mathrm D_tn
    =-\frac{nM\nabla n}{T}\operatorname{div}\bm u,
  \qquad
  \bm R_{\mathrm{rev}}
    \coloneqq-\frac{nM\nabla n\cdot\nabla T}{T}\bm I.
\end{equation}
By construction, this pair satisfies $\bm R_{\mathrm{rev}}:\nabla\bm u=\bm j_{\mathrm{rev}}\cdot\nabla T$.
The bulk phenomenological and variational constraints for \eqref{eq:action} are given by
\begin{subequations}\label{eq:bulk-constraints}
\begin{align}
 T\overline{\mathrm D}_t\Sigma
   &=(\bm\sigma+\bm R_{\mathrm{th}}):\nabla\bm u
    -\bm j\cdot\nabla(\mathrm D_t\Upsilon),\label{eq:bulk-constraints1}\\
 T\overline{\mathrm D}_\delta\Sigma
   &=(\bm\sigma+\bm R_{\mathrm{th}}):\nabla\bm v
    -\bm j\cdot\nabla(\mathrm D_\delta\Upsilon).\label{eq:bulk-constraints2}
\end{align}
\end{subequations}
Here $T=-\delta\mathcal L_{S}/\delta\hat S=\partial_{\hat S}\hat e$ is the thermodynamic variable conjugate to the entropy density $\hat S$, $\bm j\coloneqq\bm j_{\mathrm{irr}}+\bm j_{\mathrm{rev}}$ is defined as the bulk entropy flux, where $\bm j_{\mathrm{irr}}$ is the irreversible part, and $\bm R_{\mathrm{th}}\coloneqq -nM(\nabla n\cdot\nabla(\mathrm D_t\Upsilon))\bm I/T$.
Thus $\bm R_{\mathrm{th}}=\bm R_{\mathrm{rev}}$ if $\mathrm D_t\Upsilon=T$.

We note that the variational constraint \eqref{eq:bulk-constraints2} is prescribed by pairing the same coefficients with the corresponding virtual rates.
It is not the linearization of the phenomenological constraint \eqref{eq:bulk-constraints1} \cite{gay2019perspective}.
Furthermore, the reversible stress and entropy flux terms cancel for the actual rates in the phenomenological constraint \eqref{eq:bulk-constraints1}.
This cancellation does not generally hold for the virtual rates in the variational constraint \eqref{eq:bulk-constraints2}.

\paragraph{Boundary thermodynamic constraints.}
On $\Gamma$, let $\bm f_w$ be the tangential frictional resistance per unit area conjugate to the slip velocity $\bm u_\tau^{\mathrm{slip}}$, $f_r$ be the relaxation force conjugate to the material rate $\mathrm D_t^\Gamma n$, $\bm q_w$ denote the heat flux accounting for heat exchange with the wall at temperature $T_w$, and define $q_{w,\nu}\coloneqq\bm q_w\cdot\bm\nu$ as its outward normal component.
Hence $q_{w,\nu}>0$ corresponds to heat transfer from the fluid to the wall.
Let $\bm j_s$ be the tangential surface entropy flux and $j_w\coloneqq q_{w,\nu}/T_w$ the entropy flux associated with the wall heat exchange.

Assuming $\mathrm D_t^\Gamma\Upsilon>0$, the boundary phenomenological and variational constraints are
\begin{subequations}\label{eq:wall-constraints}
\begin{align}
 (\mathrm D_t^\Gamma\Upsilon)\overline{\mathrm D}_t^\Gamma\Sigma_\Gamma
   &=\bm f_w\cdot\bm u_\tau^{\mathrm{slip}}+f_r\mathrm D_t^\Gamma n
     -\bm j_s\cdot\nabla_\tau(\mathrm D_t^\Gamma\Upsilon)
     +j_w(\mathrm D_t^\Gamma\Upsilon-T_w),\label{eq:wall-constraints1}\\
 (\mathrm D_t^\Gamma\Upsilon)\overline{\mathrm D}_\delta^\Gamma\Sigma_\Gamma
   &=\bm f_w\cdot\bm v_\tau+f_r\mathrm D_\delta^\Gamma n
     -\bm j_s\cdot\nabla_\tau(\mathrm D_\delta^\Gamma\Upsilon)
     +j_w\mathrm D_\delta^\Gamma\Upsilon.\label{eq:wall-constraints2}
\end{align}
\end{subequations}
Here, the prescribed wall velocity $\bm u_w$ and temperature $T_w$ in \eqref{eq:wall-constraints1} belong to the external affine part of the phenomenological constraint, and therefore have no associated virtual rates \cite{gay2019perspective}.

\begin{proposition}[Constrained variational representation]
\label{prop:variational-proposition}
The variational principle \eqref{eq:action}, with constraints \eqref{eq:bulk-constraints} and \eqref{eq:wall-constraints}, yields
\begin{subequations}\label{eq:physical-bulk}
\begin{align}
 \overline{\mathrm D}_tn&=0,\label{eq:physical-bulk-mass}\\
 mn\mathrm D_t\bm u
   &=\operatorname{div}\hat{\bm\sigma}+\bm f_{\mathrm{ex}},\label{eq:physical-bulk-m}\\
 T(\overline{\mathrm D}_t\hat S+\operatorname{div}\bm j)
   &=\bm\sigma:\nabla\bm u-\bm j_{\mathrm{irr}}\cdot\nabla T,\label{eq:physical-bulk-S}
\end{align}
\end{subequations}
and $\mathrm D_t\Upsilon=T$, with $\mathrm D_t^\Gamma\Upsilon=T$ on $\Gamma$ by taking the trace.
On $\Gamma$, we obtain
\begin{subequations}\label{eq:physical-wall}
\begin{align}
 f_r&=-L,\label{eq:physical-wall1}\\
 \bm f_w&=
  \nabla_\tau f_s
  -\bm P_\tau\hat{\bm\sigma}\bm\nu,\label{eq:physical-wall2}\\
 T(\overline{\mathrm D}_t^\Gamma\sigma_s
          +\operatorname{div}_\Gamma\bm j_s-\bm j\cdot\bm\nu)
   &=\bm f_w\cdot\bm u_\tau^{\mathrm{slip}}+f_r\mathrm D_t^\Gamma n
     -\bm j_s\cdot\nabla_\tau T-q_{w,\nu}.\label{eq:physical-wall_T}
\end{align}
\end{subequations}
\end{proposition}
The proof is given in Appendix~\ref{app:variational-proof}.
Notice that the constrained variational principle determines the balance and force relations, whereas the dissipative constitutive closures are specified in \cref{subsec:closure} to ensure nonnegative entropy production both in the bulk and on the boundary.

\subsection{Free-energy-based variational principle}
\label{subsec:free-energy-principle}

Using the thermodynamic relations established above, we now formulate the variational principle with temperature as the thermal state variable \cite{gay2017freeenergy}.

\paragraph{Lagrangian and variational condition.}
We retain the gradient contributions, the reversible pair \eqref{eq:rev-pair} and the boundary constraints \eqref{eq:wall-constraints} in \cref{subsec:principle}.
In this formulation, we define the temperature a priori as the material rate of the thermal displacement: $T\coloneqq\mathrm D_t\Upsilon>0$ and $\mathrm D_t^\Gamma\Upsilon=T$ on $\Gamma$.
We define the corresponding Lagrangian by
\begin{equation*}
 \mathcal L_T(\bm u,n,T)
 \coloneqq\int_\Omega
 \Bigl(\frac{mn}{2}|\bm u|^2-\hat f(n,\nabla n,T)\Bigr)\rmd\bm x
 -\int_\Gamma f_s(n,T)\rmd A.
\end{equation*}
The variational condition is
\begin{equation}\label{eq:free-energy-action}
 \delta\int_{t_0}^{t_1}\biggl\{
 \mathcal L_T(\bm u,n,T)
 -\int_\Omega\Sigma T\rmd\bm x
 -\int_\Gamma\Sigma_\Gamma T\rmd A
 \biggr\}\rmd t+\int_{t_0}^{t_1}\int_\Omega
 \bm f_{\mathrm{ex}}\cdot\bm v\rmd\bm x\rmd t=0.
\end{equation}
Here the variations of the flow map and temporal endpoint conditions are those specified in \cref{subsec:principle}, whereas the physical entropy is not varied independently.
The induced temperature variation is $\delta T=\mathrm D_t(\delta\Upsilon)+\delta\bm u\cdot\nabla\Upsilon=\mathrm D_t(\mathrm D_\delta\Upsilon)-\bm v\cdot\nabla T$.
The bulk phenomenological and variational constraints are given by
\begin{subequations}\label{eq:free-energy-constraints}
\begin{align}
 T\overline{\mathrm D}_t\Sigma
   &=(\bm\sigma+\bm R_{\mathrm{rev}}):\nabla\bm u
         -\bm j\cdot\nabla T,\label{eq:free-energy-constraints1}\\
 T\overline{\mathrm D}_\delta\Sigma
   &=(\bm\sigma+\bm R_{\mathrm{rev}}):\nabla\bm v
         -\bm j\cdot\nabla(\mathrm D_\delta\Upsilon),\label{eq:free-energy-constraints2}
\end{align}
\end{subequations}
with the same boundary constraints \eqref{eq:wall-constraints}.

\begin{proposition}[Equivalent thermal representations]
\label{prop:free-energy-equivalence}
The variational condition \eqref{eq:free-energy-action}, with constraints \eqref{eq:free-energy-constraints} and \eqref{eq:wall-constraints}, yields the same mass and momentum equations \eqref{eq:physical-bulk-mass}--\eqref{eq:physical-bulk-m} and boundary relations \eqref{eq:physical-wall}.
The thermal equation takes the form
\begin{equation}\label{eq:free-temperature-balance}
 -T\overline{\mathrm D}_t
       (\partial_T\hat f)
       +T\operatorname{div}\bm j_{\mathrm{rev}}
   =\bm\sigma:\nabla\bm u-\operatorname{div}\bm q.
\end{equation}
Here $\bm q=T\bm j_{\mathrm{irr}}$ is the heat flux and $\partial_T\hat f=-\hat S$ is evaluated at fixed $(n,\nabla n)$.
Hence, the two variational principles describe the same physical solutions.
\end{proposition}

The proof is given in Appendix~\ref{app:free-energy-proof}.

\begin{remark}
We note that the action based on free energy is the thermal Legendre representation of the action based on entropy.
Indeed, using $\hat f=\hat e-T\hat S$ and $f_s=e_s-T\sigma_s$, the corresponding thermal terms satisfy
\begin{align*}
-\hat e+(\hat S-\Sigma)\mathrm D_t\Upsilon
&=
-\hat f-\Sigma\mathrm D_t\Upsilon
+\hat S(\mathrm D_t\Upsilon-T),
\\
-e_s+(\sigma_s-\Sigma_\Gamma)\mathrm D_t^\Gamma\Upsilon
&=
-f_s-\Sigma_\Gamma\mathrm D_t^\Gamma\Upsilon
+\sigma_s(\mathrm D_t^\Gamma\Upsilon-T)
\quad\text{on }\Gamma.
\end{align*}
Hence, on stationary curves of the entropy-based principle, the thermal part of its action agrees with that of the action based on free energy.
The equivalence of the resulting equations is established in Proposition~\ref{prop:free-energy-equivalence} under the invertible thermal coordinate change.
\end{remark}

\paragraph{Explicit temperature equation.}
Combining \eqref{eq:free-temperature-balance} with \eqref{eq:energy-chain} and the reversible pair identities, we can recover the internal energy balance \eqref{eq:internal}.
To express this balance explicitly in terms of $T$, we use $\hat e=e(n,T)+K|\nabla n|^2/2$.
The gradient contribution satisfies
\begin{equation*}
 \overline{\mathrm D}_t\Bigl(\frac{K}{2}|\nabla n|^2\Bigr)
 =-\frac{K+n\partial_n K}{2}|\nabla n|^2\operatorname{div}\bm u
  -nK\nabla n\cdot\nabla(\operatorname{div}\bm u)
  -K(\nabla n\otimes\nabla n):\nabla\bm u.
\end{equation*}
Substituting this identity into \eqref{eq:internal} and using the mass equation, $\partial_T e=dnk_B/2$ and $M=K+CT$, we obtain
\begin{equation}\label{eq:general-temperature-equation}
\begin{aligned}
 \frac{dnk_B}{2}\mathrm D_tT&+\Bigl(\hat p+\epsilon v_0n^2
-\frac{K+n\partial_{n}K}{2}|\nabla n|^2\Bigr)\operatorname{div}\bm u\\
   &-nK\nabla n\cdot\nabla(\operatorname{div}\bm u)
     +CT(\nabla n\otimes\nabla n):\nabla\bm u
  =\bm\sigma:\nabla\bm u-\operatorname{div}\bm q.
\end{aligned}
\end{equation}
When $K=C=0$, \eqref{eq:general-temperature-equation} reduces to
\begin{equation*}
  T(n\partial_T s\mathrm D_tT+\partial_T p\operatorname{div}\bm u)
=\bm\sigma:\nabla\bm u-\operatorname{div}\bm q,
\end{equation*}
which is the standard temperature equation for the Navier--Stokes--Fourier system \cite{gay2017freeenergy}.

\subsection{Constitutive relations and thermodynamic laws}
\label{subsec:closure}

From the entropy equations \eqref{eq:physical-bulk-S} and \eqref{eq:physical-wall_T}, to ensure nonnegative entropy production both in $\Omega$ and on $\Gamma$, we introduce the following dissipative constitutive relations: a Newtonian compressible viscous stress $\bm \sigma$, Fourier heat conduction for bulk heat flux $\bm{q}$ and surface heat flux $\bm{q}_s$, and linear boundary relations \cite{onuki2007dynamic,liu2012hydrodynamic,wang2026jcp}
\begin{equation}\label{eq:constitutive}
\begin{gathered}
 \bm\sigma=\eta D(\bm u)
   +\Bigl(\xi-\frac{2\eta}{d}\Bigr)(\operatorname{div}\bm u)\bm I,
 \qquad 
 f_r=\alpha\mathrm D_t^\Gamma n,\qquad
 \bm f_w=\beta\bm u_\tau^{\mathrm{slip}},\\
 \bm q=T\bm j_{\mathrm{irr}}=-\lambda\nabla T,\qquad
 \bm q_s\coloneqq T\bm j_s=-\lambda_s\nabla_\tau T,\qquad
 q_{w,\nu}=\frac{1}{\kappa}\Bigl(\frac{1}{T_w}-\frac{1}{T}\Bigr).
\end{gathered}
\end{equation}
Here $D(\bm u)\coloneqq\nabla\bm u+(\nabla\bm u)^{\mathsf T}$, $\eta,\xi\geq0$ are the shear and bulk viscosities, $\lambda,\lambda_s>0$ are the bulk and surface thermal conductivities, $\alpha,\beta>0$ are the relaxation and slip coefficients and $\kappa>0$ is the thermal resistance coefficient.

Thus \eqref{eq:physical-wall1} becomes $\alpha\mathrm D_t^\Gamma n=-L$, the dynamic relaxation boundary condition for $n$.
Similarly, \eqref{eq:physical-wall2} takes the form $\beta\bm u_\tau^{\mathrm{slip}}=\nabla_\tau f_s-\bm P_\tau\hat{\bm\sigma}\bm\nu$, the extension of the GNBC in non-isothermal flows.
Combining \eqref{eq:physical-wall_T} with the surface Gibbs relation \eqref{eq:Gibbs-typeequation} and \eqref{eq:physical-wall1}, we obtain the thermal boundary condition
\begin{equation}\label{eq:thermal-wall}
 \bm q\cdot\bm\nu
  =q_{w,\nu}+\operatorname{div}_\Gamma\bm q_s
    +\partial_n e_s\mathrm D_t^\Gamma n
    +T\sigma_s\operatorname{div}_\Gamma\bm u_\tau
    -\beta|\bm u_\tau^{\mathrm{slip}}|^2.
\end{equation}
These three relations constitute the hydrodynamic boundary conditions.
Indeed, \eqref{eq:physical-wall_T} also yields the surface internal energy balance in \cite{liu2012hydrodynamic,wang2026jcp}.

The total energy and entropy are $\mathcal E_{\text{tot}}\coloneqq\mathcal E_k+\mathcal E_b+\mathcal E_s$ and $\mathcal S_{\text{tot}}\coloneqq\mathcal S_b+\mathcal S_s$, respectively, where $\mathcal E_k\coloneqq\int_\Omega mn|\bm u|^2/2\rmd\bm x$ is the kinetic energy.
The bulk and surface entropy production rates are defined by
\begin{align*}
 \gamma_\Omega
  &\coloneqq\frac{\bm\sigma:\nabla\bm u}{T}
      +\bm q\cdot\nabla\frac{1}{T}=\frac{2\eta}{T}\left|\frac{D(\bm u)}{2}-\frac{\operatorname{div}\bm u}{d}\bm I\right|^2
      +\frac{\xi}{T}(\operatorname{div}\bm u)^2
      +\frac{\lambda|\nabla T|^2}{T^2}\geq0,\\
 \gamma_\Gamma
  &\coloneqq\frac{\bm f_w\cdot\bm u_\tau^{\mathrm{slip}}+f_r\mathrm D_t^\Gamma n}{T}
      +\bm q_s\cdot\nabla_\tau\frac{1}{T}+q_{w,\nu}\Bigl(\frac{1}{T_w}-\frac{1}{T}\Bigr)\\
  &=\frac{\beta|\bm u_\tau^{\mathrm{slip}}|^2+\alpha(\mathrm D_t^\Gamma n)^2}{T}
      +\frac{\lambda_s|\nabla_\tau T|^2}{T^2}
      +\frac{1}{\kappa}\Bigl(\frac{1}{T_w}-\frac{1}{T}\Bigr)^2\geq0.
\end{align*}
\begin{proposition}[First and second laws]
\label{prop:laws-proposition}
Under the constitutive relations \eqref{eq:constitutive}, both formulations satisfy
\begin{align*}
 \frac{\rmd\mathcal E_{\text{tot}}}{\rmd t}
   &=\int_\Omega\bm f_{\mathrm{ex}}\cdot\bm u\rmd\bm x
       -\int_\Gamma(q_{w,\nu}+\bm f_w\cdot\bm u_w)\rmd A,\\
 \frac{\rmd}{\rmd t}(\mathcal S_{\text{tot}}+\mathcal S_{\mathrm{out}})
   &=\int_\Omega\gamma_\Omega\rmd\bm x
       +\int_\Gamma\gamma_\Gamma\rmd A\geq0,\qquad
  \frac{\rmd\mathcal S_{\mathrm{out}}}{\rmd t}\coloneqq\int_\Gamma\frac{q_{w,\nu}}{T_w}\rmd A.
\end{align*}
Here $\mathcal S_{\mathrm{out}}$ denotes the cumulative entropy transferred to the wall, with $\mathcal S_{\mathrm{out}}|_{t=0}=0$.
\end{proposition}
\begin{proof}
Integrating the bulk and surface energy balances and using the tangential force balance gives the energy identity.
The entropy identity follows similarly from the entropy equations divided by $T$, after canceling the normal entropy fluxes $\bm j\cdot\bm\nu$ between the bulk and the surface and accounting for the wall entropy exchange $q_{w,\nu}/T_w$ \cite{wang2026jcp}.
Nonnegativity follows from \eqref{eq:constitutive}.
\end{proof}

Here, gravity is included through $\bm f_{\mathrm{ex}}=-mng\bm e_d$, the term $q_{w,\nu}$ accounts for heat exchange with the wall, and $-\bm f_w\cdot\bm u_w$ represents the mechanical power supplied by the wall per unit area.

\begin{remark}
\label{rmk:onsager-closure}
The Onsager principle provides an alternative approach to dissipative closures, with extensions to inertial and thermal effects \cite{qiao2026unified}.
Here, the generalized Hamilton's principle incorporates mass kinematics, inertia, gradient contributions and boundary conditions within a single constrained variational framework.
In this paper, we focus on using this structure to derive weak forms and guide the construction of fully discrete schemes that preserve the energy balance and entropy inequality.
\end{remark}

\paragraph{Model specialization.}
In the DVDWT model used below, we take $K=0$ and constant $C>0$.
The internal energy density then becomes $\hat e=e$.
For clarity, we denote the homogeneous entropy density by $S\coloneqq ns=\hat S+C|\nabla n|^2/2$.
Although $\hat e$ has no explicit gradient contribution in the variables $(n,T)$, it still depends on $\nabla n$ at fixed $(n,\hat S)$ through the temperature.
The shear viscosity, bulk viscosity, and thermal conductivity are proportional to $n$, with $\eta=\xi=\nu_0mn$ and $\lambda=\nu_0k_Bn$, where $\nu_0>0$ is a reference kinematic viscosity \cite{onuki2007dynamic}.
With these constitutive choices, the temperature equation \eqref{eq:general-temperature-equation} reduces to \cite{wang2026jcp}
\begin{equation*}
 \frac{dnk_B}{2}\mathrm D_t T
 +\epsilon v_0n^2\operatorname{div}\bm u
 =\hat{\bm\sigma}:\nabla\bm u-\operatorname{div}\bm q.
\end{equation*}

On $\Gamma$, we take $e_s=0$ and $\sigma_s=-a_s(n-n_c)$, which gives $f_s=a_sT(n-n_c)$ and $L=CT\nabla_\nu n+a_sT$.
Here $n_c$ is the critical number density and $a_s$ is a constant determined by the interaction between the fluid and solid at short range.
The relaxation coefficient $\alpha$ is a constant, and the coefficients at the solid surface are extrapolated via
  \begin{equation*}
    \zeta=(\zeta_{l}-\zeta_{g})\frac{n-n_{g}}{n_{l}-n_{g}}+\zeta_{g}, \quad \zeta\in\{\beta,\kappa,\lambda_s\},
  \end{equation*}
where $\zeta_{g}$ and $\zeta_{l}$ are the values of a surface parameter in the homogeneous vapor and liquid phases, respectively.
The quantities $n_g$ and $n_l$ denote the number densities of the coexisting homogeneous vapor and liquid phases, respectively.
We assume $0<v_0n_g<v_0n_l<1$, $\zeta_g,\zeta_l>0$ and $n_g/n_l<\zeta_{g}/\zeta_{l}<(1-v_0n_g)/(1-v_0n_l)$ to ensure that $\zeta(n)>0$ for $0<v_0n<1$.

In particular, it is confirmed that the static contact angle (for liquid phase) at the surface, namely whether the solid substrate is hydrophilic or hydrophobic, is determined by $a_{s}$ \cite{xu2012droplet}.
Furthermore, other choices of $e_s$ and $\sigma_s$, including polynomial or trigonometric dependence on $n$, can be used while retaining $f_s=e_s-T\sigma_s$ \cite{qian2006molecular,wang2022energy,wang2024cicp}.

\section{Nondimensionalization and weak forms}
\label{sec:The mathematical model}

In this section, we nondimensionalize the two equivalent formulations specialized in \cref{subsec:closure}, rewrite the governing equations in forms suitable for discretization, derive the corresponding weak forms, and investigate some key physical properties.

\subsection{Dimensionless formulations}
\label{subsec:Nondimensionalization}

Let $l=\sqrt{C/(2k_Bv_0)}$, $u_0>0$, and $t_c=l/u_0$ be the characteristic length, velocity, and time, respectively.
We introduce the following scalings \cite{onuki2007dynamic,xu2010contact}:
\begin{equation*}
\begin{gathered}
\bm{x}^{\ast}\leftarrow \bm{x}/l, \qquad
  \bm{u}^{\ast}\leftarrow \bm{u}/u_{0}, \qquad 
  t^{\ast}\leftarrow t / t_{c}, \qquad
  n^{\ast}\leftarrow n v_{0}, \qquad
  \hat{\mu}^{\ast}\leftarrow \hat{\mu}/ \epsilon,\\
  \hat{\bm{\sigma}}^{\ast}\leftarrow \hat{\bm{\sigma}}v_{0} / \epsilon, \qquad
  \hat{S}^{\ast}\leftarrow \hat{S} v_{0}/ k_{B}, \qquad
  T^{\ast}\leftarrow T k_{B}/ \epsilon, \qquad
  L^{\ast}\leftarrow L/(\epsilon l).
\end{gathered}
\end{equation*}
Hereafter, we use the same symbols for the dimensionless variables.
The mass and momentum equations become
\begin{subequations}
\begin{align}
  \overline{\mathrm D}_t n&=0,\label{eq:dimensionless_mass}\\
  \rmRe\rmR n\mathrm D_t\bm u&=\operatorname{div} \hat{\bm{\sigma}}-n \rmG \bm{e}_{d}.\label{eq:dimensionless_NS}
\end{align}
\end{subequations}
Here $\hat{\bm{\sigma}}=-\bm{\Pi}+\rmR\bm{\sigma}$, $\bm{\Pi}=\hat{p}\bm I+2 T\nabla n\otimes\nabla n$ and $\bm{\sigma}=n(D(\bm{u})+(1-2/d)(\operatorname{div} \bm{u})\bm I)$, where $\hat{p}=nT/(1-n)-n^{2}- T|\nabla n|^{2}-2 nT\Delta n$ and $D(\bm{u})=\nabla\bm{u}+(\nabla\bm{u})^{\mathsf{T}}$.
The dimensionless entropy equation then reads
\begin{equation}\label{eq:entropynew}
  T(\overline{\mathrm D}_t\hat S+\operatorname{div}\bm j)
  =\rmR\bm\sigma:\nabla\bm u
  -\frac{1}{\rmRe}\bm j_{\mathrm{irr}}\cdot\nabla T.
\end{equation}
Here, the bulk entropy fluxes introduced in \cref{subsec:closure} take the form
\begin{equation}\label{eq:jandjsdefine}
  \bm j=\bm j_{\mathrm{rev}}+\frac{1}{\rmRe}\bm j_{\mathrm{irr}},\qquad
  \bm j_{\mathrm{rev}}=-2n(\operatorname{div}\bm u)\nabla n,\qquad
  \bm j_{\mathrm{irr}}=\frac{\bm q}{T}=-\frac{n\nabla T}{T},
\end{equation}
where $\bm q=-n\nabla T$.
Notice that the right-hand side in \eqref{eq:entropynew} is $T$ times the bulk entropy production rate: $\rmR\bm\sigma:\nabla\bm u\geq0$ and $-\bm j_{\mathrm{irr}}\cdot\nabla T=n|\nabla T|^2/T\geq0$.
Alternatively, the thermal evolution can be characterized by the dimensionless temperature equation
\begin{equation}\label{eq:dimensionless_Temp}
  \frac{dn}{2}\mathrm D_t T+n^{2}\operatorname{div}  \bm{u}=\hat{\bm{\sigma}}:\nabla\bm{u}-\frac{1}{\rmRe}\operatorname{div} \bm{q}.
\end{equation}
The entropy density and temperature are related by
\begin{align}
  \hat S
  &=\frac{dn}{2}\ln T
    +n\ln\frac{1-n}{n}
    -|\nabla n|^2,
    \label{eq:dimensionless_ST}\\
  T
  &=\Bigl(\frac{n}{1-n}\Bigr)^{2/d}
    \exp\Bigl(\frac{2(\hat S+|\nabla n|^2)}{dn}\Bigr).
    \label{eq:dimensionless_TS}
\end{align}
The dimensionless generalized chemical potential becomes
\begin{equation}\label{eq:dimensionless_mu}
  \hat{\mu}=\mu-2T\Delta n, \qquad
  \mu=\frac{d}{2}T(1-\ln T)
  -T\ln\Bigl(\frac{1}{n}-1\Bigr)
  +\frac{T}{1-n}-2n.
\end{equation}
With the constitutive choices, the boundary relations \eqref{eq:physical-wall} and \eqref{eq:thermal-wall} become
\begin{subequations}
\begin{align}
  \mathrm D_t^\Gamma n&=-\rmLb L,\label{eq:dimensionless_rbc}\\
  \beta \bm{u}_{\tau}^{\mathrm{slip}}&=\frac{\rmLs}{\rmR}(\nabla_{\tau}f_s-\bm P_\tau\hat{\bm\sigma}\bm\nu), \qquad 
  \bm{u}\cdot\bm{\nu}=0,\label{eq:dimensionless_gnbc}\\
  \bm q\cdot\bm\nu&=\frac{1}{\rmLk}q_{w,\nu}
  +\rmLd\operatorname{div}_\Gamma\bm q_s
  -\rmRe f_s\operatorname{div}_\Gamma\bm u_\tau
  -\frac{\rmRe\rmR}{\rmLs}\beta|\bm u_\tau^{\mathrm{slip}}|^2.\label{eq:dimensionless_Tbcd}
\end{align}
\end{subequations}
Here $L=\rmWs T+2T\nabla_{\nu}n$, $\bm q_s=-\lambda_s\nabla_\tau T$, $\bm{u}_{\tau}^{\mathrm{slip}}=\bm{u}_{\tau}-\bm{u}_{w}$ and $f_{s}=\rmWs T(n-n_{c})$.
The surface entropy flux and the normal heat flux to the wall are given by
\begin{equation*}
  \bm j_s=\frac{\bm q_s}{T}=-\frac{\lambda_s\nabla_\tau T}{T},\qquad
  q_{w,\nu}=-\frac{1}{\kappa}\Bigl(\frac{1}{T}-\frac{1}{T_w}\Bigr).
\end{equation*}
The dimensionless boundary coefficients satisfy $\zeta=(1-\zeta_{gl})(n-n_g)/(n_l-n_g)+\zeta_{gl}$, where $\zeta\in\{\beta,\kappa,\lambda_s\}$ and $\zeta_{gl}\coloneqq\zeta_g/\zeta_l$.
Other dimensionless parameters are defined as follows \cite{xu2012droplet,wang2026jcp}: $\rmRe\coloneqq u_0l/\nu_0$ is the Reynolds number and $\rmR\coloneqq m\nu_0u_0/(\epsilon l)$.
The normalized gravitational acceleration is $\rmG\coloneqq mgl/\epsilon$.
The wettability number $\rmWs\coloneqq a_{s}/(k_{B}l)$ characterizes the wettability of the solid substrate.
The dimensionless relaxation parameter is $\rmLb\coloneqq \epsilon v_{0}l^{2}/(\alpha u_{0})$.
The parameters $\rmLs\coloneqq m\nu_{0}/(\beta_{l}v_{0}l)$ and $\rmLk\coloneqq \nu_{0}\kappa_{l}\epsilon^{2}/(v_{0}k_{B}l)$ measure the strength of viscous and thermal coupling between fluid and solid, respectively.
$\rmLd\coloneqq \lambda_{s,l}v_{0}/(\nu_{0} k_{B}l)$ characterizes surface heat conduction.

\paragraph{Equivalent forms for discretization.}
Using the reversible stress divergence identity \eqref{eq:rev stress identity}, we express the momentum equation \eqref{eq:dimensionless_NS} as
\begin{equation}\label{eq:dimensionless_NS2}
  \rmRe\rmR n\mathrm D_t\bm u=-n\nabla\hat{\mu}-\hat{S}\nabla T-2(\nabla n\cdot\nabla T)\nabla n+\rmR\operatorname{div} \bm{\sigma}-n \rmG \bm{e}_{d}.
\end{equation}
Differentiating $\hat e=dnT/2-n^2$ and \eqref{eq:dimensionless_ST} in time and using \eqref{eq:dimensionless_mu}, we obtain the thermodynamic identities
\begin{align}
  \partial_t\hat e
  &=\frac{dn}{2}\partial_tT
    +\Bigl(\frac d2T-2n\Bigr)\partial_tn,
    \label{eq:dedt1}\\
  \partial_t\hat S
  &=\frac{dn}{2T}\partial_tT
    +\frac{1}{T}\Bigl(\frac d2T-2n-\hat\mu\Bigr)\partial_tn
    -2\operatorname{div}(\nabla n\partial_tn).
    \label{eq:dsdtT}
\end{align}
Using \eqref{eq:dsdtT}, \eqref{eq:dimensionless_mass} and \eqref{eq:jandjsdefine}, we rewrite the temperature equation \eqref{eq:dimensionless_Temp} in the equivalent form:
\begin{equation}\label{eq:tempnew}
  \begin{aligned}
    \frac{dn}{2}\partial_t T+T&\Bigl(\operatorname{div} (\hat{S}\bm{u})+\operatorname{div} (2(\nabla n\cdot\bm{u})\nabla n)+\frac{1}{\rmRe}\operatorname{div} \bm{j}_{\mathrm{irr}}\Bigr)\\
    &=\Bigl(\frac{d}{2}T-2n-\hat{\mu}\Bigr)\operatorname{div} (n\bm{u})+\rmR\bm{\sigma}:\nabla \bm{u}-\frac{1}{\rmRe}\bm{j}_{\mathrm{irr}}\cdot\nabla T.
  \end{aligned}
\end{equation}
To formulate the thermal boundary condition \eqref{eq:dimensionless_Tbcd} in entropy form, we use $\sigma_s=-\rmWs(n-n_c)$, \eqref{eq:dimensionless_mass} and \eqref{eq:dimensionless_rbc} to obtain the following relation:
\begin{equation}\label{eq:boundaryentropyrealtion}
  \overline{\mathrm D}_t^\Gamma\sigma_s
  =\frac{1}{\rmLb T}(\mathrm D_t^\Gamma n)^2
  -2n(\operatorname{div}\bm u)\nabla_\nu n
  -\frac{f_s}{T}\operatorname{div}_\Gamma\bm u_\tau.
\end{equation}
Combining \eqref{eq:boundaryentropyrealtion} with \eqref{eq:dimensionless_Tbcd}, we obtain the equivalent surface entropy balance
\begin{equation}\label{eq:dimensionless_Tbcd_re2}
  \overline{\mathrm D}_t^\Gamma\sigma_s
  +\frac{\rmLd}{\rmRe}\operatorname{div}_\Gamma\bm j_s
  -\bm j\cdot\bm\nu
  +\frac{q_{w,\nu}}{\rmRe\rmLk T_w}
  =
  \frac{1}{\rmLb T}(\mathrm D_t^\Gamma n)^2
  +\frac{\kappa}{\rmRe\rmLk}q_{w,\nu}^2
  +\frac{\rmLd}{\rmRe\lambda_s}|\bm j_s|^2
  +\frac{\rmR\beta}{\rmLs T}|\bm u_\tau^{\mathrm{slip}}|^2.
\end{equation}
Notice that \eqref{eq:dimensionless_Tbcd_re2} is the dimensionless form of the variational boundary entropy equation \eqref{eq:physical-wall_T} with the constitutive relations in \cref{subsec:closure}.

Finally, the resulting two formulations share the mass and momentum equations \eqref{eq:dimensionless_mass} and \eqref{eq:dimensionless_NS2}, the chemical potential \eqref{eq:dimensionless_mu}, and the boundary conditions \eqref{eq:dimensionless_rbc}, \eqref{eq:dimensionless_gnbc} and \eqref{eq:dimensionless_Tbcd_re2}.
They differ only in the primary thermal variable:
\begin{itemize}
  \item \textbf{The entropy-based model} ($S$-model): The primary variables are $n$, $\bm u$ and $\hat S$.
  The thermal evolution is governed by \eqref{eq:entropynew}, with $T$ recovered from \eqref{eq:dimensionless_TS}.
  \item \textbf{The temperature-based model} ($T$-model): The primary variables are $n$, $\bm u$ and $T$.
  The thermal evolution is described by \eqref{eq:tempnew}, with $\hat S$ determined by \eqref{eq:dimensionless_ST}.
\end{itemize}
We consider the above dimensionless models on $\overline{\Omega}\times[0,t_{f}]$, where $t_{f}>0$ denotes the final time.
To close the system, we set the initial conditions as $n(\bm{x},0)=n^{0}(\bm{x})$, $\bm{u}(\bm{x},0)=\bm{u}^{0}(\bm{x})$ and $T(\bm{x},0)=T^{0}(\bm{x})$ for all $\bm{x}\in\overline{\Omega}$, with $0<n^{0}(\bm{x})<1$ and $T^{0}(\bm{x})>0$.
The initial entropy density $\hat{S}^{0}$ in the $S$-model is determined by \eqref{eq:dimensionless_ST} with $n=n^{0}$ and $T=T^{0}$.

\begin{proposition}[Mass conservation]
Both models conserve the total particle number, and hence the total mass:
\begin{equation*}
  \int_{\Omega}n(\bm{x},t)\rmd\bm{x}=\int_{\Omega}n(\bm{x},0)\rmd\bm{x} \quad \forall t\in(0,t_{f}].
\end{equation*}
\end{proposition}

\begin{proof}
  The result follows by integrating \eqref{eq:dimensionless_mass} over $\Omega$ and using $\bm u\cdot\bm\nu=0$.
  Since the particle mass $m$ is constant, total mass is also conserved.
\end{proof}

\begin{proposition}[Number density bounds and temperature positivity]
Let $n$ and $\bm u$, together with $\hat S$ in the $S$-model and $T$ in the $T$-model, be sufficiently smooth on $\overline{\Omega}\times[0,t_f]$.
Suppose that $(n,\bm u)$ satisfies \eqref{eq:dimensionless_mass}, with $\bm u\cdot\bm\nu=0$ on $\Gamma$.
Assume that $\bm u\in L^1(0,t_f;W^{1,\infty}(\Omega)^d)$ and the initial density satisfies $0<n_{\min}^0\leq n^0\leq n_{\max}^0<1$ on $\overline{\Omega}$.
If
\begin{equation}\label{eq:integral bound}
\int_0^{t_f}
\|(\operatorname{div}\bm u)^-(\cdot,s)\|_{L^\infty(\Omega)}
\rmd s<-\ln n_{\max}^0,
\end{equation}
where $r^-\coloneqq\max\{-r,0\}$, then $0<n<1$ on $\overline{\Omega}\times[0,t_f]$.

Moreover, consider the following two cases:
\begin{enumerate}
\item[(\romannumeral1)]
In the $S$-model, $\hat S$ satisfies \eqref{eq:entropynew}, and $T$ is reconstructed by \eqref{eq:dimensionless_TS}.

\item[(\romannumeral2)]
In the $T$-model, $T$ satisfies \eqref{eq:dimensionless_Temp}, with the thermal boundary condition \eqref{eq:dimensionless_Tbcd}.
The initial fluid temperature and prescribed wall temperature satisfy $T^0\geq T_{\min}^0>0$ on $\overline{\Omega}$ and $T_w\geq T_{w,\min}>0$ on $\Gamma\times[0,t_f]$.
\end{enumerate}
Thus, $T>0$ on $\overline{\Omega}\times[0,t_f]$ in both cases.
\end{proposition}

\begin{proof}
\textbf{Boundedness of number density.}
Along the flow map $\mathcal X_t$ introduced in \cref{subsec:principle}, with $\mathcal X_0=\operatorname{id}$, we integrate the mass equation to obtain
\begin{equation*}
n(\mathcal X_t(\bm X),t)
=
n^0(\bm X)\exp\Bigl(
-\int_0^t
\operatorname{div}\bm u(\mathcal X_s(\bm X),s)\rmd s
\Bigr).
\end{equation*}
Since $\mathcal X_t$ maps $\overline{\Omega}$ onto itself, the initial bounds, velocity regularity, and assumed integral bound \eqref{eq:integral bound} ensure uniform separation of $n$ from $0$ and $1$.

\textbf{Positivity of temperature.}
The temperature in the $S$-model is positive by \eqref{eq:dimensionless_TS} and $0<n<1$.
For the $T$-model, using the constitutive relations, we rewrite \eqref{eq:dimensionless_Temp} as
\begin{equation}\label{eq:T_parabolic}
\partial_tT-a_T\Delta T+\bm b_T\cdot\nabla T+c_TT=f_T,
\end{equation}
where $a_T\coloneqq2/(d\rmRe)>0$, $\bm b_T\coloneqq\bm u-2\nabla(\ln n)/(d\rmRe)$, $c_T\coloneqq2\Psi/(dn)$, and $f_T\coloneqq2\rmR\bm\sigma:\nabla\bm u/(dn)\geq0$, with
\begin{equation*}
\Psi\coloneqq
\Bigl(\frac{n}{1-n}-|\nabla n|^2-2n\Delta n\Bigr)
\operatorname{div}\bm u
+2(\nabla n\otimes\nabla n):\nabla\bm u.
\end{equation*}
The uniform density bounds and the assumed regularity ensure that $\bm b_T$ and $c_T$ are bounded.
Choose $\Lambda\geq\|c_T^-\|_{L^\infty(\Omega\times(0,t_f))}$ and define $T^{\star}\coloneqq\exp(-\Lambda t)T$.
Then \eqref{eq:T_parabolic} becomes
\begin{equation*}
\partial_t T^{\star}-a_T\Delta T^{\star}
+\bm b_T\cdot\nabla T^{\star}
+(c_T+\Lambda) T^{\star}
=\exp(-\Lambda t)f_T.
\end{equation*}
Both the zeroth-order coefficient $c_T+\Lambda$ and the source $\exp(-\Lambda t)f_T$ are nonnegative.

Suppose, to the contrary, that $t_\ast\in(0,t_f]$ is the first time at which $\min_{\overline{\Omega}}T(\cdot,t_\ast)=0$, attained at $\bm x_\ast\in\overline{\Omega}$.
Then $T>0$ for $t<t_\ast$ and $T^{\star}\geq0$ up to $t_\ast$.
If $\bm x_\ast\in\Omega$, the parabolic strong maximum principle forces $T^{\star}$ to vanish at earlier times, contradicting the strictly positive initial condition.
If $\bm x_\ast\in\Gamma$, the Kapitza relation implies $q_{w,\nu}(\bm x_\ast,t)\to-\infty$ as $t\to t_\ast^-$, since $T(\bm x_\ast,t)\to0^+$, $T_w\geq T_{w,\min}>0$, and $\kappa$ is positive and bounded above.
This contradicts \eqref{eq:dimensionless_Tbcd}, whose remaining terms stay bounded under the assumed regularity.
Thus no first zero can occur, and $T$ remains positive.
\end{proof}

\subsection{Weak forms and thermodynamic consistency}
\label{subsec:weak_form}

We now present weak forms of the two dimensionless formulations.
Let $L^{2}(\Omega)$ and $L^{2}(\Gamma)$ denote the spaces of square-integrable functions defined on the domain $\Omega$ and boundary $\Gamma$, respectively.
The corresponding inner products are defined as $(\phi,\psi) \coloneqq \int_{\Omega} \phi \psi \rmd \bm{x}$ and $\la\phi,\psi\ra \coloneqq \int_{\Gamma} \phi \psi \rmd A$.
For sufficiently regular arguments, we define the following bulk forms:
\begin{alignat*}{2}
\mathcal B(\phi,\psi,\bm v)
&\coloneqq-(\bm v\cdot\nabla\phi,\psi),
&\quad
\mathcal D(\phi,\psi,\bm v)
&\coloneqq
(2\phi(\operatorname{div}\bm v)\nabla\phi,\nabla\psi),
\\
\mathcal H(\phi,\psi,\bm v)
&\coloneqq
(2(\nabla\phi\cdot\nabla\psi)\nabla\phi,\bm v),
&\quad
\mathcal A^T(a,\phi,\psi,\theta)
&\coloneqq-\Bigl(
a\frac{\phi\nabla\psi}{\psi},\nabla\theta
\Bigr),\\
\mathcal A^n(\phi,\psi)
&\coloneqq(\nabla\phi,\nabla\psi),
&\quad
\mathcal A^u(\phi,\bm u,\bm v)
&\coloneqq\frac12(\phi D(\bm u),D(\bm v))+\Bigl(1-\frac2d\Bigr)
(\phi\operatorname{div}\bm u,\operatorname{div}\bm v),
\end{alignat*}
and the boundary forms:
\begin{alignat*}{2}
\mathcal B_\Gamma\la\phi,\psi,\bm v\ra
&\coloneqq
-\la\bm v_\tau\cdot\nabla_\tau\phi,\psi\ra,
&\qquad
\mathcal D_\Gamma\la\phi,\psi,\bm v\ra
&\coloneqq
\la2\phi(\operatorname{div}\bm v)\nabla_\nu\phi,\psi\ra,
\\
\mathcal H_\Gamma\la\phi,\psi,\bm v\ra
&\coloneqq
\la2\psi\nabla_\nu\phi\nabla_\tau\phi,\bm v_\tau\ra,
&\qquad
\mathcal A_\Gamma^T\la a,\phi,\psi,\theta\ra
&\coloneqq-\Bigl\la
a\frac{\lambda_s(\phi)\nabla_\tau\psi}{\psi},
\nabla_\tau\theta
\Bigr\ra.
\end{alignat*}
Notice that the form $\mathcal D$ represents the reversible entropy flux, while $\mathcal A^T$ and $\mathcal A_\Gamma^T$ represent bulk and surface heat conduction.
We also define the transport form using the Lie bracket:
\begin{equation*}
  \mathcal O(\bm w,\bm u,\bm v)
  \coloneqq-(\bm w,[\bm u,\bm v])
  =(\bm w,(\bm v\cdot\nabla)\bm u-(\bm u\cdot\nabla)\bm v),
\end{equation*}
which satisfies $\mathcal{O}(\bm{w},\bm{u},\bm{v})=-\mathcal{O}(\bm{w},\bm{v},\bm{u})$, namely $\mathcal{O}(\bm{w},\bm{u},\bm{u})=0$.
Using the mass equation \eqref{eq:dimensionless_mass} and $\bm u\cdot\bm\nu=0$, we obtain
\begin{equation}\label{eq:material-inertia-identity}
  (n\mathrm D_t\bm u,\bm v)
  =(\partial_t(n\bm u),\bm v)
  +\mathcal O(n\bm u,\bm u,\bm v)
  +\frac12\mathcal B(|\bm u|^2,n,\bm v).
\end{equation}
Thus, the weak form for the $S$-model reads: find $(n,\hat{\mu},\bm{u},\hat{S})$, with $\bm u\cdot\bm\nu=0$ on $\Gamma$, such that
\begin{subequations}\label{eq:weak_equations}
  \begin{align}
    &(\partial_t n,\phi)+\mathcal{B}(\phi,n,\bm{u})=0,\label{eq:weak_mass}\\
    &(\hat{\mu},\psi)-(\mu,\psi)-2 \mathcal{A}^{n}( n,T\psi)-\rmWs\la T,\psi\ra-\frac{1}{\rmLb}\la \mathrm D_t^\Gamma n,\psi\ra=0,\label{eq:weak_mu}\\
    &\rmRe\rmR\Bigl((\partial_t(n\bm u),\bm v)
    +\mathcal O(n\bm u,\bm u,\bm v)
    +\frac12\mathcal B(|\bm u|^2,n,\bm v)\Bigr)\notag \\
    &\quad -\mathcal{B}(\hat{\mu},n,\bm{v})-\mathcal{B}(T,\hat{S},\bm{v})+\mathcal{H}(n,T,\bm{v})
    +\rmR\mathcal{A}^{u}(n,\bm{u},\bm{v})+(n\rmG \bm{e}_{d},\bm{v})\notag \\
    &\quad +\frac{\rmR}{\rmLs}\la\beta\bm{u}_{\tau}^{\mathrm{slip}},\bm{v}_{\tau}\ra-\la \nabla_{\tau}f_{s},\bm{v}_{\tau}\ra-\mathcal{H}_{\Gamma}\la n,T,\bm{v}\ra=0,\label{eq:weak_NS}\\
    &(\partial_t\hat{S},T\theta)+\mathcal{B}(T\theta,\hat{S},\bm{u})+\mathcal{D}(n,T\theta,\bm{u})
    -\frac{1}{\rmRe}\mathcal{A}^{T}(1,n,T,T\theta)+\frac{1}{\rmRe}\mathcal{A}^{T}(\theta,n,T,T)\notag \\
    &\quad -\rmR\mathcal{A}^{u}(n\theta,\bm{u},\bm{u})+\la\partial_t\sigma_{s},T\theta\ra +\mathcal{B}_{\Gamma}\la T\theta,\sigma_{s},\bm{u}\ra
    -\frac{\rmLd}{\rmRe}\mathcal{A}_{\Gamma}^{T}\la 1,n,T,T\theta\ra+\frac{\rmLd}{\rmRe}\mathcal{A}_{\Gamma}^{T}\la \theta,n,T,T\ra\notag \\
    &\quad -\frac{1}{\rmLb}\la (\mathrm D_t^\Gamma n)^2,\theta\ra
    -\frac{\rmR}{\rmLs}\la \beta |\bm{u}_{\tau}^{\mathrm{slip}}|^{2},\theta\ra -\frac{1}{\rmRe\rmLk}\la\kappa q_{w,\nu}^{2},T\theta\ra
    +\frac{1}{\rmRe\rmLk}\la T_{w}^{-1}q_{w,\nu},T\theta\ra=0,\label{eq:weak_S}
  \end{align}
\end{subequations}
for all admissible test functions $(\phi,\psi,\bm v,\theta)$ with $\bm v\cdot\bm\nu=0$ on $\Gamma$.
The temperature is recovered from \eqref{eq:dimensionless_TS}, and $\mu$ is defined in \eqref{eq:dimensionless_mu}.

For the $T$-model, find $(n,\hat\mu,\bm u,T)$, with $\bm u\cdot\bm\nu=0$ on $\Gamma$, such that \eqref{eq:weak_mass}--\eqref{eq:weak_NS} and
\begin{equation}\label{eq:weak_T}
  \begin{aligned}
    &\frac{d}{2}(n\partial_t T,\theta)+\mathcal{B}(T\theta,\hat{S},\bm{u})-\mathcal{H}(n,T\theta,\bm{u})
    -\frac{1}{\rmRe}\mathcal{A}^{T}(1,n,T,T\theta)+\frac{1}{\rmRe}\mathcal{A}^{T}(\theta,n,T,T)\\
    &\quad -\rmR\mathcal{A}^{u}(n\theta,\bm{u},\bm{u})-\mathcal{B}((\hat{e}_{n}-\hat{\mu})\theta,n,\bm{u})
    +\mathcal{D}_{\Gamma}\la n,T\theta,\bm{u}\ra+\mathcal{H}_{\Gamma}\la n,T\theta,\bm{u}\ra\\
    &\quad +\la\partial_t\sigma_{s},T\theta\ra+\mathcal{B}_{\Gamma}\la T\theta,\sigma_{s},\bm{u}\ra
    -\frac{\rmLd}{\rmRe}\mathcal{A}_{\Gamma}^{T}\la 1,n,T,T\theta\ra+\frac{\rmLd}{\rmRe}\mathcal{A}_{\Gamma}^{T}\la\theta,n,T,T\ra\\
    &\quad -\frac{1}{\rmLb}\la (\mathrm D_t^\Gamma n)^2,\theta\ra-\frac{\rmR}{\rmLs}\la \beta |\bm{u}_{\tau}^{\mathrm{slip}}|^{2},\theta\ra-\frac{1}{\rmRe\rmLk}\la\kappa q_{w,\nu}^{2},T\theta\ra
    +\frac{1}{\rmRe\rmLk}\la T_{w}^{-1}q_{w,\nu},T\theta\ra=0,
  \end{aligned}
\end{equation}
for all admissible test functions $(\phi,\psi,\bm v,\theta)$ with $\bm v\cdot\bm\nu=0$ on $\Gamma$, where $\hat{e}_{n}\coloneqq\partial_n\hat e(n,T)= dT/2-2n$ and $\hat S$ is evaluated by \eqref{eq:dimensionless_ST}.

We next establish both thermodynamic laws in Proposition~\ref{prop:laws-proposition} based on the weak forms, using the cancellations that will be retained in the fully discrete schemes in \cref{subsec:Physics-preserving properties}.
\begin{theorem}[Energy balance]\label{thm:weak_firstlaw}
  Both weak forms satisfy the first law of thermodynamics:
  \begin{equation}\label{eq:firstlawweak}
    \frac{\rmd \mathcal{E}_{\text{tot}}}{\rmd t}=-(n\rmG \bm{e}_{d},\bm{u})-\frac{\rmR}{\rmLs}\la\beta\bm{u}_{\tau}^{\mathrm{slip}},\bm{u}_{w}\ra-\frac{1}{\rmRe\rmLk}\la q_{w,\nu},1\ra,
  \end{equation}
where $\mathcal{E}_{\text{tot}}=\mathcal{E}_{k}+\mathcal{E}_{b}$, $\mathcal{E}_{k}=\rmRe\rmR\int_{\Omega}\frac{1}{2}n|\bm{u}|^{2} \rmd \bm{x}$, $\mathcal{E}_{b}=\int_{\Omega}\hat{e} \rmd \bm{x}$ and $\hat e=dnT/2-n^2$.
\end{theorem}
\begin{proof}
Taking $\phi=|\bm u|^2/2$ in \eqref{eq:weak_mass} and $\bm v=\bm u$ in \eqref{eq:weak_NS}, and using $\mathcal O(n\bm u,\bm u,\bm u)=0$, we obtain the kinetic energy balance for both models:
\begin{equation}\label{eq:firstlawweakEk}
\begin{aligned}
\frac{\rmd\mathcal E_k}{\rmd t}
&=\mathcal B(\hat\mu,n,\bm u)+\mathcal B(T,\hat S,\bm u)
-\rmR\mathcal A^u(n,\bm u,\bm u)-\mathcal H(n,T,\bm u)\\
&\quad -(n\rmG\bm e_d,\bm u)
-\frac{\rmR}{\rmLs}\la\beta\bm u_\tau^{\mathrm{slip}},\bm u_\tau\ra
+\la\nabla_\tau f_s,\bm u_\tau\ra
+\mathcal H_\Gamma\la n,T,\bm u\ra.
\end{aligned}
\end{equation}

In the $S$-model, we combine \eqref{eq:dedt1} and \eqref{eq:dsdtT} to obtain $\partial_t\hat e =T\partial_t\hat S+\hat\mu\partial_t n +2T\operatorname{div}(\nabla n\partial_t n)$.
Integrating over $\Omega$, using $\partial_t n=-\bm u\cdot\nabla n-n\operatorname{div}\bm u$, and taking $\phi=\hat\mu$ in \eqref{eq:weak_mass} and $\theta=1$ in \eqref{eq:weak_S}, we have
  \begin{equation}\label{eq:firstlawweakEb}
    \begin{aligned}
      \frac{\rmd \mathcal{E}_{b}}{\rmd t}&=-\mathcal{B}(\hat{\mu},n,\bm{u})-\mathcal{B}(T,\hat{S},\bm{u})
      +\rmR\mathcal{A}^{u}(n,\bm{u},\bm{u})+\mathcal{H}(n,T,\bm{u})
      -\la\partial_t\sigma_{s},T\ra-\mathcal{B}_{\Gamma}\la T,\sigma_{s},\bm{u}\ra\\
      &\quad -\mathcal{D}_{\Gamma}\la n,T,\bm{u}\ra-\mathcal{H}_{\Gamma}\la n,T,\bm{u}\ra
      +\frac{1}{\rmLb}\la(\mathrm D_t^\Gamma n)^2,1\ra
+\frac{\rmR}{\rmLs}\la\beta|\bm u_\tau^{\mathrm{slip}}|^2,1\ra
-\frac{1}{\rmRe\rmLk}\la q_{w,\nu},1\ra.
    \end{aligned}
  \end{equation}
Here the bulk and surface heat conduction terms cancel, and the wall heat terms combine by the Kapitza relation.
The surface entropy identity \eqref{eq:boundaryentropyrealtion}, with surface integration by parts, yields
\begin{equation*}
  -\la\partial_t\sigma_s,T\ra
   -\mathcal B_\Gamma\la T,\sigma_s,\bm u\ra
   -\mathcal D_\Gamma\la n,T,\bm u\ra
   +\frac{1}{\rmLb}\la(\mathrm D_t^\Gamma n)^2,1\ra=\la f_s,\operatorname{div}_\Gamma\bm u_\tau\ra
   =-\la\nabla_\tau f_s,\bm u_\tau\ra.
\end{equation*}
Adding \eqref{eq:firstlawweakEk} cancels the bulk couplings and the remaining surface terms.
Then we use $\bm u_\tau=\bm u_\tau^{\mathrm{slip}}+\bm u_w$ to obtain \eqref{eq:firstlawweak}.

In the $T$-model, we use \eqref{eq:dedt1} and take $\phi=\hat{e}_{n}$ in \eqref{eq:weak_mass} and $\theta=1$ in \eqref{eq:weak_T} to obtain the same internal energy balance \eqref{eq:firstlawweakEb}.
The preceding cancellations therefore apply.
\end{proof}

\begin{theorem}[Weak entropy inequality]
  Both weak forms satisfy the second law of thermodynamics:
  \begin{equation}\label{eq:secondlawweak}
    \begin{aligned}
      &(\partial_t \hat{S},T\theta)+\mathcal{B}(T\theta,\hat{S},\bm{u})-\frac{1}{\rmRe}\mathcal{A}^{T}(1,n,T,T\theta)+\mathcal{D}(n,T\theta,\bm{u})\\
      &\quad +\la\partial_t\sigma_{s},T\theta\ra+\mathcal{B}_{\Gamma}\la T\theta,\sigma_{s},\bm{u}\ra
      -\frac{\rmLd}{\rmRe}\mathcal{A}_{\Gamma}^{T}\la 1,n,T,T\theta\ra
      +\frac{1}{\rmRe\rmLk}\la T_{w}^{-1}q_{w,\nu},T\theta\ra\geq 0,
    \end{aligned}
  \end{equation}
  for every admissible test function $\theta\geq0$.
\end{theorem}
\begin{proof}
  Using \eqref{eq:weak_S} for the $S$-model, we express the left-hand side of \eqref{eq:secondlawweak} as
  \begin{align*}
    \mathrm{LHS}&=\rmR\mathcal{A}^{u}(n\theta,\bm{u},\bm{u})-\frac{1}{\rmRe}\mathcal{A}^{T}(\theta,n,T,T)-\frac{\rmLd}{\rmRe}\mathcal{A}_{\Gamma}^{T}\la \theta,n,T,T\ra \\
    &\quad +\frac{1}{\rmLb}\la (\mathrm D_t^\Gamma n)^2,\theta\ra+\frac{\rmR}{\rmLs}\la \beta |\bm{u}_{\tau}^{\mathrm{slip}}|^{2},\theta\ra +\frac{1}{\rmRe\rmLk}\la\kappa q_{w,\nu}^{2},T\theta\ra.
  \end{align*}
Based on the positivity of $n$, $T$ and the physical coefficients, and $\theta\geq0$, each term is nonnegative by the definitions of the dissipative forms.
This proves \eqref{eq:secondlawweak} for the $S$-model.

To prove the inequality for the $T$-model, we multiply \eqref{eq:dsdtT} by $T\theta$ and integrate by parts.
Using the mass equation \eqref{eq:weak_mass} and taking $\phi=(\hat{e}_{n}-\hat\mu)\theta$ in \eqref{eq:weak_mass}, we obtain
  \begin{equation}\label{eq:T_entropy_compatibility}
    \begin{aligned}
    \frac{d}{2}
    (n\partial_t T,\theta)
    &=(\partial_t\hat{S},T\theta)
    +\mathcal{B}
    ((\hat{e}_{n}-\hat{\mu})\theta,n,\bm{u})
    +\mathcal{D}
    (n,T\theta,\bm{u})\\
    &\quad +\mathcal{H}
    (n,T\theta,\bm{u})
    -\mathcal{D}_{\Gamma}\la
    n,T\theta,\bm{u}\ra
    -\mathcal{H}_{\Gamma}\la
    n,T\theta,\bm{u}\ra.
    \end{aligned}
  \end{equation}
Substituting \eqref{eq:T_entropy_compatibility} into \eqref{eq:weak_T} recovers \eqref{eq:weak_S}.
The same nonnegative production terms therefore establish \eqref{eq:secondlawweak}.
\end{proof}

We note that the temperature-weighted entropy inequality \eqref{eq:secondlawweak} retains the pairing of temperature with entropy rates in the bulk and surface phenomenological constraints \eqref{eq:bulk-constraints1} and \eqref{eq:wall-constraints1} of the generalized Hamilton's principle.
Taking $\theta=1/T$ in \eqref{eq:secondlawweak} recovers $\rmd(\mathcal S_{\mathrm{tot}}+\mathcal S_{\mathrm{out}})/\rmd t\geq0$, where $\rmd\mathcal S_{\mathrm{out}}/\rmd t =(\rmRe\rmLk)^{-1}\int_\Gamma(q_{w,\nu}/T_w)\rmd A$.
Furthermore, taking $\theta=1$ in \eqref{eq:weak_S} or \eqref{eq:weak_T} and combining the resulting relation with the mass and momentum equations through the thermodynamic identities yields the total energy balance \eqref{eq:firstlawweak}.
Thus, the weak forms identify the bulk and boundary couplings required for both energy balance and nonnegative entropy production.

At the discrete level, the reciprocal of the discrete temperature generally lies outside the finite element test space, and the continuous thermodynamic chain rules do not automatically carry over to discretization.
We therefore construct compatible spatial operators and discrete thermodynamic chain rules to preserve an exact total energy balance and a weighted weak entropy inequality, with nonnegative production in the bulk and on the boundary.
The variational formulation thus serves not only to derive the governing equations but also to guide the construction of the fully discrete approximation.

\section{Numerical methods}
\label{sec:Numerical methods}

In this section, the $S$-scheme and $T$-scheme discretize the corresponding weak forms in \cref{subsec:weak_form} using common spatial operators and different thermal updates.
We then prove mass conservation, the number density bounds, the temperature positivity, the total energy balance, and the weighted entropy inequality for both fully discrete schemes.

\subsection{Time and space discretizations}

For the temporal discretization, let $\{t^{k}=k\delta t: k=0,1,\ldots,M_{t}\}$, where $\delta t\coloneqq t_{f}/M_{t}$, be an equidistant partition of the time interval $[0, t_{f}]$.
We denote the approximation to a function $\phi(\bm{x},t)$ at time $t^{k}$ by $\phi^{k}(\bm{x})$.
The temporal difference and midpoint operators are given by
\begin{equation*}
  \delta_{t}\phi^{k+1}\coloneqq\frac{\phi^{k+1}-\phi^{k}}{\delta t},\qquad 
  \phi^{k+1/2}\coloneqq\frac{1}{2}(\phi^{k}+\phi^{k+1}).
\end{equation*}

The spatial discretization combines standard DG methods for the scalar variables with $\bm H(\operatorname{div},\Omega)$-conforming finite elements for velocity and incorporates interior penalty terms \cite{cockburn2007note,hiptmair2018fully,gawlik2024variational,wang2026jcp}.
An upwind strategy is employed for the convective terms.
Let $\Th$ be a shape-regular, quasi-uniform simplicial partition of $\Omega$, consisting of triangles for $d=2$ and tetrahedra for $d=3$.
Denote the local mesh size by $h_{K}\coloneqq \operatorname{diam}(K)$ and the global mesh size by $h\coloneqq \max_{K\in\Th} h_{K}$.
Let $\mathcal{E}_h$ be the set of all facets of $\Th$, namely edges for $d=2$ and faces for $d=3$.
The sets of interior and boundary facets are denoted by $\EI$ and $\ED$, respectively.
Thus, $\mathcal{E}_{h}=\EI\cup\ED$.

For two neighboring elements $K_+,K_-\in\Th$, let $E=\partial K_+\cap\partial K_-\in\EI$ be their common facet, and let $\bm{\nu}_E$ be the unit normal vector on $E$ directed outward from $K_+$.
Let $\phi$ be a scalar function that is piecewise smooth on $\Th$, whose average $\avg{\cdot}$ and jump $\jump{\cdot}$ on $E\in\EI$ are defined by
\begin{equation*}
  \avg{\phi}\coloneqq\frac{1}{2}(\phi_{+}+\phi_{-}),\qquad 
  \jump{\phi}\coloneqq\phi_{+}-\phi_{-} \quad \text{on} \ E,
\end{equation*}
where $\phi_{\pm}$ denote the traces of $\phi$ on $E$ from the interiors of $K_{\pm}$, respectively.
On any boundary facet $E \in \ED$, the average and jump of $\phi$ on $E$ are defined by $\avg{\phi}=\jump{\phi}=\phi_{+}$.
The average and jump operators act componentwise on vector and tensor fields.

To define the DG discretization on $\Gamma$, we regard $\ED$ as a $(d-1)$-dimensional simplicial partition of $\Gamma$.
The skeleton of the boundary mesh is defined by
\begin{equation*}
  \EV
  \coloneqq
  \{
  e=E_{+}\cap E_{-}:
  E_{+},E_{-}\in\ED,
  E_{+}\neq E_{-},
  \dim(e)=d-2
  \}.
\end{equation*}
Thus, $e\in\EV$ is a vertex for $d=2$ and an edge for $d=3$.
For each $e\in\EV$ shared by $E_{+}$ and $E_{-}$, let $\bm{\nu}_{e}^{\pm}$ be the unit vectors tangent to $E_{\pm}$, orthogonal to $e$, and directed outward from $E_{\pm}$.
These vectors are referred to as the outward conormals of $e$ relative to $E_{\pm}$.
Notice that the conormals $\bm{\nu}_{e}^{+}$ and $\bm{\nu}_{e}^{-}$ need not be opposite when the adjacent boundary facets meet at an angle.
For a scalar function $\phi$ that is piecewise smooth on each facet of $\ED$, let $\phi_{\pm}$ denote its traces on $e$ from $E_{\pm}$, and define
\begin{equation*}
  \avgbd{\phi}
  \coloneqq
  \frac{1}{2}(\phi_{+}+\phi_{-}),
  \qquad
  \jumpbd{\phi}
  \coloneqq
  \phi_{+}-\phi_{-},
\end{equation*}
and the oriented average of the conormal flux of a piecewise tangential vector field $\bm v$ by $\avgbd{\bm v\cdot\bm{\nu}_{e}} \coloneqq ( \bm v_{+}\cdot\bm{\nu}_{e}^{+} - \bm v_{-}\cdot\bm{\nu}_{e}^{-})/2$.
We denote the $L^{2}(K)$ and $L^{2}(E)$ inner products by $(\cdot,\cdot)_{K}$ and $\la\cdot,\cdot\ra_{E}$, respectively.
We use the notation
\begin{align*}
  (\phi,\psi)_{\Th}
  &\coloneqq\sum_{K\in\Th}(\phi,\psi)_{K},\qquad
  \la\phi,\psi\ra_{\EI}
  \coloneqq\sum_{E\in\EI}\la\phi,\psi\ra_{E},
  \\
  \la\phi,\psi\ra_{\ED}
  &\coloneqq\sum_{E\in\ED}\la\phi,\psi\ra_{E},\qquad
  \la\phi,\psi\ra_{\EV}
  \coloneqq\sum_{e\in\EV}\la\phi,\psi\ra_{e}.
\end{align*}
Here, $\la\phi,\psi\ra_{e}=\phi(e)\psi(e)$ for $d=2$, while it denotes the $L^{2}(e)$ inner product for $d=3$.

For each $K\in\Th$ and integer $r\geq0$, let $\mathbb{P}_{r}(K)$ denote the space of polynomials of degree at most $r$ on $K$.
The Raviart--Thomas (RT) space of order $r$ and the Brezzi--Douglas--Marini (BDM) space of order $r+1$ are defined by
\begin{align*}
  \mathbb{RT}_{r}(\Th)
  &\coloneqq
  \{
  \bm{v}\in\bm{H}(\text{div},\Omega):
  \bm{v}\vert_{K}
  \in
  (\mathbb{P}_{r}(K))^{d}
  +
  \bm{x}\mathbb{P}_{r}(K)
  \quad\forall K\in\Th
  \},\\
  \mathbb{BDM}_{r+1}(\Th)
  &\coloneqq
  \{
  \bm{v}\in\bm{H}(\text{div},\Omega):
  \bm{v}\vert_{K}
  \in
  (\mathbb{P}_{r+1}(K))^{d}
  \quad\forall K\in\Th
  \}.
\end{align*}
The subspace of $\bm{H}(\text{div},\Omega)$ with vanishing normal trace on $\Gamma$ is $\bm{H}_{0}(\text{div},\Omega) \coloneqq \{ \bm{v}\in\bm{H}(\text{div},\Omega): \bm{v}\cdot\bm{\nu}=0 \ \text{on}\ \Gamma \}$.
Therefore, we use the finite element spaces
\begin{align*}
  \Phi_{h}^{q}
  &=
  \mathbb{DG}_{q}(\Th)
  \coloneqq
  \{
  \phi\in L^{2}(\Omega):
  \phi\vert_{K}\in\mathbb{P}_{q}(K)
  \quad\forall K\in\Th
  \},\\
  \bm{V}_{h}^{r}
  &\in
  \{
  \mathbb{RT}_{r}(\Th)\cap\bm{H}_{0}(\text{div},\Omega),
  \mathbb{BDM}_{r+1}(\Th)\cap\bm{H}_{0}(\text{div},\Omega)
  \},
\end{align*}
where $q\geq0$ is an integer that is not necessarily equal to $r$.
We also denote by $\bm{W}_{h}$ the infinite-dimensional space of vector fields on $\Omega$ that are piecewise smooth on each element of $\Th$.

We denote by $\nabla_h$ and $\operatorname{div}_h$ the elementwise gradient and divergence, and by $\nabla_{\tau,h}$ the tangential gradient on each boundary facet.
Using the Lie bracket elementwise, we set $[\bm u,\bm v]_h\coloneqq(\bm u\cdot\nabla_h)\bm v-(\bm v\cdot\nabla_h)\bm u$.
The discrete counterparts of $\mathcal B$ and $\mathcal O$ are
\begin{align*}
  \mathcal{B}_{h}(\phi,\psi,\bm{v})&\coloneqq
  -(\bm{v}\cdot\nabla_{h}\phi,\psi)_{\Th}
  +\la (\bm{v}\cdot\bm{\nu}_{E})\jump{\phi},\avg{\psi}\ra_{\EI},\\
  \mathcal{O}_{h}(\bm{w},\bm{u},\bm{v})&\coloneqq
  -(\bm w,[\bm u,\bm v]_h)_{\Th}
  +\la \bm{\nu}_{E}\times \avg{\bm{w}},\jump{\bm{u}\times\bm{v}}\ra_{\EI},
\end{align*}
with the identities $\mathcal O_h(\bm w,\bm u,\bm v)=-\mathcal O_h(\bm w,\bm v,\bm u)$, i.e., $\mathcal O_h(\bm w,\bm u,\bm u)=0$ for all $(\bm{w},\bm{u},\bm{v}) \in \bm{W}_{h}\times\bm{V}_{h}^r\times\bm{V}_{h}^r$.
The facet term in $\mathcal{O}_{h}$ follows from the identity $-[\bm u,\bm v]=\nabla\times(\bm u\times\bm v)-\bm u\operatorname{div}\bm v+\bm v\operatorname{div}\bm u$.
When $d=2$, we define cross products by embedding the vectors into $\mathbb{R}^{3}$ with zero third components.

Following \cite{gawlik2020conservative,gawlik2024variational}, we define the $\bm{u}$-dependent upwind operators, which preserve the cancellations needed for mass conservation and the discrete thermodynamic laws:
\begin{align*}
  \widetilde{\mathcal{B}}_{h}(\bm{u};\phi,\psi,\bm{v})&\coloneqq
  \mathcal{B}_{h}(\phi,\psi,\bm{v})
  +\la \omega_{E}(\bm{u})(\bm{v}\cdot\bm{\nu}_{E})\jump{\phi},\jump{\psi}\ra_{\EI},\\
  \widetilde{\mathcal{O}}_{h}(\bm{u};\bm{w},\bm{u},\bm{v})&\coloneqq
  \mathcal{O}_{h}(\bm{w},\bm{u},\bm{v})
  +\la \omega_{E}(\bm{u})(\bm{\nu}_{E}\times\jump{\bm{w}}),\jump{\bm{u}\times\bm{v}}\ra_{\EI},
\end{align*}
where $\omega_E(\bm u)\coloneqq \pi^{-1}\arctan(c_{\mathrm{up}}\bm u\cdot\bm\nu_E)$ and $c_{\mathrm{up}}>0$ is a smoothing parameter for the upwind weight.
The resulting weight $\omega_{E}(\bm{u})(\bm{u}\cdot\bm{\nu}_{E})$ smoothly approximates the standard upwind weight $|\bm{u}\cdot\bm{\nu}_{E}|/2$.
The remaining bulk forms are discretized as
\begin{align*}
  \mathcal{A}^{n}_{h}(\phi,\psi)&\coloneqq
  (\nabla_{h} \phi,\nabla_{h} \psi)_{\Th}
  +\alpha_{n}\la h_{E}^{-1}\jump{\phi}, \jump{\psi}\ra_{\EI}\\
  &\quad -\la\avg{\nabla_{h} \phi\cdot\bm{\nu}_{E}}, \jump{\psi}\ra_{\EI}
  -\la\avg{\nabla_{h}\psi\cdot\bm{\nu}_{E}}, \jump{\phi}\ra_{\EI},\\
  \mathcal{A}^{u}_{h}(\phi,\bm{u},\bm{v}) &\coloneqq
  \frac{1}{2}(\bm{\sigma}_{h}(\phi,\bm{u}),D_{h}(\bm{v}) )_{\Th}
  +\alpha_{u}\la h_{E}^{-1}\avg{\phi}\jump{\bm{u}}, \jump{\bm{v}}\ra_{\EI}\\
  &\quad -\la  \avg{\bm{\sigma}_{h}(\phi,\bm{u})\bm{\nu}_{E}}, \jump{\bm{v}}\ra_{\EI}
  +\la  \avg{\bm{\sigma}_{h}(\phi,\bm{v})\bm{\nu}_{E}}, \jump{\bm{u}}\ra_{\EI},\\
  \mathcal{A}^{T}_{h}(a,\phi,\psi,\theta)&\coloneqq
  -\Bigl(a\frac{\phi}{\psi}\nabla_{h}\psi,\nabla_{h}\theta\Bigr)_{\Th}
  -\alpha_{T}\Bigl\la h_{E}^{-1}\frac{\avg{a\phi}}{\avg{\psi}}\jump{\psi}, \jump{\theta}\Bigr\ra_{\EI}\\
  &\quad +\Bigl\la\frac{\avg{a\phi\nabla_{h}\psi\cdot\bm{\nu}_{E}}}{\avg{\psi}}, \jump{\theta}\Bigr\ra_{\EI}
  -\Bigl\la\frac{\avg{a\phi\nabla_{h}\theta\cdot\bm{\nu}_{E}}}{\avg{\psi}}, \jump{\psi}\Bigr\ra_{\EI},
\end{align*}
where $\bm{\sigma}_{h}(\phi,\bm{v})\coloneqq \phi(D_{h}(\bm{v})+(1-2/d)(\operatorname{div}_{h} \bm{v})\bm I)$, $D_{h}(\bm{v})=\nabla_{h} \bm{v}+(\nabla_{h}\bm{v})^{\mathsf{T}}$, $h_{E}\coloneqq \operatorname{diam}(E)$ and $\alpha_{n}$, $\alpha_{u}$, $\alpha_{T}>0$ are penalty parameters.
Here, $\mathcal{A}^{n}_{h}$ is the symmetric interior penalty Galerkin (SIPG) form for the Laplacian of $n$, while $\mathcal{A}^{u}_{h}$ and $\mathcal{A}^{T}_{h}$ are nonsymmetric interior penalty Galerkin (NIPG) forms for the viscous and thermal diffusion terms, respectively \cite{cockburn2007note,hiptmair2018fully,gawlik2024variational,wang2026jcp}.
The forms associated with the reversible entropy flux are discretized as
\begin{align*}
  \mathcal{D}_{h}(\phi,\psi,\bm{v})
  &\coloneqq
  (2\phi(\operatorname{div}_{h} \bm{v})\nabla_{h} \phi,\nabla_{h} \psi)_{\Th}
  -\la \avg{2\phi(\operatorname{div}_{h} \bm{v})\nabla_{h}\phi \cdot \bm{\nu}_{E}}, \jump{\psi}\ra_{\EI}, \\
  \mathcal{H}_{h}(\phi,\psi,\bm{v})
  &\coloneqq
  (2(\nabla_{h}\phi\cdot\nabla_{h}\psi)\nabla_{h}\phi,\bm{v})_{\Th}
  -\la\avg{2(\bm{v}\cdot\nabla_{h}\phi)\nabla_{h}\phi\cdot\bm{\nu}_{E}},\jump{\psi}\ra_{\EI}.
\end{align*}
The boundary forms are discretized as
\begin{align*}
  \mathcal{B}_{\Gamma,h}\la\phi,\psi,\bm{v}\ra
  &\coloneqq -\la \bm{v}_{\tau}\cdot\nabla_{\tau,h} \phi,\psi\ra_{\ED}
  +\la\avgbd{\psi\bm v_\tau\cdot\bm\nu_e},\jumpbd{\phi}\ra_{\EV}, \\
  \mathcal{C}_{\Gamma,h}\la\phi,\psi,\bm{v}\ra
  &\coloneqq
  -\la\nabla_{\tau,h}(\phi\psi),\bm{v}_{\tau}\ra_{\ED}
  +\la
  \avgbd{\psi\bm{v}_{\tau}\cdot\bm{\nu}_{e}},
  \jumpbd{\phi}
  \ra_{\EV}.
\end{align*}
The form $\mathcal C_{\Gamma,h}\la T,\sigma_s,\bm v\ra$ discretizes $\la\nabla_\tau f_s,\bm v_\tau\ra$ and uses the same numerical flux on the boundary mesh as $\mathcal B_{\Gamma,h}$ to retain the boundary cancellation in the energy balance.
The NIPG form for surface thermal diffusion in \eqref{eq:dimensionless_Tbcd_re2} is defined by
\begin{align*}
  \mathcal{A}^{T}_{\Gamma,h}\la a,\phi,\psi,\theta\ra&\coloneqq
  -\Bigl\la a\frac{\lambda_{s}(\phi)}{\psi}\nabla_{\tau,h}\psi,\nabla_{\tau,h}\theta\Bigr\ra_{\ED}
  -\alpha_{\Gamma}
  \Bigl\la h_{e}^{-1}\frac{\avgbd{a\lambda_{s}(\phi)}}{\avgbd{\psi}}
  \jumpbd{\psi}, \jumpbd{\theta}\Bigr\ra_{\EV}\\
  &\quad +\Bigl\la\frac{\avgbd{a\lambda_{s}(\phi)\nabla_{\tau,h}\psi\cdot\bm{\nu}_{e}}}{\avgbd{\psi}},\jumpbd{\theta}\Bigr\ra_{\EV}
  -\Bigl\la\frac{\avgbd{a\lambda_{s}(\phi)\nabla_{\tau,h}\theta\cdot\bm{\nu}_{e}}}{\avgbd{\psi}},\jumpbd{\psi}\Bigr\ra_{\EV},
\end{align*}
where $h_e\coloneqq 2h_{E_{+}}h_{E_{-}}/(h_{E_{+}}+h_{E_{-}})$ for each $e\in\EV$, and $\alpha_{\Gamma}>0$ is a penalty parameter.

\subsection{Fully discrete scheme for the \texorpdfstring{$S$}{S}-model}
\label{subsec:Fully discrete scheme for the entropy-based model}

To enforce the number density bounds in both fully discrete schemes, we introduce a logit variable $\varphi_h^k\in\Phi_h^{r_n}$ and define \cite{huang2022bound}
\begin{equation*}
  n_h^k\coloneqq\mathcal{N}(\varphi_h^k),
  \qquad
  \mathcal{N}(s)\coloneqq\frac{1}{1+\exp(-s)}.
\end{equation*}
Thus, $\varphi_h^k$ is the primary discrete density variable, and $n_h^k$ is the corresponding physical number density evaluated pointwise.
We further define the pointwise logistic secant $\chi_h^{k\star}\coloneqq(\mathcal{N}(\varphi_h^{k+1})-\mathcal{N}(\varphi_h^k))/(\varphi_h^{k+1}-\varphi_h^k)$, which yields $\delta_t n_h^{k+1}=\chi_h^{k\star}\delta_t\varphi_h^{k+1}$.
The superscript $k\star$ distinguishes two-point quantities from arithmetic midpoint values.
In both schemes, $n_h^{k+1/2}=(n_h^k+n_h^{k+1})/2$ denotes the arithmetic midpoint of the physical densities and is generally different from $\mathcal N(\varphi_h^{k+1/2})$.

For the $S$-model, the homogeneous entropy density $S$ allows us to express the temperature $T(n,\nabla n,\hat{S})$ and internal energy density $\hat{e}(n,\nabla n,\hat{S})$ in terms of $(n,S)$.
Retaining the notation $T$ and $\hat e$ after the substitution $n=\mathcal N(\varphi)$, we have
\begin{equation*}
  T(\varphi,S)
  =\exp\biggl(
  \frac{2}{d}\Bigl(
  \varphi+\frac{S}{\mathcal N(\varphi)}
  \Bigr)\biggr),\qquad
  \hat e(\varphi,S)
  =\frac{d}{2}\mathcal N(\varphi)
  T(\varphi,S)-\mathcal N(\varphi)^2.
\end{equation*}
The Gibbs relation \eqref{eq:threemaineq} gives $\partial_\varphi\hat e(\varphi,S)=\mathcal N'(\varphi)\mu$ and $\partial_S\hat e(\varphi,S)=T$.
We approximate these derivatives by the symmetric two-point difference quotients \cite{gawlik2024variational,gawlik2025structure}
\begin{align*}
  \delta_{\varphi}\hat{e}&\coloneqq\frac{1}{2}\biggl(\frac{\hat{e}(\varphi_{h}^{k+1},S_{h}^{k+1})-\hat{e}(\varphi_{h}^{k},S_{h}^{k+1})}{\varphi_{h}^{k+1}-\varphi_{h}^{k}}+\frac{\hat{e}(\varphi_{h}^{k+1},S_{h}^{k})-\hat{e}(\varphi_{h}^{k},S_{h}^{k})}{\varphi_{h}^{k+1}-\varphi_{h}^{k}}\biggr),\\
  \delta_{S}\hat{e}&\coloneqq\frac{1}{2}\biggl(\frac{\hat{e}(\varphi_{h}^{k+1},S_{h}^{k+1})-\hat{e}(\varphi_{h}^{k+1},S_{h}^{k})}{S_{h}^{k+1}-S_{h}^{k}}+\frac{\hat{e}(\varphi_{h}^{k},S_{h}^{k+1})-\hat{e}(\varphi_{h}^{k},S_{h}^{k})}{S_{h}^{k+1}-S_{h}^{k}}\biggr).
\end{align*}
We denote $T_{h}^{k\star}\coloneqq\delta_{S}\hat{e}$.
These definitions ensure the exact discrete chain rule
\begin{equation*}
  \hat{e}(\varphi_{h}^{k+1},S_{h}^{k+1})-\hat{e}(\varphi_{h}^{k},S_{h}^{k})
  =\delta_{\varphi}\hat{e}(\varphi_{h}^{k+1}-\varphi_{h}^{k})+T_{h}^{k\star}(S_{h}^{k+1}-S_{h}^{k}).
\end{equation*}
The quantities $\delta_{\varphi}\hat e$ and $T_h^{k\star}$ are evaluated pointwise and are not additional unknowns.
Furthermore, we discretize the inertia terms in \eqref{eq:material-inertia-identity} using
\begin{equation}\label{eq:discrete-kinetic-chain}
  \frac12\delta_t( n_h|\bm u_h|^2)^{k+1}
  =\bm u_h^{k+1/2}\cdot\delta_t(n_h\bm u_h)^{k+1}
  -\frac12(\bm u_h^k\cdot\bm u_h^{k+1})\delta_t n_h^{k+1}.
\end{equation}
The weighted $L^2$ projection \eqref{eq:fully_discret_u2} ensures that the mass and momentum equations recover \eqref{eq:discrete-kinetic-chain}.

The resulting scheme of Crank--Nicolson type is defined as follows: Given initial conditions $n^{0}(\bm{x})$, $\bm{u}^{0}(\bm{x})$ and $T^{0}(\bm{x})$ satisfying $0<n^0<1$ and $T^0>0$, choose $\varphi_h^0\in\Phi_h^{r_n}$, $\bm u_h^0\in\bm V_h^{r_u}$ and $T_{h,\mathrm{in}}^0\in\Phi_h^{r_s}$ as finite element approximations of $\ln(n^0(\bm{x})/(1-n^0(\bm{x})))$, $\bm u^0(\bm{x})$ and $T^0(\bm{x})$, respectively, with $T_{h,\mathrm{in}}^0>0$.
Set $n_h^0=\mathcal N(\varphi_h^0)$ and define $S_h^0,\hat S_h^0\in\Phi_h^{r_s}$ by
\begin{equation*}
\begin{alignedat}{2}
  (S_h^0,\ell_h)_{\Th}
  &=
  -(n_h^0\varphi_h^0,\ell_h)_{\Th}
  +\frac{d}{2}
  (n_h^0\ln T_{h,\mathrm{in}}^0,\ell_h)_{\Th}
  &\qquad &\forall\ell_h\in\Phi_h^{r_s},\\
  (\hat S_h^0,\theta_h)_{\Th}
  &=
  (S_h^0,\theta_h)_{\Th}
  -(|\nabla_h n_h^0|^2,\theta_h)_{\Th}
  &\qquad &\forall\theta_h\in\Phi_h^{r_s}.
\end{alignedat}
\end{equation*}
Here, $T_{h,\mathrm{in}}^0$ is used only to initialize $S_h^0$.
For $k=0,\ldots,M_t-1$, find $\varphi_{h}^{k+1},\hat{\mu}_{h}^{k\star},\mathcal U_h^{k\star}\in\Phi_{h}^{r_{n}}$, $\hat{S}_{h}^{k+1},S_{h}^{k+1}\in\Phi_{h}^{r_{s}}$ and $\bm{u}_{h}^{k+1}\in\bm{V}_{h}^{r_{u}}$, such that
\begin{subequations}\label{eq:fully_discret_S_eq}
\begin{align}
  &(\delta_t n_h^{k+1},\phi_h)_{\Th}
  +\widetilde{\mathcal B}_h
  (\bm u_h^{k+1/2};\phi_h,n_h^{k+1/2},
   \bm u_h^{k+1/2})
  =0,
  \label{eq:fully_discret_mass}\\
  &\rmRe\rmR
  (\delta_t(n_h\bm u_h)^{k+1},\bm v_h)_{\Th}
  +\rmRe\rmR\widetilde{\mathcal O}_h
  (\bm u_h^{k+1/2};(n_h\bm u_h)^{k+1/2},
   \bm u_h^{k+1/2},\bm v_h)
  \notag\\
  &\quad
  +\rmRe\rmR\widetilde{\mathcal B}_h
  (\bm u_h^{k+1/2};\mathcal U_h^{k\star},
   n_h^{k+1/2},\bm v_h)
  -\widetilde{\mathcal B}_h
  (\bm u_h^{k+1/2};\hat\mu_h^{k\star},
   n_h^{k+1/2},\bm v_h)
  -\widetilde{\mathcal B}_h
  (\bm u_h^{k+1/2};T_h^{k\star},
   \hat S_h^{k+1/2},\bm v_h)
  \notag\\
  &\quad
  +\mathcal H_h
  (n_h^{k+1/2},T_h^{k\star},\bm v_h)
  +\rmR\mathcal A_h^u
  (n_h^{k+1/2},\bm u_h^{k+1/2},\bm v_h)
  +(n_h^{k+1/2}\rmG\bm e_d,\bm v_h)_{\Th}
  \notag\\
  &\quad
  +\frac{\rmR}{\rmLs}
  \la\beta_h^{k+1/2}\bm u_{\tau,h}^{\mathrm{slip},k+1/2},
      \bm v_{\tau,h}\ra_{\ED}
  -\mathcal C_{\Gamma,h}
  \la T_h^{k\star},\sigma_{s,h}^{k+1/2},\bm v_h\ra
  -\mathcal H_{\Gamma,h}
  \la n_h^{k+1/2},T_h^{k\star},\bm v_h\ra
  =0,
  \label{eq:fully_discret_ns}\\
  &(\delta_t\hat S_h^{k+1},T_h^{k\star}\theta_h)_{\Th}
  +\widetilde{\mathcal B}_h
  (\bm u_h^{k+1/2};T_h^{k\star}\theta_h,
   \hat S_h^{k+1/2},\bm u_h^{k+1/2})
  +\mathcal D_h
  (n_h^{k+1/2},T_h^{k\star}\theta_h,
   \bm u_h^{k+1/2})
  \notag\\
  &\quad
  -\frac{1}{\rmRe}\mathcal A_h^T
  (1,n_h^{k+1/2},T_h^{k\star},T_h^{k\star}\theta_h)
  +\frac{1}{\rmRe}\mathcal A_h^T
  (\theta_h,n_h^{k+1/2},T_h^{k\star},T_h^{k\star})
  -\rmR\mathcal A_h^u
  (n_h^{k+1/2}\theta_h,\bm u_h^{k+1/2},
   \bm u_h^{k+1/2})
  \notag\\
  &\quad
  +\la\delta_t\sigma_{s,h}^{k+1},
      T_h^{k\star}\theta_h\ra_{\ED}
  +\mathcal B_{\Gamma,h}
  \la T_h^{k\star}\theta_h,\sigma_{s,h}^{k+1/2},
      \bm u_h^{k+1/2}\ra
  -\frac{\rmLd}{\rmRe}\mathcal A_{\Gamma,h}^T
  \la1,n_h^{k+1/2},T_h^{k\star},
      T_h^{k\star}\theta_h\ra
  \notag\\
  &\quad
  +\frac{\rmLd}{\rmRe}\mathcal A_{\Gamma,h}^T
  \la\theta_h,n_h^{k+1/2},T_h^{k\star},
      T_h^{k\star}\ra
  -\frac{1}{\rmLb}
  \la(\mathrm D_{\delta t}^{\Gamma} n_h^{k+1/2})^2,\theta_h\ra_{\ED}
  -\frac{\rmR}{\rmLs}
  \la\beta_h^{k+1/2}
      |\bm u_{\tau,h}^{\mathrm{slip},k+1/2}|^2,
      \theta_h\ra_{\ED}
  \notag\\
  &\quad
  -\frac{1}{\rmRe\rmLk}
  \la\kappa_h^{k+1/2}(q_{w,\nu,h}^{k\star})^2,
      T_h^{k\star}\theta_h\ra_{\ED}
  +\frac{1}{\rmRe\rmLk}
  \la T_w^{-1}q_{w,\nu,h}^{k\star},
      T_h^{k\star}\theta_h\ra_{\ED}
  =0,
  \label{eq:fully_discret_S}\\
  &(\delta_t S_h^{k+1},T_h^{k\star}\ell_h)_{\Th}
  -(\delta_t\hat S_h^{k+1},T_h^{k\star}\ell_h)_{\Th}
  -2\mathcal A_h^n
  (n_h^{k+1/2},T_h^{k\star}\ell_h\delta_t n_h^{k+1})
  \notag\\
  &\quad
  -\mathcal D_h
  (n_h^{k+1/2},T_h^{k\star}\ell_h,\bm u_h^{k+1/2})
  -\mathcal H_h
  (n_h^{k+1/2},T_h^{k\star}\ell_h,\bm u_h^{k+1/2})
  +\mathcal J_h(\varpi_h^{k+1/2},\ell_h)
  =0,
  \label{eq:fully_discret_chi}\\
  &(\hat\mu_h^{k\star},\chi_h^{k\star}\psi_h)_{\Th}
  -(\delta_{\varphi}\hat e,\psi_h)_{\Th}
  -2\mathcal A_h^n
  (n_h^{k+1/2},T_h^{k\star}\chi_h^{k\star}\psi_h)
  +\la\hat F_{\nu,h}^{k\star},
      \chi_h^{k\star}\psi_h\ra_{\ED}
  =0,
  \label{eq:fully_discret_mu}\\
  &(\mathcal U_h^{k\star},
    \chi_h^{k\star}\varrho_h)_{\Th}
  -\frac12
  (\bm u_h^k\cdot\bm u_h^{k+1},
   \chi_h^{k\star}\varrho_h)_{\Th}
  =0,
  \label{eq:fully_discret_u2}
\end{align}
\end{subequations}
  for all $\phi_{h},\psi_{h},\varrho_h\in\Phi_{h}^{r_{n}}$, $\theta_{h},\ell_{h}\in\Phi_{h}^{r_{s}}$ and $\bm{v}_{h}\in\bm{V}_{h}^{r_{u}}$.
  We define the remaining discrete quantities and boundary forms by
\begin{align*}
  &\mathrm D_{\delta t}^{\Gamma} n_h^{k+1/2}\coloneqq\delta_{t}n_{h}^{k+1}+\bm{u}^{k+1/2}_{\tau,h}\cdot\nabla_{\tau,h}n_{h}^{k+1/2},\\
  &\bm{u}_{\tau,h}^{\mathrm{slip},k+1/2}=\bm{u}_{\tau,h}^{k+1/2}-\bm{u}_{w},\qquad
  \sigma_{s,h}^{k}= -\rmWs (n_{h}^{k}-n_{c}),\\
  &\mathcal{H}_{\Gamma,h}\la n_{h}^{k+1/2},T_{h}^{k\star},\bm{v}_{h}\ra\coloneqq\la\hat{F}_{\nu,h}^{k\star}\nabla_{\tau,h}n_{h}^{k+1/2},\bm{v}_{\tau,h}\ra_{\ED},\qquad
  \hat{F}_{\nu,h}^{k\star}\coloneqq-\rmWs T_{h}^{k\star}-\frac{1}{\rmLb}\mathrm D_{\delta t}^{\Gamma} n_h^{k+1/2},\\
  &\mathcal{J}_h(\varpi,\ell)
  \coloneqq\alpha_\varpi^1(\nabla_{h}\varpi,\nabla_{h}\ell)_{\Th}+\alpha_\varpi^2\la h_E^{-1}\jump{\varpi},\jump{\ell}\ra_{\EI},\qquad
  \varpi_h^k\coloneqq S_h^k-\hat S_h^k-|\nabla_{h} n_h^k|^2,\\
  &\kappa_{h}^{k+1/2}q_{w,\nu,h}^{k\star}=-\Bigl(\frac{1}{T_{h}^{k\star}}-\frac{1}{T_w}\Bigr),\qquad
  \zeta_{h}^{k+1/2}=(1-\zeta_{gl})\frac{n_{h}^{k+1/2}-n_{g}}{n_{l}-n_{g}}+\zeta_{gl},\quad \zeta\in\{\beta,\kappa,\lambda_{s}\}.
\end{align*}
Here, the quantity $\hat F_{\nu,h}^{k\star}$ approximates $2T\nabla_\nu n$ through the relaxation boundary condition \eqref{eq:dimensionless_rbc}.
Notice that \eqref{eq:fully_discret_u2} defines $\mathcal U_h^{k\star}$ as the weighted $L^2$ projection of $\bm u_h^k\cdot\bm u_h^{k+1}/2$ onto $\Phi_h^{r_n}$.
Since $\mathcal N$ is strictly increasing, $\chi_h^{k\star}>0$, which ensures that this weighted $L^2$ projection is well defined.
All two-point difference quotients in \cref{sec:Numerical methods} are evaluated pointwise and understood by continuous extension when denominators vanish.
In particular,
\begin{equation*}
  \frac{\phi(a^{k+1},b)-\phi(a^k,b)}{a^{k+1}-a^k}
  \coloneqq\partial_a\phi(a^k,b)
  \quad\text{if }a^{k+1}=a^k.
\end{equation*}

The form $\mathcal J_h$ is a stabilization term with parameters $\alpha_\varpi^1,\alpha_\varpi^2\geq0$.
Combining the relation $S=\hat S+|\nabla n|^2$ with the mass equation \eqref{eq:dimensionless_mass}, we obtain
\begin{equation*}
(\partial_t S-\partial_t\hat S,\ell)
=
2(\ell\nabla n,\nabla\partial_t n)
=
2\mathcal A^n(n,\ell\partial_t n)
+\mathcal D(n,\ell,\bm u)
+\mathcal H(n,\ell,\bm u).
\end{equation*}
This identity motivates \eqref{eq:fully_discret_chi}, with consistent DG terms weighted by $T_h^{k\star}$.
Notice that the compatibility relation is approximated through this evolution update rather than imposed pointwise.
The entropy density reconstructed from $n_h^k$ and $T_h^k$ is $S_h^k-|\nabla_h n_h^k|^2=\hat S_h^k+\varpi_h^k$, where $\varpi_h^k$ measures the deviation from the evolved variable $\hat S_h^k$.
The discrete entropy inequality established below applies to $\hat S_h$ rather than the reconstructed quantity.

The stabilization term $\mathcal J_h$ is introduced to reduce spatial variations of $\varpi_h$.
It vanishes for the exact solution and is therefore consistent with the continuous compatibility relation.
Moreover, $\mathcal J_h(\varpi_h^{k+1/2},1)=0$, and hence the stabilization term does not alter the discrete energy balance.

\subsection{Fully discrete scheme for the \texorpdfstring{$T$}{T}-model}
\label{subsec:Fully discrete scheme for the temperature-based model}

The $T$-scheme inherits the density parametrization $n_h^k=\mathcal N(\varphi_h^k)$, the mass and momentum equations, and the weighted projection for $\mathcal U_h^{k\star}$ from the $S$-scheme.
To retain a weighted weak entropy inequality, we couple the discrete temperature equation with an auxiliary entropy update \eqref{eq:fully_discret_T_S} based on the continuous compatibility relation \eqref{eq:T_entropy_compatibility}.

To ensure positivity of the temperature, we introduce the logarithmic temperature $\vartheta_h^k\in\Phi_h^{r_s}$ and define
\begin{equation*}
  T_{h}^{k}\coloneqq\exp(\vartheta_{h}^{k}).
\end{equation*}
Thus, $\varphi_h^k$ and $\vartheta_h^k$ are the computational number density and temperature variables, and $n_h^k$ and $T_h^k$ are the physical counterparts.
Retaining the notation $S$ after the variable substitutions, the homogeneous entropy density becomes $S(\varphi,\vartheta) = \mathcal N(\varphi) (-\varphi+d\vartheta/2)$.
In this scheme, $S_h^k\coloneqq S(\varphi_h^k,\vartheta_h^k)$ is evaluated pointwise rather than treated as an independent unknown as in the $S$-scheme.
We define its symmetric two-point difference quotient
\begin{equation*}
  \delta_{\varphi}S
  \coloneqq
  \frac12\biggl(
    \frac{
      S(\varphi_h^{k+1},\vartheta_h^{k+1})
      -S(\varphi_h^k,\vartheta_h^{k+1})
    }{\varphi_h^{k+1}-\varphi_h^k}
    +
    \frac{
      S(\varphi_h^{k+1},\vartheta_h^k)
      -S(\varphi_h^k,\vartheta_h^k)
    }{\varphi_h^{k+1}-\varphi_h^k}
  \biggr),
\end{equation*}
and the logarithmic mean of the temperature
\begin{equation*}
  T_{h}^{k\star}
  \coloneqq
  \frac{T_{h}^{k+1}-T_{h}^{k}}
  {\vartheta_{h}^{k+1}-\vartheta_{h}^{k}}
  =
  \frac{\exp(\vartheta_{h}^{k+1})-\exp(\vartheta_{h}^{k})}
  {\vartheta_{h}^{k+1}-\vartheta_{h}^{k}}.
\end{equation*}
When $\vartheta_h^{k+1}=\vartheta_h^k$, we have $T_h^{k\star}=T_h^k$.
These definitions yield the discrete chain rule
\begin{equation*}
T_h^{k\star}(S_h^{k+1}-S_h^k)
  =
  T_h^{k\star}\delta_{\varphi}S
  (\varphi_h^{k+1}-\varphi_h^k)
  +\frac d2 n_h^{k+1/2}
  (T_h^{k+1}-T_h^k).
\end{equation*}
The internal energy density becomes $\hat e(\varphi,\vartheta) = d\mathcal N(\varphi)\exp(\vartheta)/2 -\mathcal N(\varphi)^2$, which satisfies the midpoint identity
\begin{equation*}
  \hat e(\varphi_h^{k+1},\vartheta_h^{k+1})
  -\hat e(\varphi_h^k,\vartheta_h^k)=
  \frac d2 n_h^{k+1/2}T_h^{k\star}
  (\vartheta_h^{k+1}-\vartheta_h^k)
  +\hat{e}_{n,h}^{k+1/2}\chi_h^{k\star}
  (\varphi_h^{k+1}-\varphi_h^k),
\end{equation*}
where $\hat{e}_{n,h}^{k+1/2}= dT_h^{k+1/2}/2-2n_h^{k+1/2}$, $T_h^{k+1/2}=(T_h^{k+1}+T_h^k)/2$, and we have used $n_h^{k+1}-n_h^k =\chi_h^{k\star}(\varphi_h^{k+1}-\varphi_h^k)$ and $T_h^{k\star}\delta_t\vartheta_h^{k+1} =\delta_tT_h^{k+1}$.
These identities provide the thermodynamic coupling used below to prove the discrete first and second laws.

The resulting Crank--Nicolson type scheme is defined as follows: Given initial conditions $n^{0}(\bm{x})$, $\bm{u}^{0}(\bm{x})$ and $T^{0}(\bm{x})$ with $0<n^{0}(\bm{x})<1$ and $T^{0}(\bm{x})>0$, choose $\varphi_h^0\in\Phi_h^{r_n}$, $\bm u_h^0\in\bm V_h^{r_u}$ and $\vartheta_h^0\in\Phi_h^{r_s}$ as finite element approximations of $\ln(n^0/(1-n^0))$, $\bm u^0$ and $\ln T^0$, respectively.
Set $n_h^0=\mathcal N(\varphi_h^0)$ and $T_h^0=\exp(\vartheta_h^0)$, and define $\hat S_h^0\in\Phi_h^{r_s}$ by
\begin{equation*}
  (\hat S_h^0,\ell_h)_{\Th}
  =
  -(n_h^0\varphi_h^0,\ell_h)_{\Th}
  +\frac{d}{2}(n_h^0\vartheta_h^0,\ell_h)_{\Th}
  -(|\nabla_h n_h^0|^2,\ell_h)_{\Th}
  \quad
  \forall\ell_h\in\Phi_h^{r_s}.
\end{equation*}
For $k=0,\ldots,M_t-1$, find $\varphi_h^{k+1},\hat\mu_h^{k\star}, \mathcal U_h^{k\star},\Xi_h^{k\star}\in\Phi_h^{r_n}$, $\vartheta_h^{k+1},\hat S_h^{k+1}\in\Phi_h^{r_s}$ and $\bm u_h^{k+1}\in\bm V_h^{r_u}$ satisfying \eqref{eq:fully_discret_mass}, \eqref{eq:fully_discret_ns} and \eqref{eq:fully_discret_u2}, together with
\begin{subequations}\label{eq:fully_discret_T_eq}
\begin{align}
  &\frac{d}{2}
  (n_h^{k+1/2}T_h^{k\star}\delta_t\vartheta_h^{k+1},
   \theta_h)_{\Th}
  +\widetilde{\mathcal B}_h
  (\bm u_h^{k+1/2};T_h^{k\star}\theta_h,
  \hat S_h^{k+1/2},\bm u_h^{k+1/2})
  \notag\\
  &\quad
  -\mathcal H_h
  (n_h^{k+1/2},T_h^{k\star}\theta_h,
   \bm u_h^{k+1/2})
  -\frac{1}{\rmRe}\mathcal A_h^T
  (1,n_h^{k+1/2},T_h^{k\star},T_h^{k\star}\theta_h)
  +\frac{1}{\rmRe}\mathcal A_h^T
  (\theta_h,n_h^{k+1/2},T_h^{k\star},T_h^{k\star})
  \notag\\
  &\quad
  -\rmR\mathcal A_h^u
  (n_h^{k+1/2}\theta_h,\bm u_h^{k+1/2},
   \bm u_h^{k+1/2})
  -\widetilde{\mathcal B}_h
  (\bm u_h^{k+1/2};\Xi_h^{k\star}\theta_h,
   n_h^{k+1/2},\bm u_h^{k+1/2})
  \notag\\
  &\quad
  +\mathcal D_{\Gamma,h}
  \la n_h^{k+1/2},T_h^{k\star},
      \bm u_h^{k+1/2},\theta_h\ra
  +\mathcal H_{\Gamma,h}
  \la n_h^{k+1/2},T_h^{k\star},
      \bm u_h^{k+1/2},\theta_h\ra
  \notag\\
  &\quad
  +\la\delta_t\sigma_{s,h}^{k+1},
      T_h^{k\star}\theta_h\ra_{\ED}
  +\mathcal B_{\Gamma,h}
  \la T_h^{k\star}\theta_h,\sigma_{s,h}^{k+1/2},
      \bm u_h^{k+1/2}\ra
  -\frac{\rmLd}{\rmRe}\mathcal A_{\Gamma,h}^T
  \la1,n_h^{k+1/2},T_h^{k\star},
      T_h^{k\star}\theta_h\ra
  \notag\\
  &\quad
  +\frac{\rmLd}{\rmRe}\mathcal A_{\Gamma,h}^T
  \la\theta_h,n_h^{k+1/2},T_h^{k\star},
      T_h^{k\star}\ra
  -\frac{1}{\rmLb}
  \la(\mathrm D_{\delta t}^{\Gamma} n_h^{k+1/2})^2,\theta_h\ra_{\ED}
  -\frac{\rmR}{\rmLs}
  \la\beta_h^{k+1/2}
      |\bm u_{\tau,h}^{\mathrm{slip},k+1/2}|^2,
      \theta_h\ra_{\ED}
  \notag\\
  &\quad
  -\frac{1}{\rmRe\rmLk}
  \la\kappa_h^{k+1/2}(q_{w,\nu,h}^{k\star})^2,
      T_h^{k\star}\theta_h\ra_{\ED}
  +\frac{1}{\rmRe\rmLk}
  \la T_w^{-1}q_{w,\nu,h}^{k\star},
      T_h^{k\star}\theta_h\ra_{\ED}
  +\mathcal J_h(\varpi_h^{k+1/2},\theta_h)
  =0,
  \label{eq:fully_discret_T_T}\\
  &(\delta_t\hat S_h^{k+1},T_h^{k\star}\ell_h)_{\Th}
  -\frac{d}{2}
  (n_h^{k+1/2}T_h^{k\star}\delta_t\vartheta_h^{k+1},
   \ell_h)_{\Th}
  +\widetilde{\mathcal B}_h
  (\bm u_h^{k+1/2};\Xi_h^{k\star}\ell_h,
   n_h^{k+1/2},\bm u_h^{k+1/2})
  \notag\\
  &\quad
  +\mathcal D_h
  (n_h^{k+1/2},T_h^{k\star}\ell_h,
   \bm u_h^{k+1/2})
  +\mathcal H_h
  (n_h^{k+1/2},T_h^{k\star}\ell_h,
   \bm u_h^{k+1/2})
  \notag\\
  &\quad
  -\mathcal D_{\Gamma,h}
  \la n_h^{k+1/2},T_h^{k\star},
      \bm u_h^{k+1/2},\ell_h\ra
  -\mathcal H_{\Gamma,h}
  \la n_h^{k+1/2},T_h^{k\star},
      \bm u_h^{k+1/2},\ell_h\ra
  -\mathcal J_h(\varpi_h^{k+1/2},\ell_h)
  =0,
  \label{eq:fully_discret_T_S}\\
  &(\hat\mu_h^{k\star},
    \chi_h^{k\star}\psi_h)_{\Th}
  -(\hat{e}_{n,h}^{k+1/2},
    \chi_h^{k\star}\psi_h)_{\Th}
  +(T_h^{k\star}\delta_{\varphi}S,\psi_h)_{\Th}
  \notag\\
  &\quad
  -2\mathcal A_h^n
  (n_h^{k+1/2},T_h^{k\star}\chi_h^{k\star}\psi_h)
  +\la\hat F_{\nu,h}^{k\star},
      \chi_h^{k\star}\psi_h\ra_{\ED}
  =0,
  \label{eq:fully_discret_T_mu}\\
  &(\Xi_h^{k\star},\chi_h^{k\star}z_h)_{\Th}
  -(\hat{e}_{n,h}^{k+1/2},\chi_h^{k\star}z_h)_{\Th}
  +(\hat\mu_h^{k\star},\chi_h^{k\star}z_h)_{\Th}
  =0,
  \label{eq:fully_discret_T_Xi}
\end{align}
\end{subequations}
for all $\phi_{h},\psi_{h},\varrho_h,z_h\in\Phi_{h}^{r_{n}}$, $\theta_{h},\ell_{h}\in\Phi_{h}^{r_{s}}$ and $\bm{v}_{h}\in\bm{V}_{h}^{r_{u}}$.
In the shared equations, $T_h^{k\star}$ is the logarithmic mean defined above, and $\hat S_h$ is the auxiliary entropy unknown updated by \eqref{eq:fully_discret_T_S}.
The quantities $\mathrm D_{\delta t}^{\Gamma} n_h^{k+1/2}$, $\bm u_{\tau,h}^{\mathrm{slip},k+1/2}$, $\hat F_{\nu,h}^{k\star}$, $\sigma_{s,h}^{k+1/2}$, $\zeta_h^{k+1/2}$, $q_{w,\nu,h}^{k\star}$  and $\mathcal J_h$ are defined as in the $S$-scheme.
The weighted boundary forms are
\begin{align*}
  \mathcal D_{\Gamma,h}
  \la n_h^{k+1/2},T_h^{k\star},
      \bm u_h^{k+1/2},\theta_h\ra
  &\coloneqq
  -\la\hat F_{\nu,h}^{k\star}\mathrm D_{\delta t}^{\Gamma} n_h^{k+1/2},
      \theta_h\ra_{\ED},\\
  \mathcal H_{\Gamma,h}
  \la n_h^{k+1/2},T_h^{k\star},
      \bm v_h,\theta_h\ra
  &\coloneqq
  \la\hat F_{\nu,h}^{k\star}\theta_h
      \nabla_{\tau,h}n_h^{k+1/2},
      \bm v_{\tau,h}\ra_{\ED}.
\end{align*}
Setting $\theta_h=1$ recovers the boundary form $\mathcal H_{\Gamma,h}$ defined in \cref{subsec:Fully discrete scheme for the entropy-based model}.
In \eqref{eq:fully_discret_T_eq}, $\mathcal U_h^{k\star}$ and $\Xi_h^{k\star}$ are weighted $L^2$ projections of $\bm u_h^k\cdot\bm u_h^{k+1}/2$ and $\hat{e}_{n,h}^{k+1/2}-\hat\mu_h^{k\star}$ onto $\Phi_h^{r_n}$, respectively.

\subsection{Physics-preserving properties}
\label{subsec:Physics-preserving properties}

To summarize, both schemes solve for the logit density $\varphi_h$ and velocity $\bm u_h$, with the physical number density evaluated as $n_h=\mathcal N(\varphi_h)$.
The thermal variables are treated as follows:
\begin{itemize}
  \item $S$-scheme: The primary thermal unknown is the entropy density $\hat S_h$, coupled with an auxiliary homogeneous entropy unknown $S_h$.
  The temperature is recovered from $(n_h,S_h)$ by
  \begin{equation}\label{eq:T_S-model}
    T_h^k
    =
    \biggl(
      \frac{n_h^k}{1-n_h^k}
      \exp\Bigl(\frac{S_h^k}{n_h^k}\Bigr)
    \biggr)^{2/d}.
  \end{equation}
  The quantity $T_h^{k\star}=\delta_S\hat e$ is defined through the discrete internal energy chain rule.

  \item $T$-scheme: The primary thermal unknown is the logarithmic temperature $\vartheta_h$, which is coupled with the auxiliary entropy unknown $\hat S_h$.
  The physical temperature $T_h=\exp(\vartheta_h)$ and homogeneous entropy density $S_h=S(\varphi_h,\vartheta_h)$ are evaluated pointwise.
  Here, $T_h^{k\star}$ is the logarithmic mean of $T_h^k$ and $T_h^{k+1}$.
\end{itemize}
In both schemes, the generalized chemical potential $\hat\mu_h^{k\star}$ is an auxiliary unknown.
Assuming the discrete systems are solvable, the following properties hold.
Pointwise bounds apply elementwise, including facet traces.

\begin{proposition}[Discrete mass conservation]
Both fully discrete schemes conserve the total particle number and hence the total mass:
\begin{equation}\label{eq:massbalance_fullydiscret}
  (n_h^{k},1)_{\Th}
  =
  (n_h^0,1)_{\Th},
  \quad k=0,\ldots,M_t.
\end{equation}
\end{proposition}
\begin{proof}
Taking $\phi_{h}=\mathbf{1}_{K}\in\Phi_{h}^{r_{n}}$ in \eqref{eq:fully_discret_mass}, we obtain the local mass balance on any $K\in\Th$.
The interior fluxes cancel upon summation over all elements, and the boundary flux vanishes by the impermeability boundary condition.
This proves \eqref{eq:massbalance_fullydiscret}, and since the particle mass $m$ is constant, the total mass is conserved as well.
\end{proof}

\begin{proposition}[Discrete number density bounds and temperature positivity]\label{prop:fully_discrete_admissibility}
Suppose that the initial conditions satisfy $0<n_{h}^{0}<1$ and $T_{h}^{0}>0$.
Then, both fully discrete schemes satisfy
\begin{equation*}
  0<n_{h}^{k}<1,\qquad T_{h}^{k}>0,
  \quad k=0,\ldots,M_{t},\quad\text{ and }\quad
  T_{h}^{k\star}>0,
  \quad k=0,\ldots,M_{t}-1.
\end{equation*}
\end{proposition}

\begin{proof}
For both schemes, the parametrization $n_h^k=\mathcal N(\varphi_h^k)$ ensures $0<n_h^k<1$ for every finite $\varphi_h^k$.
In the $S$-scheme, the recovered temperature $T_h^k$ is positive by \eqref{eq:T_S-model}.
The identity $\partial_S\hat e(\varphi,S)=T(\varphi,S)>0$ shows that $\hat e(\varphi,S)$ is strictly increasing in $S$ for fixed $\varphi$.
Hence, $(\hat e(\varphi,b)-\hat e(\varphi,a))/(b-a)>0$ for $a\neq b$.
At $a=b$, its continuous extension equals $T(\varphi,a)>0$.
Therefore, we obtain $T_h^{k\star}=\delta_S\hat e>0$.

In the $T$-scheme, $T_h^k=\exp(\vartheta_h^k)>0$.
Since the exponential function is strictly increasing, the logarithmic mean $T_h^{k\star}>0$ when $\vartheta_h^{k+1}\neq\vartheta_h^k$.
When $\vartheta_h^{k+1}=\vartheta_h^k$, we have $T_h^{k\star}=\exp(\vartheta_h^k)>0$ by continuous extension.
The proof is complete.
\end{proof}

At time $t^k$, the discrete total and kinetic energies are
\begin{equation*}
  \mathcal{E}_{\text{tot},h}^{k}
  \coloneqq
  \mathcal{E}_{k,h}^{k}
  +
  \mathcal{E}_{b,h}^{k},
  \qquad
  \mathcal{E}_{k,h}^{k}
  \coloneqq
  \frac{\rmRe\rmR}{2}
  (
  n_{h}^{k}|\bm{u}_{h}^{k}|^{2},1
  )_{\Th},
\end{equation*}
where the bulk internal energy is given by $\mathcal{E}_{b,h}^{k}\coloneqq(\hat{e}(\varphi_{h}^{k},S_{h}^{k}),1)_{\Th}$ in the $S$-scheme and $\mathcal{E}_{b,h}^{k}\coloneqq(\hat{e}(\varphi_{h}^{k},\vartheta_{h}^{k}),1)_{\Th}$ in the $T$-scheme.

\begin{theorem}[Discrete energy balance]\label{thm:fully_1st_law}
Both fully discrete schemes satisfy the first law of thermodynamics:
\begin{equation}\label{eq:fully_1st_law}
  \delta_{t}\mathcal{E}_{\text{tot},h}^{k+1}
  =
  -(n_{h}^{k+1/2}\rmG\bm{e}_{d},
  \bm{u}_{h}^{k+1/2})_{\Th}
  -\frac{\rmR}{\rmLs}
  \la\beta_{h}^{k+1/2}
  \bm{u}_{\tau,h}^{\mathrm{slip},k+1/2},
  \bm{u}_{w}\ra_{\ED}
  -\frac{1}{\rmRe\rmLk}
  \la q_{w,\nu,h}^{k\star},1\ra_{\ED}.
\end{equation}
\end{theorem}

\begin{proof}
Taking $\phi_{h}=\mathcal{U}_{h}^{k\star}$ in \eqref{eq:fully_discret_mass}, $\bm{v}_{h}=\bm{u}_{h}^{k+1/2}$ in \eqref{eq:fully_discret_ns} and $\varrho_h=\delta_t\varphi_h^{k+1}$ in \eqref{eq:fully_discret_u2}, we obtain
\begin{equation}\label{eq:E_K^{k}+1-E_K^{k}_fully}
  \begin{aligned}
    \delta_{t}\mathcal{E}_{k,h}^{k+1}
    &=\widetilde{\mathcal{B}}_{h}
    (\bm{u}_{h}^{k+1/2};\hat{\mu}_{h}^{k\star},
    n_{h}^{k+1/2},\bm{u}_{h}^{k+1/2})
    +\widetilde{\mathcal{B}}_{h}
    (\bm{u}_{h}^{k+1/2};T_{h}^{k\star},
    \hat{S}_{h}^{k+1/2},\bm{u}_{h}^{k+1/2})
    \\
    &\quad-\mathcal{H}_{h}(n_{h}^{k+1/2},T_{h}^{k\star},\bm{u}_{h}^{k+1/2})
    -\rmR\mathcal{A}^{u}_{h}
    (n_{h}^{k+1/2},\bm{u}_{h}^{k+1/2},\bm{u}_{h}^{k+1/2})\\
    &\quad -(n_{h}^{k+1/2}\rmG\bm{e}_{d},
    \bm{u}_{h}^{k+1/2})_{\Th}
    +\mathcal{H}_{\Gamma,h}\la
    n_{h}^{k+1/2},T_{h}^{k\star},\bm{u}_{h}^{k+1/2}\ra\\
    &\quad
    -\frac{\rmR}{\rmLs}
    \la \beta_{h}^{k+1/2}
    \bm{u}_{\tau,h}^{\mathrm{slip},k+1/2},
    \bm{u}_{\tau,h}^{k+1/2}\ra_{\ED}
    +\mathcal{C}_{\Gamma,h}\la
    T_{h}^{k\star},
    \sigma_{s,h}^{k+1/2},
    \bm{u}_{h}^{k+1/2}
    \ra.
  \end{aligned}
\end{equation}
For the $S$-scheme, the discrete chain rule gives $\delta_t\mathcal E_{b,h}^{k+1} = (\delta_{\varphi}\hat e, \delta_t\varphi_h^{k+1})_{\Th} +(T_h^{k\star},\delta_t S_h^{k+1})_{\Th}$.
Taking $\phi_{h}=\hat{\mu}_{h}^{k\star}$ in \eqref{eq:fully_discret_mass}, $\psi_{h}=\delta_t\varphi_h^{k+1}$ in \eqref{eq:fully_discret_mu} and $\ell_{h}=1$ in \eqref{eq:fully_discret_chi} gives
\begin{align*}
  (\delta_{\varphi}\hat e,
   \delta_t\varphi_h^{k+1})_{\Th}
  &=
  -\widetilde{\mathcal B}_h
  (\bm u_h^{k+1/2};\hat\mu_h^{k\star},
   n_h^{k+1/2},\bm u_h^{k+1/2})
  \\
  &\quad-2\mathcal A_h^n
  (n_h^{k+1/2},T_h^{k\star}\delta_t n_h^{k+1})
  +\la\hat F_{\nu,h}^{k\star},
      \delta_t n_h^{k+1}\ra_{\ED},
  \\
  (T_h^{k\star},\delta_t S_h^{k+1})_{\Th}
  &=
  (\delta_t\hat S_h^{k+1},T_h^{k\star})_{\Th}
  +2\mathcal A_h^n
  (n_h^{k+1/2},T_h^{k\star}\delta_t n_h^{k+1})
  \\
  &\quad +\mathcal D_h
  (n_h^{k+1/2},T_h^{k\star},\bm u_h^{k+1/2})
  +\mathcal H_h
  (n_h^{k+1/2},T_h^{k\star},\bm u_h^{k+1/2}).
\end{align*}
Substituting the above relations into $\delta_t\mathcal E_{b,h}^{k+1}$ and taking $\theta_{h}=1$ in \eqref{eq:fully_discret_S} yields
\begin{equation}\label{eq:E_b_common_fully}
  \begin{aligned}
    \delta_{t}\mathcal{E}_{b,h}^{k+1}
    &=
    -\widetilde{\mathcal{B}}_{h}
    (\bm{u}_{h}^{k+1/2};
    \hat{\mu}_{h}^{k\star},
    n_{h}^{k+1/2},
    \bm{u}_{h}^{k+1/2})
    +\mathcal{H}_{h}
    (n_{h}^{k+1/2},
    T_{h}^{k\star},
    \bm{u}_{h}^{k+1/2})
    \\
    &\quad
    +\rmR\mathcal{A}^{u}_{h}
    (n_{h}^{k+1/2},
    \bm{u}_{h}^{k+1/2},
    \bm{u}_{h}^{k+1/2})
    -\widetilde{\mathcal{B}}_{h}
    (\bm{u}_{h}^{k+1/2};
    T_{h}^{k\star},
    \hat{S}_{h}^{k+1/2},
    \bm{u}_{h}^{k+1/2})
    \\
    &\quad
    +\frac{1}{\rmLb}
    \la
    \mathrm D_{\delta t}^{\Gamma} n_h^{k+1/2},
    \bm{u}_{\tau,h}^{k+1/2}
    \cdot\nabla_{\tau,h}n_{h}^{k+1/2}
    \ra_{\ED}
    -\mathcal{B}_{\Gamma,h}\la
    T_{h}^{k\star},
    \sigma_{s,h}^{k+1/2},
    \bm{u}_{h}^{k+1/2}\ra
    \\
    &\quad
    +\frac{\rmR}{\rmLs}
    \la
    \beta_{h}^{k+1/2}
    |\bm{u}_{\tau,h}^{\mathrm{slip},k+1/2}|^{2},
    1\ra_{\ED}
    -\frac{1}{\rmRe\rmLk}
    \la q_{w,\nu,h}^{k\star},1\ra_{\ED},
  \end{aligned}
\end{equation}
where the definitions of $\hat{F}_{\nu,h}^{k\star}$, $q_{w,\nu,h}^{k\star}$, $\mathrm D_{\delta t}^{\Gamma} n_h^{k+1/2}$ and $\delta_{t}\sigma_{s,h}^{k+1}=-\rmWs\delta_{t}n_{h}^{k+1}$ have been used.
Adding \eqref{eq:E_K^{k}+1-E_K^{k}_fully} and \eqref{eq:E_b_common_fully} cancels the bulk couplings.
The remaining surface terms cancel by
\begin{equation*}
\begin{aligned}
  &\mathcal C_{\Gamma,h}\la T_h^{k\star},\sigma_{s,h}^{k+1/2},\bm u_h^{k+1/2}\ra
   -\mathcal B_{\Gamma,h}\la T_h^{k\star},\sigma_{s,h}^{k+1/2},\bm u_h^{k+1/2}\ra\\
  &\quad=-\mathcal H_{\Gamma,h}\la n_h^{k+1/2},T_h^{k\star},\bm u_h^{k+1/2},1\ra
   -\frac{1}{\rmLb}\la\mathrm D_{\delta t}^\Gamma n_h^{k+1/2},
     \bm u_{\tau,h}^{k+1/2}\cdot\nabla_{\tau,h}n_h^{k+1/2}\ra_{\ED}.
\end{aligned}
\end{equation*}
This identity follows from the definitions of $\sigma_{s,h}^{k+1/2}$, $\mathcal H_{\Gamma,h}$ and $\hat F_{\nu,h}^{k\star}$.
Thus, we obtain \eqref{eq:fully_1st_law}.

Using the midpoint identity for the $T$-scheme and taking $z_h=\delta_t\varphi_h^{k+1}$ in \eqref{eq:fully_discret_T_Xi}, we obtain
\begin{equation*}
\delta_{t}\mathcal{E}_{b,h}^{k+1}=
  \frac{d}{2}
  (n_{h}^{k+1/2}T_{h}^{k\star}
  \delta_{t}\vartheta_{h}^{k+1},1)_{\Th}
  +
  (\Xi_{h}^{k\star}+\hat{\mu}_{h}^{k\star},
  \delta_{t}n_{h}^{k+1})_{\Th}.
\end{equation*}
Taking $\theta_h=1$ in \eqref{eq:fully_discret_T_T} and $\phi_h=\Xi_h^{k\star}+\hat\mu_h^{k\star}$ in \eqref{eq:fully_discret_mass} gives the bulk terms in \eqref{eq:E_b_common_fully}.
The boundary terms agree by
\begin{align*}
  \frac{1}{\rmLb}
  \la
  \mathrm D_{\delta t}^{\Gamma} n_h^{k+1/2},
  \bm{u}_{\tau,h}^{k+1/2}
  \cdot\nabla_{\tau,h}n_{h}^{k+1/2}
  \ra_{\ED}&=-\mathcal{D}_{\Gamma,h}\la
  n_{h}^{k+1/2},
  T_{h}^{k\star},
  \bm{u}_{h}^{k+1/2},1\ra
  -\la\delta_t\sigma_{s,h}^{k+1},T_h^{k\star}\ra_{\ED}
  \\
  &\quad-\mathcal H_{\Gamma,h}\la
  n_h^{k+1/2},T_h^{k\star},\bm u_h^{k+1/2},1\ra
  +\frac{1}{\rmLb}
  \la
  (\mathrm D_{\delta t}^{\Gamma} n_h^{k+1/2})^{2},1
  \ra_{\ED}.
\end{align*}
Thus the $T$-scheme also satisfies \eqref{eq:E_b_common_fully}, and the same boundary cancellations yield \eqref{eq:fully_1st_law}.
\end{proof}

\begin{theorem}[Discrete weak entropy inequality]\label{thm:fully_2nd_law}
Both schemes satisfy the fully discrete second law of thermodynamics:
\begin{equation}\label{eq:fully_2nd_law}
  \begin{aligned}
    &(\delta_{t}\hat{S}_{h}^{k+1},
    T_{h}^{k\star}\theta_{h})_{\Th}
    +\widetilde{\mathcal{B}}_{h}
    (\bm{u}_{h}^{k+1/2};
    T_{h}^{k\star}\theta_{h},
    \hat{S}_{h}^{k+1/2},
    \bm{u}_{h}^{k+1/2})
    +\mathcal{D}_{h}
    (n_{h}^{k+1/2},
    T_{h}^{k\star}\theta_{h},
    \bm{u}_{h}^{k+1/2})
    \\
    &\quad
    -\frac{1}{\rmRe}
    \mathcal{A}^{T}_{h}
    (1,n_{h}^{k+1/2},
    T_{h}^{k\star},
    T_{h}^{k\star}\theta_{h})
    +\la
    \delta_{t}\sigma_{s,h}^{k+1},
    T_{h}^{k\star}\theta_{h}\ra_{\ED}
    +\mathcal{B}_{\Gamma,h}\la
    T_{h}^{k\star}\theta_{h},
    \sigma_{s,h}^{k+1/2},
    \bm{u}_{h}^{k+1/2}\ra
    \\
    &\quad
    -\frac{\rmLd}{\rmRe}
    \mathcal{A}^{T}_{\Gamma,h}\la
    1,n_{h}^{k+1/2},
    T_{h}^{k\star},
    T_{h}^{k\star}\theta_{h}\ra
    +\frac{1}{\rmRe\rmLk}
    \la
    T_{w}^{-1}
    q_{w,\nu,h}^{k\star},
    T_{h}^{k\star}\theta_{h}
    \ra_{\ED}
    \geq 0,
  \end{aligned}
\end{equation}
for every $\theta_h\in\Phi_h^{r_s}$ with $\theta_h\geq0$.
\end{theorem}

\begin{proof}
In the $S$-scheme, \eqref{eq:fully_2nd_law} follows directly from \eqref{eq:fully_discret_S}.
We obtain the same discrete entropy balance for the $T$-scheme by taking $\ell_{h}=\theta_{h}$ in \eqref{eq:fully_discret_T_S} and adding the resulting equation to \eqref{eq:fully_discret_T_T}.
In both schemes, the left-hand side of \eqref{eq:fully_2nd_law} can be expressed as
\begin{align*}
  \mathrm{LHS}&=\rmR\mathcal{A}^{u}_{h}
  (n_{h}^{k+1/2}\theta_{h},
  \bm{u}_{h}^{k+1/2},
  \bm{u}_{h}^{k+1/2})
  -\frac{1}{\rmRe}
  \mathcal{A}^{T}_{h}
  (\theta_{h},
  n_{h}^{k+1/2},
  T_{h}^{k\star},
  T_{h}^{k\star})
  \\
  &\quad
  -\frac{\rmLd}{\rmRe}
  \mathcal{A}^{T}_{\Gamma,h}\la
  \theta_{h},
  n_{h}^{k+1/2},
  T_{h}^{k\star},
  T_{h}^{k\star}\ra
  +\frac{1}{\rmLb}
  \la
  (\mathrm D_{\delta t}^{\Gamma} n_h^{k+1/2})^{2},
  \theta_{h}\ra_{\ED}
  \\
  &\quad
  +\frac{\rmR}{\rmLs}
  \la
  \beta_{h}^{k+1/2}
  |\bm{u}_{\tau,h}^{\mathrm{slip},k+1/2}|^{2},
  \theta_{h}\ra_{\ED}
  +\frac{1}{\rmRe\rmLk}
  \la
  \kappa_{h}^{k+1/2}
  (q_{w,\nu,h}^{k\star})^{2},
  T_{h}^{k\star}\theta_{h}
  \ra_{\ED}.
\end{align*}
Setting $\bm v=\bm u$ in the viscous form $\mathcal{A}^{u}_{h}$, and $\theta=\psi$ in the thermal forms $\mathcal{A}^{T}_{h}$ and $\mathcal{A}^{T}_{\Gamma,h}$, we have
\begin{align*}
  \mathcal{A}^{u}_{h}(\phi,\bm{u},\bm{u})
  &=
  \frac{1}{2}
  (\phi|D_{h}(\bm{u})|^{2},1)_{\Th}
  +
  \Bigl(1-\frac{2}{d}\Bigr)
  (\phi(\operatorname{div}_{h} \bm{u})^{2},1)_{\Th}
  +\alpha_{u}
  \la
  h_{E}^{-1}\avg{\phi}\jump{\bm{u}},
  \jump{\bm{u}}
  \ra_{\EI}
  \geq0,
  \\
  -\mathcal{A}^{T}_{h}(a,\phi,\psi,\psi)
  &=
  \Bigl(
  a\frac{\phi}{\psi}|\nabla_{h}\psi|^{2},1
  \Bigr)_{\Th}
  +
  \alpha_{T}
  \Bigl\la
  h_{E}^{-1}
  \frac{\avg{a\phi}}{\avg{\psi}}
  \jump{\psi},
  \jump{\psi}
  \Bigr\ra_{\EI}
  \geq0,
  \\
  -\mathcal{A}^{T}_{\Gamma,h}\la a,\phi,\psi,\psi\ra
  &=
  \Bigl\la
  a\frac{\lambda_{s}(\phi)}{\psi}
  |\nabla_{\tau,h}\psi|^{2},1
  \Bigr\ra_{\ED}+
  \alpha_{\Gamma}
  \Bigl\la
  h_{e}^{-1}
  \frac{\avgbd{a\lambda_{s}(\phi)}}
       {\avgbd{\psi}}
  \jumpbd{\psi},
  \jumpbd{\psi}
  \Bigr\ra_{\EV}
  \geq0.
\end{align*}
For $0<n_h^{k+1/2}<1$, $T_h^{k\star}>0$ and $\theta_h\geq0$, these identities show that every term in the entropy production is nonnegative, which proves \eqref{eq:fully_2nd_law}.
\end{proof}

\section{Numerical examples}
\label{sec:num_ex}

In this section, we test the accuracy, number density bounds and temperature positivity of both the $S$- and $T$-schemes in \cref{subsec:convergence_structure_tests}.
The subsequent physical examples use the $T$-scheme, as the two schemes show comparable behavior in these tests.

The finite element discretization is implemented with FEniCS \cite{logg2012automated}.
We use uniform triangular ($d=2$) or tetrahedral ($d=3$) meshes and choose $\bm V_h^{r_u}$ to be the RT space with vanishing normal trace on $\Gamma$.
We set the penalty parameters to $\alpha_n=10r_n^2$, $\alpha_u=10(r_u+1)^2$ and $\alpha_T=\alpha_\Gamma=10r_s^2$, and the upwind parameter to $c_{\mathrm{up}}=10$.
The relative and absolute tolerances for Newton's method are $(10^{-9},10^{-10})$ in \cref{subsec:convergence_structure_tests} and $(10^{-10},10^{-11})$ in the physical examples.
We use midpoint derivatives to approximate secants with absolute increments below $10^{-6}$.

In the physical examples, we report the relative mass error $(M^k-M^0)/|M^0|$, with $M^k\coloneqq(n_h^k,1)_{\Th}$, and the normalized energy balance residual $\mathrm{RES}_E^k\coloneqq (E_h^k-E_h^0+Q_h^k-W_h^k)/|E_h^0|$.
Here, $E_h^k\coloneqq\mathcal E_{\text{tot},h}^k+(\rmG(\bm x\cdot\bm e_d)n_h^k,1)_{\Th}$ includes gravitational potential energy.
The cumulative outward heat transfer and work of the additional body force $\bm f_h$ are
\begin{equation*}
Q_h^k\coloneqq
\delta t\rmRe^{-1}\sum_{j=0}^{k-1}
\la\rmLk^{-1}q_{w,\nu,h}^{j\star},1\ra_{\ED},
\qquad
W_h^k\coloneqq
\delta t\sum_{j=0}^{k-1}
(\bm f_h^{j+1/2},\bm u_h^{j+1/2})_{\Th},
\end{equation*}
respectively, with $\bm f_h=\bm0$ when no additional forcing is applied.
We also report the entropy sum $\mathcal S_{\mathrm{tot},h}^k+\mathcal S_{\mathrm{out},h}^k$, where the total entropy and net cumulative entropy transferred to the walls are
\begin{equation*}
\mathcal S_{\mathrm{tot},h}^k\coloneqq
(\hat S_h^k,1)_{\Th}+\la\sigma_{s,h}^k,1\ra_{\ED},
\qquad
\mathcal S_{\mathrm{out},h}^k\coloneqq
\delta t\rmRe^{-1}\sum_{j=0}^{k-1}
\la\rmLk^{-1}T_w^{-1}q_{w,\nu,h}^{j\star},1\ra_{\ED}.
\end{equation*}
The temperature-weighted entropy production $\gamma_{\mathrm{tot},h}^{k\star}$ is the sum of the nonnegative production terms in \cref{thm:fully_2nd_law} for $\theta_h=1$.

\subsection{Convergence, number density bounds and temperature positivity}
\label{subsec:convergence_structure_tests}

\paragraph{Accuracy test.}
We test the convergence of both schemes using manufactured solutions on $\Omega=(0,1)^d$, $d=2,3$.
Each coordinate interval is divided into $N$ equal subintervals.
Let $s_\phi\coloneqq \sin(\pi\phi)$ and $c_\phi\coloneqq \cos(\pi\phi)$.
We set
\begin{align*}
  \text{for } d=2: \quad &n_e=0.5+0.1\exp(-t)(s_x+s_y),\qquad
  T_e=0.3+0.1\exp(-t)c_xc_y,\\
  &\bm u_e=0.05\exp(-t)((1+y)s_xc_y,(1+x)s_yc_x)^{\mathsf T},\\
  \text{for } d=3: \quad &n_e=0.5+0.1\exp(-t)(s_x+s_y+s_z),\qquad
  T_e=0.3+0.1\exp(-t)c_xc_yc_z,\\
  &\bm u_e=0.05\exp(-t)((1+y)s_xc_yc_z,
  (1+z)s_yc_zc_x,(1+x)s_zc_xc_y)^{\mathsf T},
\end{align*}
which satisfy $\bm u_e\cdot\bm\nu=0$ on $\Gamma$, while $\operatorname{div}\bm u_e$ is generally nonzero.
We prescribe the tangential wall velocity $\bm u_w=0.4\bm P_\tau\bm u_e$ on $\Gamma$.
The exact entropy density $\hat{S}_{e}$ follows from \eqref{eq:dimensionless_ST}, and the transformed variables are $\varphi_e=\ln(n_e/(1-n_e))$ and $\vartheta_e=\ln T_e$.

The bulk and boundary source terms are calculated by substituting the above manufactured solutions into the continuous equations \eqref{eq:dimensionless_mass}, \eqref{eq:dimensionless_NS2} and \eqref{eq:entropynew} and the boundary conditions \eqref{eq:dimensionless_rbc}--\eqref{eq:dimensionless_Tbcd}.
The mass source term also induces bulk and boundary terms in the auxiliary equations in both schemes.
Likewise, corrections due to the sources are included when the thermal boundary condition is rewritten in entropy form.
All source terms are evaluated at $t^{k+1/2}$.

The physical and stabilization parameters are set to $\rmRe=\rmR=\rmG=\rmLb=\rmLs=\rmLd=\rmLk=1$, $\rmWs=0.1$, $T_w=0.5$, $n_c=1/3$, $n_g=0.2$, $n_l=0.8$, $\beta_{gl}=\kappa_{gl}=\lambda_{s,gl}=0.5$ and $\alpha_\varpi^1=\alpha_\varpi^2=0.1$.
In the spatial refinement tests, we take $N=2,4,8,16$ and $\delta t=10^{-4}$.
We use $r_n=r_s=r_u=1,2,3$ and $t_f=0.5$ for $d=2$, and $r_n=r_s=r_u=1$ and $t_f=0.1$ for $d=3$.
To assess temporal convergence, we fix $d=2$, $N=16$, $r_n=r_s=r_u=4$, $t_f=0.5$ and take $\delta t=1/10,1/20,1/40,1/80$.
All tabulated errors are measured in $L^2(\Omega)$ at $t_f$, with rates $\log_2(E_{\rm coarse}/E_{\rm fine})$.
The spatial temperature errors for the $S$-scheme use the $L^2$ projection of the recovered temperature onto $\Phi_h^{r_s}$, while the temporal errors use the pointwise thermodynamic reconstruction.

The results in \cref{tab:S_model_spatial_convergence,tab:T_model_spatial_convergence,tab:spatial_convergence3d} show the spatial accuracy for both schemes in two and three dimensions.
The density, velocity and temperature variables generally exhibit rates close to the expected approximation orders, while the entropy density converges more slowly.
This behavior may be related to the approximation of the density gradient and the discrete entropy compatibility relation.
\cref{tab:temporal_convergence} shows the second-order temporal accuracy for all reported variables in both schemes.
The mean number of nonlinear iterations per time step is approximately $3$ in the spatial tests and $4$ to $6$ in the temporal tests.

\begin{table}[!htbp]
  \centering
  \caption{Spatial errors and rates for the $S$-scheme: $d=2$, $r_n=r_s=r_u=r$, $t_f=0.5$, $\delta t=10^{-4}$.}
  \label{tab:S_model_spatial_convergence}
  \footnotesize
  \setlength{\tabcolsep}{3.5pt}
  \renewcommand{\arraystretch}{1.10}
  \begin{tabular*}{\textwidth}{@{\extracolsep{\fill}}cc r@{\;}l r@{\;}l r@{\;}l r@{\;}l@{}}
    \toprule
    $r$ & $N$
    & \multicolumn{2}{c}{$\|\varphi_h-\varphi_e\|_{L^2}$}
    & \multicolumn{2}{c}{$\|\bm u_h-\bm u_e\|_{L^2}$}
    & \multicolumn{2}{c}{$\|\hat S_h-\hat S_e\|_{L^2}$}
    & \multicolumn{2}{c}{$\|T_{h,\mathrm{proj}}-T_e\|_{L^2}$}\\
    \midrule
    \multirow{4}{*}{$1$} & 2 & \num{1.124e-1} & ($-$) & \num{3.467e-3} & ($-$) & \num{1.178e-1} & ($-$) & \num{1.408e-2} & ($-$)\\
     & 4 & \num{2.657e-2} & (2.08) & \num{1.794e-3} & (0.95) & \num{3.004e-2} & (1.97) & \num{2.692e-3} & (2.39)\\
     & 8 & \num{6.593e-3} & (2.01) & \num{4.521e-4} & (1.99) & \num{8.067e-3} & (1.90) & \num{6.264e-4} & (2.10)\\
     & 16 & \num{1.645e-3} & (2.00) & \num{1.125e-4} & (2.01) & \num{2.452e-3} & (1.72) & \num{1.469e-4} & (2.09)\\
    \midrule
    \multirow{4}{*}{$2$} & 2 & \num{5.093e-3} & ($-$) & \num{1.021e-3} & ($-$) & \num{7.166e-3} & ($-$) & \num{1.117e-3} & ($-$)\\
     & 4 & \num{5.811e-4} & (3.13) & \num{1.536e-4} & (2.73) & \num{1.649e-3} & (2.12) & \num{3.239e-4} & (1.79)\\
     & 8 & \num{7.246e-5} & (3.00) & \num{2.131e-5} & (2.85) & \num{4.066e-4} & (2.02) & \num{4.505e-5} & (2.85)\\
     & 16 & \num{9.753e-6} & (2.89) & \num{2.908e-6} & (2.87) & \num{1.014e-4} & (2.00) & \num{7.640e-6} & (2.56)\\
    \midrule
    \multirow{4}{*}{$3$} & 2 & \num{5.901e-4} & ($-$) & \num{2.388e-4} & ($-$) & \num{1.680e-3} & ($-$) & \num{5.801e-4} & ($-$)\\
     & 4 & \num{2.992e-5} & (4.30) & \num{1.492e-5} & (4.00) & \num{9.166e-5} & (4.20) & \num{3.020e-5} & (4.26)\\
     & 8 & \num{1.833e-6} & (4.03) & \num{8.976e-7} & (4.06) & \num{6.617e-6} & (3.79) & \num{1.706e-6} & (4.15)\\
     & 16 & \num{1.142e-7} & (4.00) & \num{5.498e-8} & (4.03) & \num{6.439e-7} & (3.36) & \num{1.023e-7} & (4.06)\\
    \bottomrule
  \end{tabular*}
\end{table}

\begin{table}[!htbp]
  \centering
  \caption{Spatial errors and rates for the $T$-scheme: $d=2$, $r_n=r_s=r_u=r$, $t_f=0.5$, $\delta t=10^{-4}$.}
  \label{tab:T_model_spatial_convergence}
  \footnotesize
  \setlength{\tabcolsep}{3.5pt}
  \renewcommand{\arraystretch}{1.10}
  \begin{tabular*}{\textwidth}{@{\extracolsep{\fill}}cc r@{\;}l r@{\;}l r@{\;}l r@{\;}l@{}}
    \toprule
    $r$ & $N$
    & \multicolumn{2}{c}{$\|\varphi_h-\varphi_e\|_{L^2}$}
    & \multicolumn{2}{c}{$\|\bm u_h-\bm u_e\|_{L^2}$}
    & \multicolumn{2}{c}{$\|\vartheta_h-\vartheta_e\|_{L^2}$}
    & \multicolumn{2}{c}{$\|\hat S_h-\hat S_e\|_{L^2}$}\\
    \midrule
    \multirow{4}{*}{$1$} & 2 & \num{1.124e-1} & ($-$) & \num{3.464e-3} & ($-$) & \num{4.587e-2} & ($-$) & \num{1.684e-1} & ($-$)\\
     & 4 & \num{2.651e-2} & (2.08) & \num{1.792e-3} & (0.95) & \num{8.938e-3} & (2.36) & \num{3.844e-2} & (2.13)\\
     & 8 & \num{6.574e-3} & (2.01) & \num{4.516e-4} & (1.99) & \num{2.037e-3} & (2.13) & \num{9.896e-3} & (1.96)\\
     & 16 & \num{1.641e-3} & (2.00) & \num{1.124e-4} & (2.01) & \num{4.774e-4} & (2.09) & \num{2.834e-3} & (1.80)\\
    \midrule
    \multirow{4}{*}{$2$} & 2 & \num{4.930e-3} & ($-$) & \num{1.018e-3} & ($-$) & \num{3.415e-3} & ($-$) & \num{6.986e-3} & ($-$)\\
     & 4 & \num{5.628e-4} & (3.13) & \num{1.535e-4} & (2.73) & \num{9.514e-4} & (1.84) & \num{1.660e-3} & (2.07)\\
     & 8 & \num{6.988e-5} & (3.01) & \num{2.130e-5} & (2.85) & \num{1.284e-4} & (2.89) & \num{4.101e-4} & (2.02)\\
     & 16 & \num{9.141e-6} & (2.93) & \num{2.906e-6} & (2.87) & \num{1.980e-5} & (2.70) & \num{1.025e-4} & (2.00)\\
    \midrule
    \multirow{4}{*}{$3$} & 2 & \num{5.698e-4} & ($-$) & \num{2.388e-4} & ($-$) & \num{1.706e-3} & ($-$) & \num{1.588e-3} & ($-$)\\
     & 4 & \num{2.952e-5} & (4.27) & \num{1.492e-5} & (4.00) & \num{9.246e-5} & (4.21) & \num{8.952e-5} & (4.15)\\
     & 8 & \num{1.830e-6} & (4.01) & \num{8.976e-7} & (4.06) & \num{5.124e-6} & (4.17) & \num{6.671e-6} & (3.75)\\
     & 16 & \num{1.142e-7} & (4.00) & \num{5.499e-8} & (4.03) & \num{3.102e-7} & (4.05) & \num{6.468e-7} & (3.37)\\
    \bottomrule
  \end{tabular*}
\end{table}

\begin{table}[!htbp]
  \centering
  \caption{Spatial errors and rates for both schemes: $d=3$, $r_n=r_s=r_u=1$, $t_f=0.1$, $\delta t=10^{-4}$.}
  \label{tab:spatial_convergence3d}
  \footnotesize
  \setlength{\tabcolsep}{3.5pt}
  \renewcommand{\arraystretch}{1.10}
  \begin{tabular*}{\textwidth}{@{\extracolsep{\fill}}cc r@{\;}l r@{\;}l r@{\;}l r@{\;}l@{}}
    \toprule
    $S$-scheme & $N$
    & \multicolumn{2}{c}{$\|\varphi_h-\varphi_e\|_{L^2}$}
    & \multicolumn{2}{c}{$\|\bm u_h-\bm u_e\|_{L^2}$}
    & \multicolumn{2}{c}{$\|\hat S_h-\hat S_e\|_{L^2}$}
    & \multicolumn{2}{c}{$\|T_{h,\mathrm{proj}}-T_e\|_{L^2}$}\\
    \midrule
     & 2 & \num{1.740e-1} & ($-$) & \num{9.680e-3} & ($-$) & \num{2.414e-1} & ($-$) & \num{9.637e-3} & ($-$)\\
     & 4 & \num{4.329e-2} & (2.01) & \num{3.324e-3} & (1.54) & \num{6.557e-2} & (1.88) & \num{3.485e-3} & (1.47)\\
     & 8 & \num{1.080e-2} & (2.00) & \num{1.011e-3} & (1.72) & \num{1.779e-2} & (1.88) & \num{9.167e-4} & (1.93)\\
     & 16 & \num{2.698e-3} & (2.00) & \num{2.933e-4} & (1.79) & \num{5.517e-3} & (1.69) & \num{2.285e-4} & (2.00)\\
    \midrule
    $T$-scheme & $N$
    & \multicolumn{2}{c}{$\|\varphi_h-\varphi_e\|_{L^2}$}
    & \multicolumn{2}{c}{$\|\bm u_h-\bm u_e\|_{L^2}$}
    & \multicolumn{2}{c}{$\|\vartheta_h-\vartheta_e\|_{L^2}$}
    & \multicolumn{2}{c}{$\|\hat S_h-\hat S_e\|_{L^2}$}\\
    \midrule
     & 2 & \num{1.739e-1} & ($-$) & \num{9.638e-3} & ($-$) & \num{3.135e-2} & ($-$) & \num{2.611e-1} & ($-$)\\
     & 4 & \num{4.323e-2} & (2.01) & \num{3.334e-3} & (1.53) & \num{1.090e-2} & (1.52) & \num{6.811e-2} & (1.94)\\
     & 8 & \num{1.078e-2} & (2.00) & \num{1.013e-3} & (1.72) & \num{2.872e-3} & (1.92) & \num{1.828e-2} & (1.90)\\
     & 16 & \num{2.694e-3} & (2.00) & \num{2.939e-4} & (1.79) & \num{7.201e-4} & (2.00) & \num{5.616e-3} & (1.70)\\
    \bottomrule
  \end{tabular*}
\end{table}

\begin{table}[!htbp]
  \centering
  \caption{Temporal errors and rates for both schemes: $d=2$, $r_n=r_s=r_u=4$, $t_f=0.5$, $N=16$.}
  \label{tab:temporal_convergence}
  \footnotesize
  \setlength{\tabcolsep}{3.5pt}
  \renewcommand{\arraystretch}{1.10}
  \begin{tabular*}{\textwidth}{@{\extracolsep{\fill}}cc r@{\;}l r@{\;}l r@{\;}l r@{\;}l@{}}
    \toprule
    $S$-scheme & $\delta t^{-1}$
    & \multicolumn{2}{c}{$\|\varphi_h-\varphi_e\|_{L^2}$}
    & \multicolumn{2}{c}{$\|\bm u_h-\bm u_e\|_{L^2}$}
    & \multicolumn{2}{c}{$\|\hat S_h-\hat S_e\|_{L^2}$}
    & \multicolumn{2}{c}{$\|T_{h,\mathrm{proj}}-T_e\|_{L^2}$}\\
    \midrule
     & \num{10} & \num{1.311e-4} & ($-$) & \num{5.426e-5} & ($-$) & \num{2.743e-4} & ($-$) & \num{1.615e-4} & ($-$)\\
     & \num{20} & \num{3.368e-5} & (1.96) & \num{1.215e-5} & (2.16) & \num{6.745e-5} & (2.02) & \num{3.897e-5} & (2.05)\\
     & \num{40} & \num{8.415e-6} & (2.00) & \num{3.018e-6} & (2.01) & \num{1.681e-5} & (2.00) & \num{9.783e-6} & (1.99)\\
     & \num{80} & \num{2.103e-6} & (2.00) & \num{7.553e-7} & (2.00) & \num{4.198e-6} & (2.00) & \num{2.449e-6} & (2.00)\\
    \midrule
    $T$-scheme & $\delta t^{-1}$
    & \multicolumn{2}{c}{$\|\varphi_h-\varphi_e\|_{L^2}$}
    & \multicolumn{2}{c}{$\|\bm u_h-\bm u_e\|_{L^2}$}
    & \multicolumn{2}{c}{$\|\vartheta_h-\vartheta_e\|_{L^2}$}
    & \multicolumn{2}{c}{$\|\hat S_h-\hat S_e\|_{L^2}$}\\
    \midrule
     & \num{10} & \num{1.374e-4} & ($-$) & \num{4.642e-5} & ($-$) & \num{2.225e-4} & ($-$) & \num{1.941e-4} & ($-$)\\
     & \num{20} & \num{3.456e-5} & (1.99) & \num{1.227e-5} & (1.92) & \num{3.308e-5} & (2.75) & \num{5.422e-5} & (1.84)\\
     & \num{40} & \num{8.634e-6} & (2.00) & \num{3.051e-6} & (2.01) & \num{8.675e-6} & (1.93) & \num{1.315e-5} & (2.04)\\
     & \num{80} & \num{2.158e-6} & (2.00) & \num{7.636e-7} & (2.00) & \num{2.198e-6} & (1.98) & \num{3.253e-6} & (2.02)\\
    \bottomrule
  \end{tabular*}
\end{table}

\paragraph{Bounds and positivity test.}
We then construct a manufactured test with number density close to the bounds and temperature close to zero.
We retain the two-dimensional velocity, wall velocity and parameters of the accuracy test, and set
\begin{equation*}
  n_e=(1+999^{-a(t)q(x,y)})^{-1},\qquad
  T_e=1000^{-\frac{1-a(t)}2-\frac{a(t)}4(c_x+c_y)^2},
\end{equation*}
where $a(t)\coloneqq (3-\cos(2\pi t))/4$ and $q(x,y)\coloneqq x+y-1+(x(1-x)+y(1-y))/4$.
At $t=0.5$, the exact number density and temperature ranges are $[10^{-3},0.999]$ and $[10^{-3},1]$, respectively, and both fields return to their initial profiles at $t=1$.
We set $N=16$, $r_n=r_s=r_u=3$, $\delta t=10^{-3}$ and $t_f=1$.

\begin{figure}[!htbp]
  \centering
  \begin{subfigure}[t]{0.33\textwidth}
    \centering
    \includegraphics[width=\linewidth]{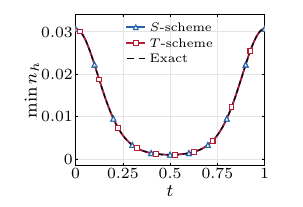}
    \caption{Minimum number density.}
    \label{fig:bounds_density_minimum}
  \end{subfigure}\hfill
  \begin{subfigure}[t]{0.33\textwidth}
    \centering
    \includegraphics[width=\linewidth]{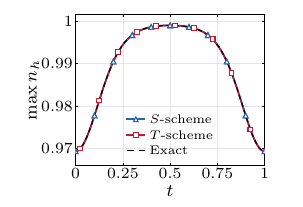}
    \caption{Maximum number density.}
    \label{fig:bounds_density_maximum}
  \end{subfigure}\hfill
  \begin{subfigure}[t]{0.33\textwidth}
    \centering
    \includegraphics[width=\linewidth]{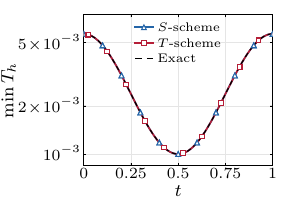}
    \caption{Minimum temperature.}
    \label{fig:bounds_temperature_minimum}
  \end{subfigure}
  \caption{Time evolution of number density extrema and minimum temperature for both schemes.}
  \label{fig:bounds_positive_history}
\end{figure}

\cref{fig:bounds_positive_history} shows the sampled $\min n_h$, $\max n_h$ and $\min T_h$ for both schemes, closely following the exact extrema.
In both schemes, the sampled values satisfy $0<n_h<1$ and $T_h,T_h^{k\star}>0$ throughout the test.
The maximum sampled absolute errors in $(n,T)$ are $(1.282\times10^{-6},2.693\times10^{-4})$ for the $S$-scheme and $(4.770\times10^{-7},2.799\times10^{-5})$ for the $T$-scheme.

\subsection{Sessile droplet dynamics}
\label{subsec:thermal_wetting}

We first examine the effects of wall temperature and wettability on a sessile droplet \cite{xu2012droplet,wang2026jcp}.
In $\Omega=(0,120)\times(0,60)$, the initial number density is
\begin{equation*}
  n^0(\bm x)=n_g+\frac{n_l-n_g}{2}
  \biggl(1-\tanh\Bigl(\frac{|\bm x-\bm x_0|-r_0}{2.5}\Bigr)\biggr),
\end{equation*}
with $\bm x_0=(60,0)$, $r_0=25$, $n_g=0.122$ and $n_l=0.58$.
The droplet is initially semicircular, with $\bm u^0=\bm0$ and $T^0=0.875T_c$, where $T_c=8/27$.
All walls are stationary and impermeable, with density relaxation, velocity slip, surface heat conduction and finite thermal resistance.
The bottom wettability is set to $\rmWs=-0.064$ (hydrophilic) or $0.1$ (hydrophobic), corresponding to equilibrium liquid contact angles of approximately $66^\circ$ and $129^\circ$, respectively, as determined by Young's relation at $T^0$ without gravity \cite{xu2014single}.
We set $\rmWs=0$ corresponding to $90^\circ$ elsewhere.
For each wettability case, we prescribe $T_w/T_c=0.875$, $0.860$ or $0.890$ on the bottom wall from $t=0$, corresponding to the reference, cooled and heated substrate cases, respectively.
On the other three walls, we prescribe $T_w=0.875T_c$.
Indeed, all six cases start from the same fluid state, and the fluid temperature at each wall is determined by the thermal boundary condition, rather than prescribed directly as $T_w$.
We set $\rmRe=1$, $\rmR=0.06$, $\rmG=0$, $\rmLb=250$, $\rmLs=2/n_l$, $\rmLk=1/(n_lT_c^2)$, $\rmLd=0.002$, $n_c=1/3$, $\beta_{gl}=\kappa_{gl}=\lambda_{s,gl}=0.5$ and $\alpha_\varpi^1=\alpha_\varpi^2=0.1$.
We use $r_n=r_s=r_u=2$ on a triangular mesh obtained from a $60\times30$ partition, with $\delta t=0.05$ and $t_f=300$.

\begin{figure}[!htbp]
  \centering
  \captionsetup[subfigure]{skip=2pt}
  \begin{subfigure}{\textwidth}
    \centering
    \includegraphics[width=0.95\linewidth]{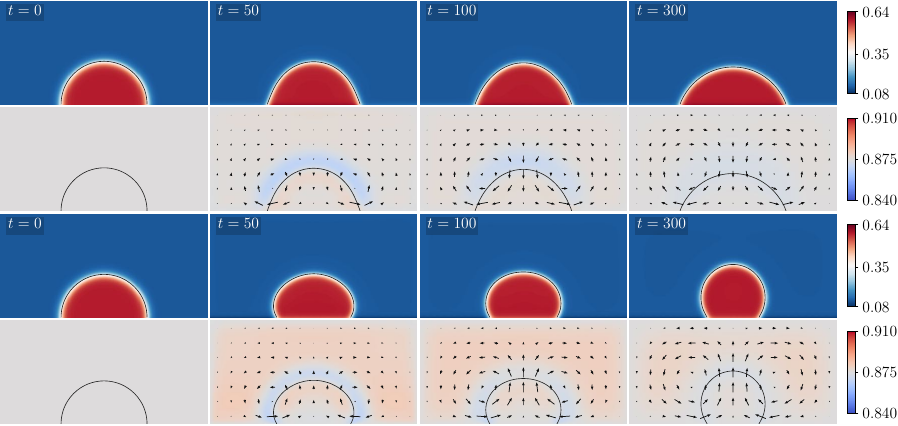}
    \caption{Reference cases, $T_w/T_c=0.875$.}
    \label{fig:wetting_2d_reference}
  \end{subfigure}
  \par\vspace{2pt}
  \begin{subfigure}{\textwidth}
    \centering
    \includegraphics[width=0.95\linewidth]{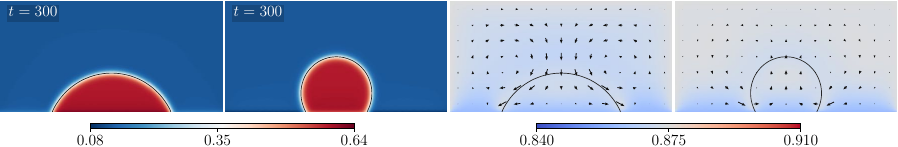}
    \caption{Cooled cases, $T_w/T_c=0.860$.}
    \label{fig:wetting_2d_cooled}
  \end{subfigure}
  \par\vspace{2pt}
  \begin{subfigure}{\textwidth}
    \centering
    \includegraphics[width=0.95\linewidth]{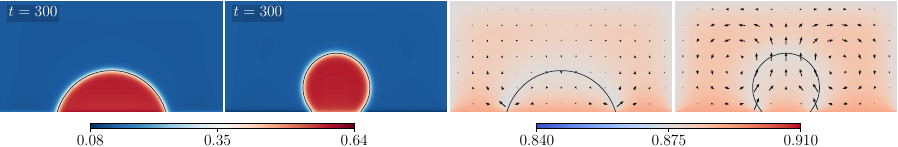}
    \caption{Heated cases, $T_w/T_c=0.890$.}
    \label{fig:wetting_2d_hot}
  \end{subfigure}
  \caption{Droplet evolution in two dimensions: (a) reference, (b) cooled and (c) heated cases.
  In (a), the first and second rows show the number density $n$ and temperature $T/T_c$ with velocity arrows on the hydrophilic substrate, respectively; the last two rows show the hydrophobic case.
  (b) and (c) show the same fields at $t=300$ on the cooled and heated substrates, respectively.
  Black curves mark $n=(n_g+n_l)/2$.}
\label{fig:wetting_2d_fields}
\end{figure}

As shown in \cref{fig:wetting_2d_reference}, the initially semicircular droplet spreads and becomes flatter on the hydrophilic substrate at the reference temperature $T_w/T_c=0.875$.
The liquid moves outward toward the advancing contact lines.
On the hydrophobic substrate, the contact region contracts as the contact lines recede, and the droplet becomes taller, accompanied by upward flow inside the droplet.
Velocity arrows are scaled separately in each snapshot.

Compared with the reference case, the cooled droplets ($T_w/T_c=0.860$) in \cref{fig:wetting_2d_cooled} have slightly wider contact regions at $t=300$.
Cooling enhances spreading on the hydrophilic substrate and reduces contact line recession on the hydrophobic substrate.
The temperature plots show a cooled region along the bottom wall extending into the adjacent fluid.

Heating produces the opposite change, as shown in \cref{fig:wetting_2d_hot}.
The contact regions are narrower than in the corresponding reference cases, although the hydrophilic droplet still spreads and the hydrophobic droplet retracts.
The fluid near the bottom wall becomes warmer, while a cooler region remains near the droplet surface.
The larger contact regions under cooling and smaller contact regions under heating are qualitatively consistent with the decrease in droplet contact area with increasing substrate temperature reported in \cite{xu2012droplet,wang2026jcp,qiao2026unified}.

\begin{figure}[!htbp]
  \centering
  \begin{subfigure}[t]{0.32\textwidth}
    \centering
    \includegraphics[width=\linewidth]{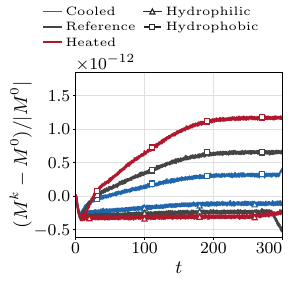}
    \caption{Mass error.}
    \label{fig:wetting_2d_mass}
  \end{subfigure}\hfill
  \begin{subfigure}[t]{0.30\textwidth}
    \centering
    \includegraphics[width=\linewidth]{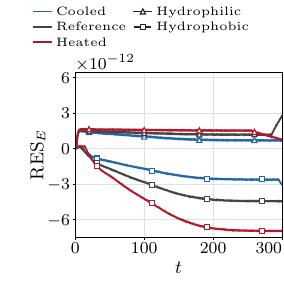}
    \caption{Energy balance residual.}
    \label{fig:wetting_2d_energy}
  \end{subfigure}\hfill
  \begin{subfigure}[t]{0.36\textwidth}
    \centering
    \includegraphics[width=\linewidth]{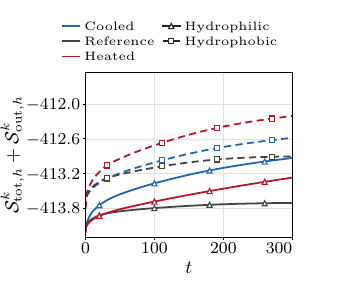}
    \caption{Entropy sum.}
    \label{fig:wetting_2d_entropy}
  \end{subfigure}
  \caption{Time evolution of mass, energy and entropy for all six two-dimensional droplet cases.}
  \label{fig:wetting_2d_balances}
\end{figure}

\begin{figure}[!htbp]
  \centering
  \begin{subfigure}[t]{0.32\textwidth}
    \centering
    \includegraphics[width=\linewidth]{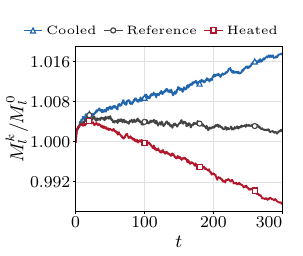}
    \caption{Hydrophilic substrate.}
    \label{fig:Hydrophilic substrate}
  \end{subfigure}\hfill
  \begin{subfigure}[t]{0.30\textwidth}
    \centering
    \includegraphics[width=\linewidth]{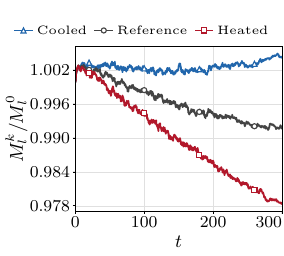}
    \caption{Hydrophobic substrate.}
    \label{fig:Hydrophobic substrate}
  \end{subfigure}\hfill
  \begin{subfigure}[t]{0.36\textwidth}
    \centering
    \includegraphics[width=\linewidth]{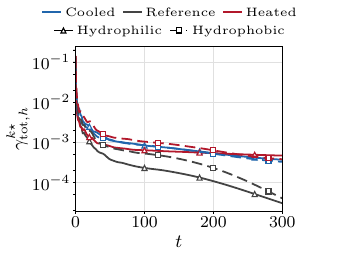}
    \caption{Entropy production.}
    \label{fig:Weighted entropy production}
  \end{subfigure}
  \caption{Time evolution of liquid mass ratio on the (a) hydrophilic and (b) hydrophobic substrates, and (c) entropy production for all six two-dimensional droplet cases.}
  \label{fig:wetting_2d_liquid_production}
\end{figure}

\cref{fig:wetting_2d_balances} shows that the relative mass errors and energy balance residuals remain on the order of $10^{-13}$--$10^{-12}$ in all six cases, with heat exchange included in the energy balance.
These results demonstrate the discrete mass conservation \eqref{eq:massbalance_fullydiscret} and energy balance in \cref{thm:fully_1st_law}.
The sum of the total entropy and the cumulative entropy transferred to the walls, denoted by $\mathcal S_{\mathrm{tot},h}^k+\mathcal S_{\mathrm{out},h}^k$, is also observed to increase monotonically in all six cases.
We also calculate the liquid mass using the density threshold $n>0.35$ and compare the ratio $M_l^k/M_l^0$ in \cref{fig:Hydrophilic substrate,fig:Hydrophobic substrate}, where $M_l^k\coloneqq\int_{\{n_h^k>0.35\}}n_h^k\rmd\bm x$.
We observe that, after the initial adjustment, heating (evaporation) reduces the liquid content, whereas cooling (condensation) produces more liquid than in the corresponding reference cases.
At $t=300$, the cooled droplets have approximately $1.74\%$ and $0.42\%$ more liquid mass than initially on the hydrophilic and hydrophobic substrates, respectively.
The heated droplets lose approximately $1.24\%$ and $2.15\%$, respectively.
This phenomenon is consistent with \cite{xu2012droplet,qiao2026unified}.
Notice that these changes concern the distribution of mass between the liquid and vapor phases, while the total mass remains conserved.
\cref{fig:Weighted entropy production} shows that the weighted entropy production $\gamma_{\mathrm{tot},h}^{k\star}$ decreases rapidly during the initial adjustment and remains positive at every time step in all six cases, consistent with \cref{thm:fully_2nd_law}.

We further simulate three-dimensional cooling on a hydrophilic substrate and heating on a hydrophobic substrate in $\Omega=(0,120)^2\times(0,60)$, with $d=3$ and the same initial profile centered at $\bm x_0=(60,60,0)$.
The wettability parameters and wall temperatures are the same as in the corresponding two-dimensional cases.
We use a tetrahedral mesh obtained from a $20\times20\times10$ partition, $r_n=r_s=r_u=1$, $\delta t=0.2$ and $t_f=300$.
The remaining parameters are unchanged.

\begin{figure}[!htbp]
  \centering
  \begin{subfigure}{\textwidth}
    \centering
    \includegraphics[width=\linewidth]{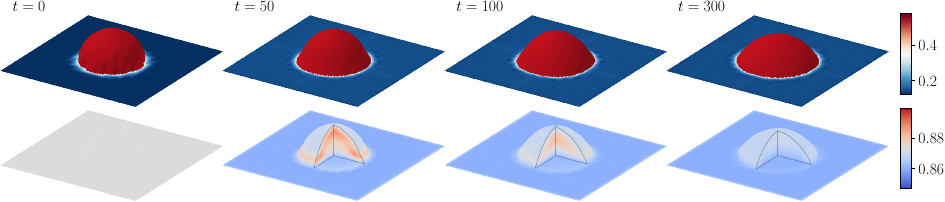}
    \caption{Cooled hydrophilic substrate.}
    \label{fig:wetting_3d_philic}
  \end{subfigure}
  \par\smallskip
  \begin{subfigure}{\textwidth}
    \centering
    \includegraphics[width=\linewidth]{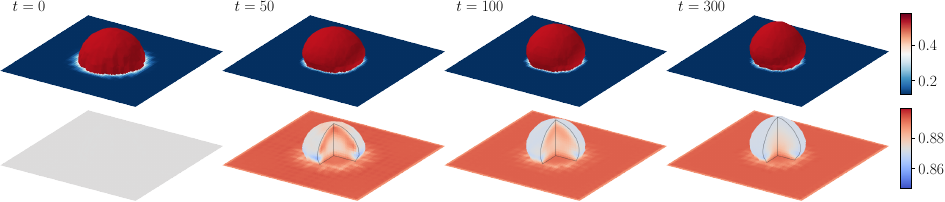}
    \caption{Heated hydrophobic substrate.}
    \label{fig:wetting_3d_phobic}
  \end{subfigure}
  \caption{Droplet evolution in three dimensions.
  The top and bottom rows of each subfigure show the number density $n$ and temperature $T/T_c$, respectively.}
  \label{fig:wetting_3d_fields}
\end{figure}

For visualization, the $n=0.35$ isosurface is reconstructed from locally averaged DG samples.
In the temperature plots, a quarter of the droplet facing the viewer is removed to expose the interior sections.
The three-dimensional results in \cref{fig:wetting_3d_philic,fig:wetting_3d_phobic} exhibit the same trends as the two-dimensional simulations.
On the cooled hydrophilic substrate, the footprint expands and the droplet flattens, while cooling spreads from the bottom wall into the liquid.
On the heated hydrophobic substrate, the contact footprint contracts and the droplet becomes taller, with warmer fluid near its base and a cooler outer surface.

\begin{figure}[!htbp]
  \centering
  \begin{subfigure}[t]{0.32\textwidth}
    \centering
    \includegraphics[width=\linewidth]{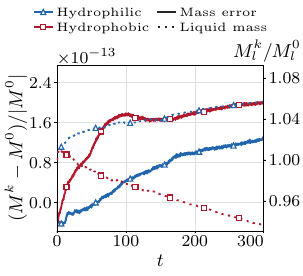}
    \caption{Mass error and $M_l^k/M_l^0$.}
    \label{fig:wetting_3d_mass}
  \end{subfigure}\hfill
  \begin{subfigure}[t]{0.30\textwidth}
    \centering
    \includegraphics[width=\linewidth]{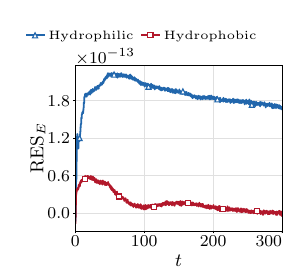}
    \caption{Energy balance residual.}
    \label{fig:wetting_3d_energy}
  \end{subfigure}\hfill
  \begin{subfigure}[t]{0.36\textwidth}
    \centering
    \includegraphics[width=\linewidth]{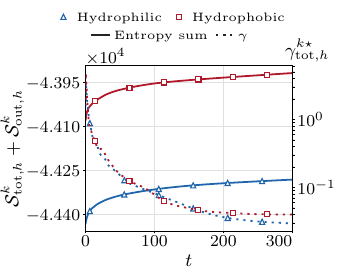}
    \caption{Entropy sum and production.}
    \label{fig:wetting_3d_entropy}
  \end{subfigure}
  \caption{Time evolution of mass, energy and entropy for the two three-dimensional droplet cases.}
  \label{fig:wetting_3d_balances}
\end{figure}

As shown in \cref{fig:wetting_3d_balances}, the liquid mass ratio reaches $1.057$ in the cooled case and $0.937$ in the heated case at $t=300$, consistent with condensation and evaporation, respectively.
The mass errors and energy balance residuals remain on the order of $10^{-14}$--$10^{-13}$, with heat exchange through all six walls included in the energy balance.
In both cases, the entropy sum increases monotonically and the weighted entropy production remains positive.

\subsection{Pool boiling}
\label{subsec:pool_boiling}

In this section, we examine the effects of substrate wettability and thermal resistance on bubble evolution in a two-dimensional pool boiling simulation \cite{xu2014single}.
The domain $\Omega=(0,200)\times(0,185)$ contains a lower liquid pool, a vapor layer and a top liquid layer of heights $85$, $90$ and $10$, respectively.
Following \cite{xu2014single}, we introduce this liquid--vapor--liquid arrangement to maintain the supersaturated state of the liquid surrounding the bubble.
A semicircular vapor bubble of radius $25$ is initially centered at $(100,0)$, with hyperbolic tangent profiles of width parameter $2.5$ across initial liquid--vapor interfaces.
We set $\bm u^0=\bm0$, $T^0=0.84T_c$, $n_g= 0.100667$, $n_l= 0.611279$, $t_f=6000$ and downward gravity $\rmG=10^{-4}$.
The remaining physical parameters follow \cref{subsec:thermal_wetting}.

All walls are stationary and impermeable.
The bottom and top wall temperatures are prescribed as $0.90T_c$ and $0.775T_c$, respectively, from $t=0$, with heat exchange governed by the finite thermal resistance condition.
The lateral walls have no external heat exchange, with the thermal coupling coefficient $\rmLk^{-1}$ set to zero.
On the bottom wall, $\rmWs=-0.154906$ and $0.039252$ define the hydrophilic and hydrophobic cases, respectively.
The corresponding planar equilibrium liquid contact angles are approximately $41^\circ$ and $101^\circ$ at $T^0$ and without gravity.
We set $\rmWs=0$ on the other walls.
We use a $50\times50$ partition subdivided into triangles, $r_n=r_s=r_u=2$ and $\delta t=0.5$.

\begin{figure}[!htbp]
  \centering
  \begin{subfigure}{\textwidth}
    \centering
    \includegraphics[width=\linewidth]{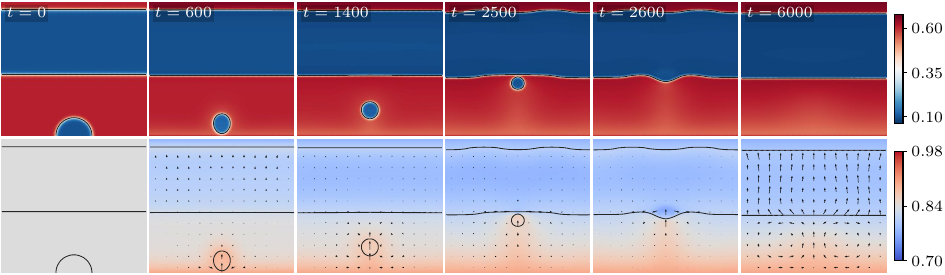}
    \caption{Hydrophilic substrate.}
    \label{fig:boiling_hydrophilic}
  \end{subfigure}
  \begin{subfigure}{\textwidth}
    \centering
    \includegraphics[width=\linewidth]{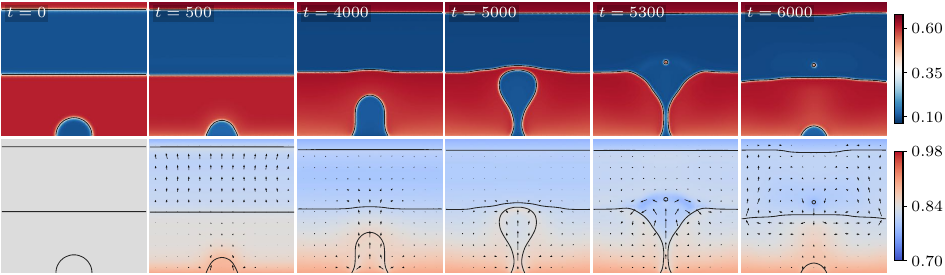}
    \caption{Hydrophobic substrate.}
    \label{fig:boiling_hydrophobic}
  \end{subfigure}
  \caption{The evolution of pool boiling on (a) hydrophilic and (b) hydrophobic substrates.
  The top and bottom rows of each subfigure show the number density $n$ and temperature $T/T_c$ with velocity arrows, respectively.}
  \label{fig:boiling_wettability}
\end{figure}

On the hydrophilic substrate (\cref{fig:boiling_hydrophilic}), the bubble contact region contracts during the initial adjustment, and the bubble has detached and risen into the liquid by $t=1400$.
Upward flow accompanies its rise, with a warm region around the bubble.
It merges with the upper vapor layer by $t=2600$.
The deformed interface gradually returns to a nearly flat profile at $t=6000$.

On the hydrophobic substrate (\cref{fig:boiling_hydrophobic}), the bubble remains attached as it grows and develops an elongated neck by $t=5000$.
At $t=5300$, the bubble joins the upper vapor layer, forming a continuous vapor connection to the bottom wall.
This connection then breaks, leaving a smaller attached bubble at $t=6000$.
The longer attachment on the hydrophobic substrate is qualitatively consistent with the wettability trend reported in \cite{xu2014single}.

\begin{figure}[!htbp]
  \centering
  \begin{subfigure}[t]{0.32\textwidth}
    \centering
    \includegraphics[width=\linewidth]{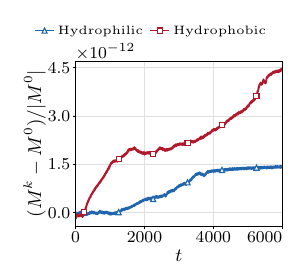}
    \caption{Mass error.}
    \label{fig:boiling_mass}
  \end{subfigure}\hfill
  \begin{subfigure}[t]{0.30\textwidth}
    \centering
    \includegraphics[width=\linewidth]{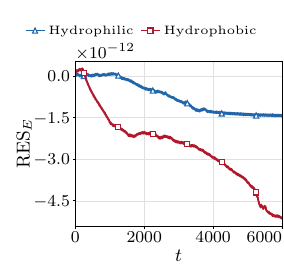}
    \caption{Energy balance residual.}
    \label{fig:boiling_energy}
  \end{subfigure}\hfill
  \begin{subfigure}[t]{0.36\textwidth}
    \centering
    \includegraphics[width=\linewidth]{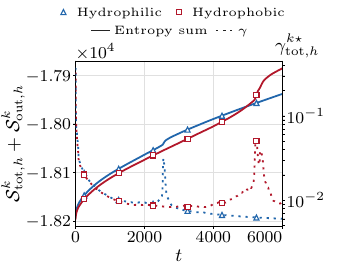}
    \caption{Entropy sum and production.}
    \label{fig:boiling_entropy}
  \end{subfigure}
  \caption{Time evolution of mass, energy and entropy for the two pool-boiling cases.}
  \label{fig:boiling_balances}
\end{figure}

\cref{fig:boiling_balances} shows the evolution of the relative mass errors, energy balance residuals and entropy sum and production in both cases.
The energy residual accounts for gravitational potential energy and heat exchanged with walls.
These results verify the discrete mass and energy balances during bubble detachment, coalescence and breakup.
The weighted entropy production remains positive, consistent with \cref{thm:fully_2nd_law}, and exhibits peaks near $t=2600$ and $t=5300$ in the hydrophilic and hydrophobic cases, respectively, coinciding with the rapid interface changes shown in \cref{fig:boiling_wettability}.
The entropy sum is also observed to increase at every time step in both cases.

\begin{figure}[!htbp]
  \centering
  \includegraphics[width=\linewidth]{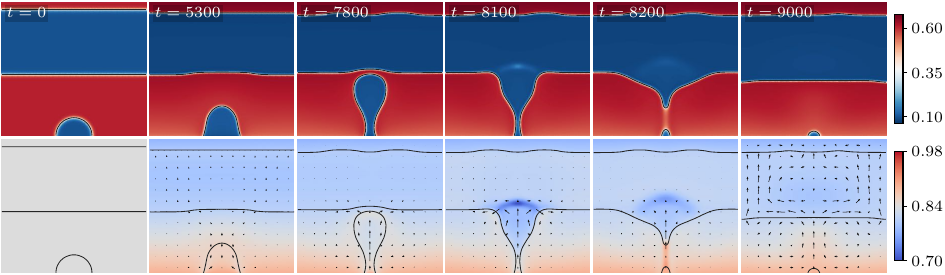}
  \caption{The evolution of pool boiling on the hydrophobic substrate, with the bottom thermal resistance increased tenfold.
  The top and bottom rows show the number density $n$ and temperature $T/T_c$ with velocity arrows, respectively.}
  \label{fig:boiling_high_Lk}
\end{figure}

To assess the effect of thermal resistance, we repeat the hydrophobic case with $\rmLk=10/(n_lT_c^2)$ on the bottom wall and $t_f=9000$, keeping all other parameters unchanged.
In \cref{fig:boiling_high_Lk}, the bubble remains attached and well below the upper vapor layer at $t=5300$, when the baseline case in \cref{fig:boiling_hydrophobic} has already formed a vapor connection.
With the larger resistance, the connection is present by $t=8100$ and has broken by $t=8200$, leaving a smaller bubble on the bottom wall.
The increased thermal resistance reduces heat transfer from the bottom wall, consistent with the slower bubble growth and delayed connection with the upper vapor layer.

\subsection{Pulse-driven droplet ejection}
\label{subsec:nozzle_ejection}

Filament pinch-off and drop formation have been investigated using phase-field methods \cite{yang2006numerical}.
In this example, we consider a slit nozzle connecting the upper reservoir $(-20,20)\times(24,40)$ to the receiving chamber $(-28,28)\times(-48,0)$ through the slit $(-6,6)\times(0,24)$.
All exterior walls are stationary and impermeable.
Initially, $\bm u^0=\bm0$, $T^0=0.84T_c$, $n_g=0.100667$ and $n_l=0.611279$.
The liquid column spans $6<y<33$ at $t=0$.

On both inner slit walls, we set $\rmWs=-0.102233$ (hydrophilic) or $0.103934$ (hydrophobic), and prescribe $T_w=0.84T_c$ for the two reference cases, $T_w=0.80T_c$ for the cooled hydrophilic case and $T_w=0.88T_c$ for the heated hydrophobic case.
We set $\rmWs=0$ and $T_w=0.84T_c$ for the remaining walls.
We use $\rmRe=2.5$, $\rmR=0.024$, $\rmG=0$, $\rmLb=250$, $\rmLs=0.8/n_l$, $\rmLk=0.4/(n_lT_c^2)$, $\rmLd=0.005$, $r_n=r_s=r_u=1$, $h=\sqrt{2}$, $\delta t=0.25$ and $t_f=130$.
The remaining coefficients follow \cref{subsec:thermal_wetting}.
The driving body force is $\bm f_{\mathrm{dri}}=-\rmRe\rmR n a(t)\sin^2(\pi y/24)\bm e_2$ inside the slit and vanishes elsewhere, where the pulse amplitude $a(t)$ is defined as
\begin{equation*}
  a(t)=
  \begin{cases}
    0.23\sin^2 (\pi(t-8)/40) & 8<t<48,\\
    -0.17\sin^2 (\pi(t-48)/8) & 48<t<56,\\
    0 & \text{otherwise}.
  \end{cases}
\end{equation*}

\begin{figure}[!htbp]
  \centering
  \captionsetup[subfigure]{skip=2pt}
  \begin{subfigure}{\textwidth}
    \centering
    \includegraphics[width=\linewidth]{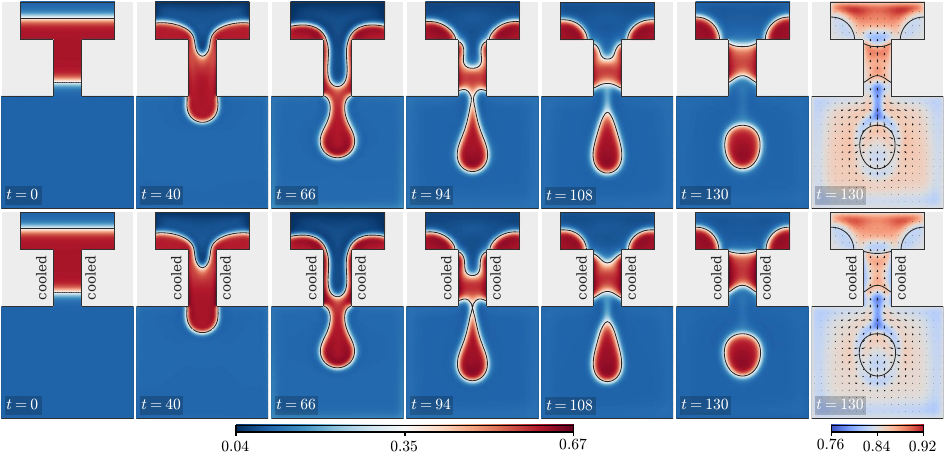}
    \caption{Hydrophilic cases.}
    \label{fig:nozzle_hydrophilic_fields}
  \end{subfigure}
  \par\vspace{3pt}
  \begin{subfigure}{\textwidth}
    \centering
    \includegraphics[width=\linewidth]{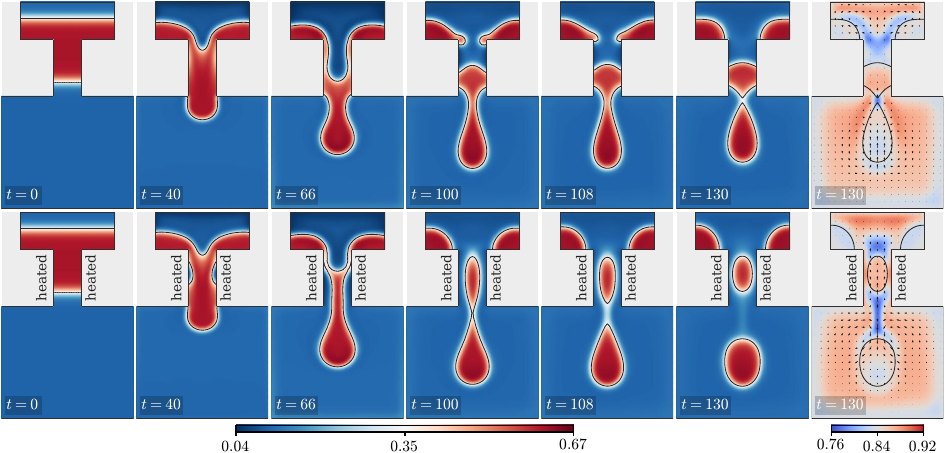}
    \caption{Hydrophobic cases.}
    \label{fig:nozzle_hydrophobic_fields}
  \end{subfigure}
  \caption{The evolution of pulse-driven droplet ejection in a two-dimensional hydrophilic (a) and hydrophobic (b) slit nozzle.
  In each subfigure, the top row is the reference case ($T_w/T_c=0.84$), while the bottom row is cooled ($T_w/T_c=0.80$) in (a) and heated ($T_w/T_c=0.88$) in (b).
  The first six columns show the number density $n$, and the last column shows final-time $T/T_c$ with velocity arrows.}
  \label{fig:nozzle_2d_fields}
\end{figure}

\cref{fig:nozzle_2d_fields} shows liquid extrusion followed by ligament thinning and pinch-off after the driving pulse has ended.
As we can see, at the reference wall temperature, the hydrophobic case retains a longer-lived ligament than the hydrophilic case.
The liquid remains attached to the slit walls longer in the hydrophilic case than in the hydrophobic case.

\begin{figure}[!htbp]
  \centering
  \begin{subfigure}[t]{0.32\textwidth}
    \centering
    \includegraphics[width=\linewidth]{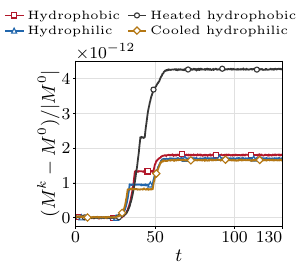}
    \caption{Mass error.}
    \label{fig:nozzle_mass}
  \end{subfigure}\hfill
  \begin{subfigure}[t]{0.30\textwidth}
    \centering
    \includegraphics[width=\linewidth]{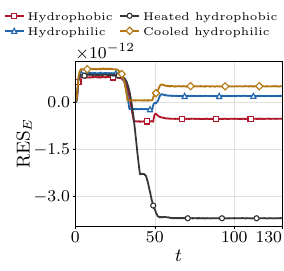}
    \caption{Energy balance residual.}
    \label{fig:nozzle_energy}
  \end{subfigure}\hfill
  \begin{subfigure}[t]{0.36\textwidth}
    \centering
    \includegraphics[width=\linewidth]{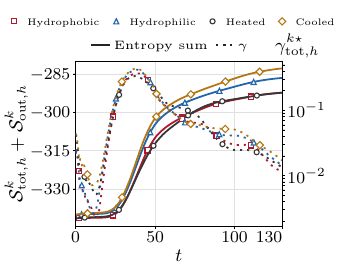}
    \caption{Entropy sum and production.}
    \label{fig:nozzle_entropy}
  \end{subfigure}
  \caption{Time evolution of mass, energy and entropy for the four two-dimensional nozzle cases.}
  \label{fig:nozzle_balances}
\end{figure}

With cooled hydrophilic walls (\cref{fig:nozzle_hydrophilic_fields}), pinch-off still occurs near $t=95$.
At $t=130$, the cooled case retains a larger liquid segment attached to both slit walls than the reference case.
Cooling lowers the liquid temperature near the slit walls, while the interior of the retained liquid remains warmer.
In contrast, heating the hydrophobic walls (\cref{fig:nozzle_hydrophobic_fields}) raises the fluid temperature near the slit walls and causes the retained liquid inside the slit to detach from both walls, forming an isolated small droplet.
The main droplet pinches off near $t=101$, earlier than in the hydrophobic reference case.
At $t=130$, the main drop is more rounded and lies farther below the outlet than in the reference case.
\cref{fig:nozzle_balances} shows the relative mass error, energy balance residual, entropy sum and production, where the work $W_h^k$ uses $\bm f_h^{j+1/2}=\bm f_{\mathrm{dri},h}^{j+1/2}$.

We also simulate the  cooled hydrophilic and heated hydrophobic cases in a three-dimensional coaxial nozzle aligned with the $z$-axis.
The upper reservoir ($24<z\leq 40$), bore ($0<z\leq 24$) and receiving chamber ($-48\leq z\leq 0$) have radii $20$, $6$ and $28$, respectively, with the bore outlet at $z=0$.
We use $d=3$, $r_n=r_s=r_u=1$ and $\delta t=0.1$ on a symmetric layered tetrahedral mesh with $h\approx4.70$.
The driving body force acts along the nozzle axis and is given by $\bm f_{\mathrm{dri}}=-1.5\rmRe\rmR na(t)\sin^2(\pi z/24)\bm e_3$ inside the cylindrical bore and zero elsewhere.

\begin{figure}[!htbp]
  \centering
  \includegraphics[width=\linewidth]{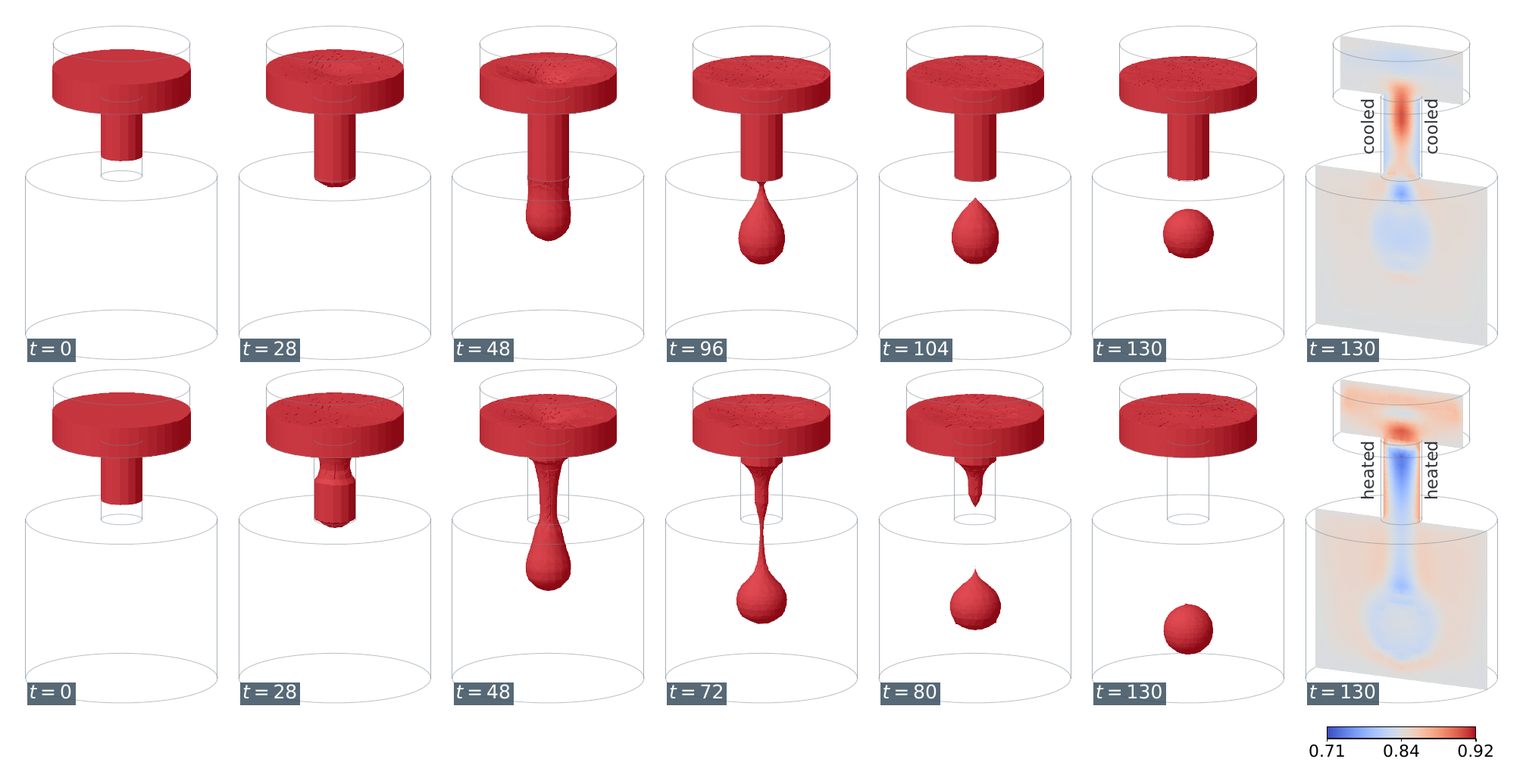}
  \caption{The evolution of pulse-driven droplet ejection in a three-dimensional coaxial nozzle with cooled hydrophilic walls (top) and heated hydrophobic walls (bottom).
  The first six columns show the liquid region $n\geq(n_g+n_l)/2$, and the last column shows $T/T_c$ on the central plane $y=0$.}
  \label{fig:nozzle_3d_fields}
\end{figure}

\begin{figure}[!htbp]
  \centering
  \begin{subfigure}[t]{0.32\textwidth}
    \centering
    \includegraphics[width=\linewidth]{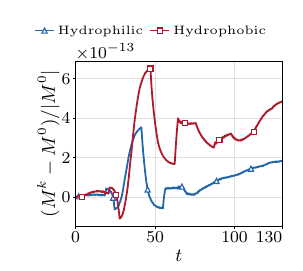}
    \caption{Mass error.}
    \label{fig:nozzle_3d_mass}
  \end{subfigure}\hfill
  \begin{subfigure}[t]{0.30\textwidth}
    \centering
    \includegraphics[width=\linewidth]{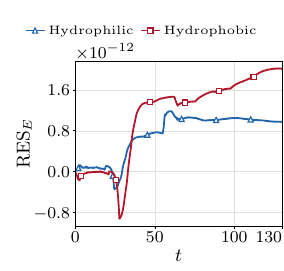}
    \caption{Energy balance residual.}
    \label{fig:nozzle_3d_energy}
  \end{subfigure}\hfill
  \begin{subfigure}[t]{0.36\textwidth}
    \centering
    \includegraphics[width=\linewidth]{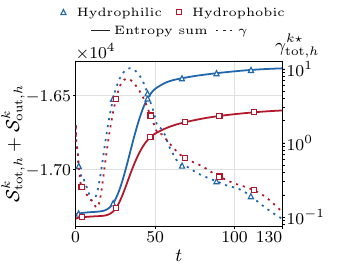}
    \caption{Entropy sum and production.}
    \label{fig:nozzle_3d_entropy}
  \end{subfigure}
  \caption{Time evolution of mass, energy and entropy for the two three-dimensional nozzle cases.}
  \label{fig:nozzle_3d_balances}
\end{figure}

As we can see in \cref{fig:nozzle_3d_fields}, both cases exhibit ligament thinning and pinch-off.
A liquid column remains attached to the bore wall in the cooled hydrophilic case, whereas no liquid remains attached to the bore wall in the heated hydrophobic case.
The temperature sections show cooler fluid near the cooled wall and warmer fluid near the heated wall in each case.
We note that the comparison is only qualitatively consistent with the two-dimensional results: setting $d=3$ alters the volumetric heat capacity $dn/2$ and viscous stress of the model, and the nozzle geometry and pulse amplitudes also differ.
\cref{fig:nozzle_3d_balances} presents the relative mass error, energy balance residual, the entropy sum and production.

\section{Conclusions}
\label{sec:conclusion}

In conclusion, we have developed two thermodynamically consistent variational formulations for non-isothermal liquid--vapor flows with moving contact lines.
The internal energy and Helmholtz free energy formulations extend the generalized Hamilton's principle for nonequilibrium thermodynamics to include density gradients and surface thermodynamics.
With a specified reversible entropy flux and stress coupling, the two variational formulations yield the same bulk equations and hydrodynamic boundary conditions, using entropy density and temperature, respectively, as the thermal state variable.
The boundary conditions account for number density relaxation, velocity slip, surface heat conduction, and heat exchange with the walls.

The weak forms of the dimensionless equations guide the construction of two fully discrete finite element schemes.
Compatible spatial operators and exact discrete chain rules are constructed to preserve the cancellation of energy exchange terms between the discrete kinetic and internal energy balances.
Both schemes strictly conserve mass and satisfy an exact total energy balance including external work and heat exchange, as well as a temperature-weighted entropy balance with nonnegative bulk and boundary production.
The number density bounds and temperature positivity are also preserved.
The density bounds follow from the logit transformation, while temperature positivity follows from entropy-based recovery in the $S$-scheme and the logarithmic temperature variable in the $T$-scheme.

Numerical experiments with manufactured solutions demonstrate the spatial and temporal accuracy of both schemes.
Additional tests near the density bounds and at low temperatures assess the preservation of admissible states.
Simulations of non-isothermal droplet wetting, pool boiling and droplet ejection in two and three dimensions illustrate the applicability of the proposed approach.
The numerical results confirm mass conservation, the energy balance accounting for heat exchange and external work, and nonnegative entropy production.
The sum of the discrete total entropy and the cumulative entropy transferred to the walls is also observed to increase monotonically in the simulations.
Future work will focus on developing linear, decoupled schemes that preserve the thermodynamic structure and establishing rigorous error estimates for the proposed schemes.

\subsection*{Acknowledgments}
This research is partially sponsored by the National Natural Science Foundation of China (No.12371434, No. 12531015, No. 12671515, No. U25A20200), the National Key R \& D Program of China (No.2022YFE03040002), Sichuan University Research Project on Teaching Reform in Graduate Education (No. SCYZTD2025003). 
\appendix
\numberwithin{equation}{section}
\section{Proofs of the variational formulations}
\label{app:variational-proofs}

\subsection{Proof of Proposition~\ref{prop:variational-proposition}}
\label{app:variational-proof}

\begin{proof}
The mass equation \eqref{eq:physical-bulk-mass} follows from the push-forward relation.
Indeed, differentiating $n(\mathcal X_t(\bm X),t)J_{\mathcal X}(\bm X,t) =n_{\mathrm{ref}}(\bm X)$ with respect to time $t$ and using $\partial_t J_{\mathcal X} =J_{\mathcal X}(\operatorname{div}\bm u)\circ\mathcal X_t$ gives $\overline{\mathrm D}_tn=0$.
The variation of bulk internal energy yields $\delta\int_\Omega\hat e\rmd\bm x =\int_\Omega (T\delta\hat S+\tilde\mu\delta n)\rmd\bm x +\int_\Gamma M\nabla_\nu n\delta n\rmd A$.
Using the kinematic relation for $\delta\bm u$, we also have $\delta(\mathrm D_t\Upsilon) =\mathrm D_t(\delta\Upsilon)+\delta\bm u\cdot\nabla\Upsilon =\mathrm D_t(\mathrm D_\delta\Upsilon) -\bm v\cdot\nabla(\mathrm D_t\Upsilon)$.
Using this identity, $\delta n=-\operatorname{div}(n\bm v)$ and $\delta\Sigma=\overline{\mathrm D}_\delta\Sigma -\operatorname{div}(\Sigma\bm v)$, we substitute the bulk variational constraint \eqref{eq:bulk-constraints2} into the first variation of \eqref{eq:action}.
After integration by parts, the bulk part is
\begin{equation}\label{eq:appendix-bulk-first-variation}
\begin{aligned}
 \int_{t_0}^{t_1}\int_\Omega
  \Bigl\{&(\mathrm D_t\Upsilon-T)\delta\hat S+\Bigl(\overline{\mathrm D}_t\Sigma
       -\overline{\mathrm D}_t\hat S
       -\operatorname{div}\Bigl(
          \frac{\mathrm D_t\Upsilon}{T}\bm j\Bigr)\Bigr)
          \mathrm D_\delta\Upsilon\\
 &+\Bigl(-mn\mathrm D_t\bm u-n\nabla\tilde\mu
       -\hat S\nabla(\mathrm D_t\Upsilon)
       +\operatorname{div}\Bigl(
          \frac{\mathrm D_t\Upsilon}{T}
          (\bm\sigma+\bm R_{\mathrm{th}})\Bigr)
       +\bm f_{\mathrm{ex}}\Bigr)\cdot\bm v\Bigr\}
  \rmd\bm x\rmd t.
\end{aligned}
\end{equation}
The endpoint conditions $\bm v(t_i)=0$ and $\delta\Upsilon(t_i)=0$ imply $\mathrm D_\delta\Upsilon(t_i)=0$ and $\mathrm D_\delta^\Gamma\Upsilon(t_i)=0$, $i=0,1$.
The temporal boundary terms therefore vanish.

Stationarity with respect to $\delta\hat S$ in \eqref{eq:appendix-bulk-first-variation} requires $\mathrm D_t\Upsilon=T$ in $\Omega$ and, by taking the trace, $\mathrm D_t^\Gamma\Upsilon=T$ on $\Gamma$.
Hence $\mathrm D_t\Upsilon/T=1$ and $\bm R_{\mathrm{th}}=\bm R_{\mathrm{rev}}$.
Since $\bm v$ and $\mathrm D_\delta\Upsilon$ can also be chosen independently with compact support in $\Omega$, the coefficients in \eqref{eq:appendix-bulk-first-variation} vanish.
Using $\operatorname{div}\widetilde{\bm\Pi}=n\nabla\tilde\mu+\hat S\nabla T$ and $\bm\Pi=\widetilde{\bm\Pi}-\bm R_{\mathrm{rev}}$, we obtain the momentum equation \eqref{eq:physical-bulk-m} and
\begin{equation}\label{eq:bulk-aux-entropy}
 \overline{\mathrm D}_t\Sigma
 =\overline{\mathrm D}_t\hat S+\operatorname{div}\bm j.
\end{equation}
Combining \eqref{eq:bulk-aux-entropy} with the bulk phenomenological constraint \eqref{eq:bulk-constraints1} yields \eqref{eq:physical-bulk-S}.

We next consider the boundary terms.
From the first variation of \eqref{eq:action}, the gradient energy term and the variations of $e_s$ and $\sigma_s$ combine into $-(M\nabla_\nu n+\partial_n e_s -T\partial_n\sigma_s)\delta n=-L\delta n =-L\mathrm D_\delta^\Gamma n +L\nabla_\tau n\cdot\bm v_\tau$.
We write $\delta\Sigma_\Gamma =\overline{\mathrm D}_\delta^\Gamma\Sigma_\Gamma -\operatorname{div}_\Gamma(\Sigma_\Gamma\bm v_\tau)$ and integrate by parts on the closed surface $\Gamma$ to obtain
\begin{equation*}
 \int_\Gamma
 (-T\delta\Sigma_\Gamma
       -(\sigma_s-\Sigma_\Gamma)
        \bm v_\tau\cdot\nabla_\tau T)\rmd A=
 \int_\Gamma
 (-T\overline{\mathrm D}_\delta^\Gamma\Sigma_\Gamma
       -\sigma_s\bm v_\tau\cdot\nabla_\tau T)\rmd A.
\end{equation*}
Since $\bm R_{\mathrm{rev}}$ is isotropic, we have $\bm P_\tau\bm R_{\mathrm{rev}}\bm\nu=0$.
We combine these terms with the stress and entropy flux boundary terms, substitute the surface variational constraint \eqref{eq:wall-constraints2}, and integrate by parts in time and on $\Gamma$ to obtain the boundary part of the first variation
\begin{equation}\label{eq:wall-first-variation}
\begin{aligned}
 \int_{t_0}^{t_1}\int_\Gamma
  \Bigl\{&-(L+f_r)\mathrm D_\delta^\Gamma n+
 (L\nabla_\tau n-\sigma_s\nabla_\tau T
       -\bm P_\tau\bm\sigma\bm\nu-\bm f_w)
       \cdot\bm v_\tau\\
 &+
 (\bm j\cdot\bm\nu
       -\overline{\mathrm D}_t^\Gamma(\sigma_s-\Sigma_\Gamma)
       -\operatorname{div}_\Gamma\bm j_s-j_w)
       \mathrm D_\delta^\Gamma\Upsilon\Bigr\}
  \rmd A\rmd t.
\end{aligned}
\end{equation}
The condition $\bm v\cdot\bm\nu=0$ leaves $\bm\nu\cdot(\nabla\bm v)\bm\nu$ unconstrained.
Since $n>0$, \eqref{eq:wall-rates} allows $\mathrm D_\delta^\Gamma n$ and $\bm v_\tau$ to be prescribed independently.
The thermal variation $\mathrm D_\delta^\Gamma\Upsilon$ is also independent.
Thus the three coefficients in \eqref{eq:wall-first-variation} vanish independently, yielding the first two relations in \eqref{eq:physical-wall} and
\begin{equation}\label{eq:surface-aux-entropy}
 \overline{\mathrm D}_t^\Gamma\Sigma_\Gamma
 =\overline{\mathrm D}_t^\Gamma\sigma_s
  +\operatorname{div}_\Gamma\bm j_s
  -\bm j\cdot\bm\nu+j_w.
\end{equation}
Similarly, substituting \eqref{eq:surface-aux-entropy} into the surface phenomenological constraint \eqref{eq:wall-constraints1} gives \eqref{eq:physical-wall_T}.
\end{proof}
For a polygonal or polyhedral boundary $\Gamma$, surface integration by parts is performed facewise.
We assume common scalar traces, continuous velocity and virtual displacement fields tangent to each incident face, and balanced surface entropy fluxes at facet interfaces.
Under these assumptions, the resulting interface terms cancel.

\subsection{Proof of Proposition~\ref{prop:free-energy-equivalence}}
\label{app:free-energy-proof}

\begin{proof}
The mass equation follows as in Appendix~\ref{app:variational-proof}.
We use the same kinematic variations and temporal endpoint conditions.
Using $\partial_T\hat f=-\hat S$, $\partial_{\nabla n}\hat f=M\nabla n$ and $\tilde\mu=\partial_n\hat f-\operatorname{div}(M\nabla n)$, we express the variation of the bulk free energy as $\delta\int_\Omega \hat f\rmd\bm x = \int_\Omega ( \tilde\mu\delta n -\hat S\delta T )\rmd\bm x + \int_\Gamma M\nabla_\nu n\delta n\rmd A$.
Hence the bulk part of the first variation of \eqref{eq:free-energy-action} is
\begin{equation*}
\int_{t_0}^{t_1}\int_\Omega
  \Bigl\{mn\bm u\cdot\delta\bm u
    +\Bigl(\frac{m}{2}|\bm u|^2-\tilde\mu\Bigr)\delta n
    +(\hat S-\Sigma)\delta T
    -T\delta\Sigma
    +\bm f_{\mathrm{ex}}\cdot\bm v
  \Bigr\}
  \rmd\bm x\rmd t.
\end{equation*}
Here $\delta T =\mathrm D_t(\mathrm D_\delta\Upsilon)-\bm v\cdot\nabla T$.
Using this relation together with $\delta n=-\operatorname{div}(n\bm v)$ and $\delta\Sigma=\overline{\mathrm D}_\delta\Sigma -\operatorname{div}(\Sigma\bm v)$, we substitute \eqref{eq:free-energy-constraints2} and integrate by parts.
The resulting bulk integrand is
\begin{equation*}
  (-mn\mathrm D_t\bm u
  -n\nabla\tilde\mu
  -\hat S\nabla T
  +\operatorname{div}
  (\bm\sigma+\bm R_{\mathrm{rev}})
  +\bm f_{\mathrm{ex}})\cdot\bm v
  -(\overline{\mathrm D}_t(\hat S-\Sigma)
  +\operatorname{div}\bm j)\mathrm D_\delta\Upsilon.
\end{equation*}
Setting the coefficients of the independent interior variations $\bm v$ and $\mathrm D_\delta\Upsilon$ to zero, and using the stress identities from Appendix~\ref{app:variational-proof}, we obtain \eqref{eq:physical-bulk-m} and
\begin{equation}\label{eq:free-energy-aux-entropy}
  \overline{\mathrm D}_t\Sigma
  =
  \overline{\mathrm D}_t\hat S
  +
  \operatorname{div}\bm j
  =
  -\overline{\mathrm D}_t(\partial_T\hat f)
  +
  \operatorname{div}\bm j.
\end{equation}
Substituting \eqref{eq:free-energy-aux-entropy} into the phenomenological constraint \eqref{eq:free-energy-constraints1} yields
\begin{equation*}
  T(
    -\overline{\mathrm D}_t(\partial_T\hat f)
    +\operatorname{div}\bm j
  )
  =
  (\bm\sigma+\bm R_{\mathrm{rev}}):\nabla\bm u
  -
  \bm j\cdot\nabla T.
\end{equation*}
Using $\bm j=\bm j_{\mathrm{irr}}+\bm j_{\mathrm{rev}}$ and $\bm R_{\mathrm{rev}}:\nabla\bm u=\bm j_{\mathrm{rev}}\cdot\nabla T$, the reversible contributions cancel.
Since $\bm q=T\bm j_{\mathrm{irr}}$, we obtain \eqref{eq:free-temperature-balance}.

The bulk gradient term and the surface terms in the action contribute the boundary integrand
\begin{equation}\label{eq:wall-first-variation_bcd}
  -M\nabla_\nu n\delta n
  -(\partial_n f_s)\delta n
  +(\sigma_s-\Sigma_\Gamma)\delta T
  -T\delta\Sigma_\Gamma=-L\delta n+(\sigma_s-\Sigma_\Gamma)\delta T
  -T\delta\Sigma_\Gamma,
\end{equation}
where $\partial_T f_s=-\sigma_s$.
Moreover, on $\Gamma$, we have $\delta T = \mathrm D_t^\Gamma (\mathrm D_\delta^\Gamma\Upsilon) - \bm v_\tau\cdot\nabla_\tau T$ and $\delta\Sigma_\Gamma = \overline{\mathrm D}_\delta^\Gamma\Sigma_\Gamma - \operatorname{div}_\Gamma (\Sigma_\Gamma\bm v_\tau)$.
We substitute these relations and the surface variational constraint \eqref{eq:wall-constraints2} into \eqref{eq:wall-first-variation_bcd} and integrate by parts in time and on $\Gamma$.
Adding the boundary contributions from the bulk stress and entropy flux gives \eqref{eq:wall-first-variation}.
The first two relations in \eqref{eq:physical-wall} and \eqref{eq:surface-aux-entropy} follow by the same independence argument.
Combining the latter with the surface phenomenological constraint \eqref{eq:wall-constraints1} yields \eqref{eq:physical-wall_T}.
\end{proof}

\bibliographystyle{plainnat-nodoi}
\bibliography{references}
\end{document}